\RequirePackage{silence}
\RequirePackage{fix-cm}
\PassOptionsToPackage{super}{natbib}
\PassOptionsToPackage{hyphens}{url}
\documentclass[sn-nature,iicol]{sn-jnl}

\usepackage{graphicx}
\usepackage{multirow}
\usepackage{amsmath,amssymb,amsfonts}
\usepackage{booktabs}
\usepackage{xcolor}
\usepackage{textcomp}
\usepackage{manyfoot}
\usepackage{array}
\usepackage{placeins}
\usepackage{needspace}
\let\oldsubsection\subsection
\renewcommand{\subsection}{\Needspace*{5\baselineskip}\oldsubsection}
\usepackage{afterpage}

\makeatletter
\expandafter\let\expandafter\snjnltablestar\csname table*\endcsname
\expandafter\let\expandafter\endsnjnltablestar\csname endtable*\endcsname
\renewenvironment{table*}[1][]%
{\begin{snjnltablestar}[#1]%
\begin{center}
\begin{threeparttable}
\tablebodyfont%
\renewcommand\footnotetext[2][]{{\removelastskip\vskip3pt%
\let\tablebodyfont\tablefootnotefont%
\hskip0pt\if!##1!\else{\smash{$^{##1}$}}\fi##2\par}}%
}{\end{threeparttable}\end{center}\end{snjnltablestar}}
\makeatother

\usepackage{microtype}
\DeclareMathSizes{12bp}{12bp}{8}{6}
\DeclareMathSizes{11bp}{11bp}{8}{6}
\DeclareMathSizes{10bp}{10bp}{7}{5}
\DeclareMathSizes{9bp}{9bp}{7}{5}
\DeclareMathSizes{8bp}{8bp}{6}{5}
\newcommand{\pbrain}{\mathbf{p}_{\mathrm{brain}}}
\newcommand{\acc}{\mathrm{acc}}

\hypersetup{pdftitle={The Platonic brain bridge hypothesis: brain networks as architectural priors for multimodal large language models},pdfauthor={Pengfei Zhang, Biao Tian, Xiangang Li, Li Liu}}

\begin{document}

\title[The Platonic brain bridge hypothesis]{The Platonic brain bridge hypothesis: brain networks as architectural priors for multimodal large language models}

\author[1]{\fnm{Pengfei} \sur{Zhang}}
\author[2]{\fnm{Biao} \sur{Tian}}
\author[2]{\fnm{Xiangang} \sur{Li}}
\author*[1]{\fnm{Li} \sur{Liu}}\email{avrillliu@hkust-gz.edu.cn (ORCID: 0000-0002-4497-0135)}

\affil[1]{\orgdiv{AI Thrust, Information Hub}, \orgname{The Hong Kong University of Science and Technology (Guangzhou)}, \orgaddress{\city{Guangzhou}, \country{China}}}
\affil[2]{\orgdiv{Alibaba Token Hub}, \orgname{Alibaba Group}, \orgaddress{\city{Beijing}, \country{China}}}

\abstract{Multimodal large language models predict brain activity, but alignment has remained a measurement, not a design tool. We propose the Platonic brain bridge hypothesis: omni models, which jointly process video, audio and text, share brain-like representations usable in both directions. From model to brain, seven omni models show stable brain-likeness across participants, and our encoding experts lead the Algonauts 2025 out-of-distribution leaderboard by mean score. From brain to model, we make three contributions. Brain-MoE ties the experts of a frozen model to the seven cortical networks, raises held-out accuracy in all 15 model-benchmark pairs and exceeds capacity-matched random experts in 14. Brain-AVQA, to our knowledge the first audiovisual benchmark with cortical-network labels for every question, supplies the pretraining questions. Brain-Scope finds sparse features that predict brain responses, steer routing and, when removed, weaken brain prediction. Human brain organization is therefore a usable architectural prior for multimodal large language models.}


\maketitle

\section*{Introduction}

Artificial networks trained in very different ways may be learning the same picture of the world. The Platonic Representation Hypothesis, named after the allegory in which prisoners see only the shadows that one set of objects casts on a wall, proposes that networks differing in architecture, training data, objective and modality converge on a shared statistical model of the reality behind their data\cite{Huh2024}. Such organization can arise without explicit design: in Claude models, the J-space, a set of verbalizable representations, is reported to show the functional properties of a global workspace\cite{Gurnee2026}. The human brain offers an independent reference. Performance-optimized visual networks predict neural responses in macaque visual cortex through fitted linear mappings\cite{Yamins2014}, and the same approach extends to language cortex\cite{Schrimpf2021,Goldstein2022}. There, brain prediction rises with capability in tested language-model panels\cite{Mischler2024} and often peaks at intermediate layers, a tendency noted in a review of 25 functional magnetic resonance imaging (fMRI) studies\cite{LopezCardona2025}. Language-derived and multimodal representations also align with high-level visual and category-selective cortex\cite{Doerig2025,Du2025}. If models and brains converge, their correspondence should be measurable and, ideally, usable.

Omni models make this testable with the input that the brain itself receives. Unlike models that see text, images, or images with text, omni models, multimodal large language models that process video, audio and language jointly, can receive the audiovisual stimulus that a participant sees and hears, together with aligned transcripts. Brains and omni models acquired this ability by different routes, evolution and training, and may nevertheless arrive at similar representations. In the Algonauts 2025 challenge, entrants predict whole-brain fMRI responses from synchronized video, audio and language\cite{Gifford2025}. TRIBE and the gated encoder MIRAGE\cite{dAscoli2025,Gokce2026}, the latter with an omni backbone, showed that multimodal representations track cortical responses during natural viewing, with the largest gains in association cortex\cite{dAscoli2025,Scotti2025}. Features learned inside one joint model outperformed concatenated features in the tested Qwen-Omni models\cite{Gokce2026}, subtitle features added nothing beyond audiovisual features in one pipeline\cite{Abdollahi2025}, and low-rank adapters on an omni backbone raised scores further\cite{Ying2026BrainLoRA}. These studies show that omni representations predict the brain, not that the brain can make the models more capable.

Using the brain to improve models has empirical support, but the contribution of brain organization itself remains untested. Reducing brain alignment can lower aggregate performance on the Holmes suite of more than two hundred linguistic tasks\cite{Merlin2026}. Mouse visual-cortex similarity has regularized vision-model training\cite{Li2019}, fMRI has tuned speech and audiovisual models\cite{Moussa2025,Policzer2025}, and targets or directions derived from task fMRI have steered reasoning\cite{Xiao2026}; brain-data-informed analysis has also motivated uniform attention in shallow BERT layers\cite{Toneva2019}. These methods transfer information from recordings and were not designed to test the functional-network division of the brain. Organizational priors have been tested in other forms: the cognitive-domain experts of MiCRo\cite{AlKhamissi2026}, the functional-network expert prior of FPED\cite{Ren2026}, which targets image reconstruction from fMRI, pathways induced by routing-cost priors\cite{Cook2025} and the topographic constraints of Topoformer and Topo-Omni\cite{Binhuraib2025,AlKhamissi2026b}. In a mixture of experts (MoE) the question is concrete: the division into experts is normally learned or set by hand, and whether experts organized and pretrained by the functional-network division of the whole brain outperform capacity-matched random experts has not been tested in multimodal large language models on public capability benchmarks.

Closing this gap requires answers to three questions. The first concerns measurability: how stable and structured is brain correspondence across omni models measured under one encoding protocol? The second concerns structure: can functional-network organization be written into an omni model as an explicit, controllable prior? The third concerns usability: does that prior add value beyond matched modular capacity, and how does the gain depend on the headroom of the base? Answering them requires intervention as well as prediction, because convergence appears conditional\cite{Koepke2026} and a high encoding score does not establish structural or mechanistic alignment\cite{Jia2026}.

\begin{figure*}[!tbp]
\centering
\includegraphics[trim=12.7786bp 0bp 12.7786bp 0bp,clip,width=\textwidth]{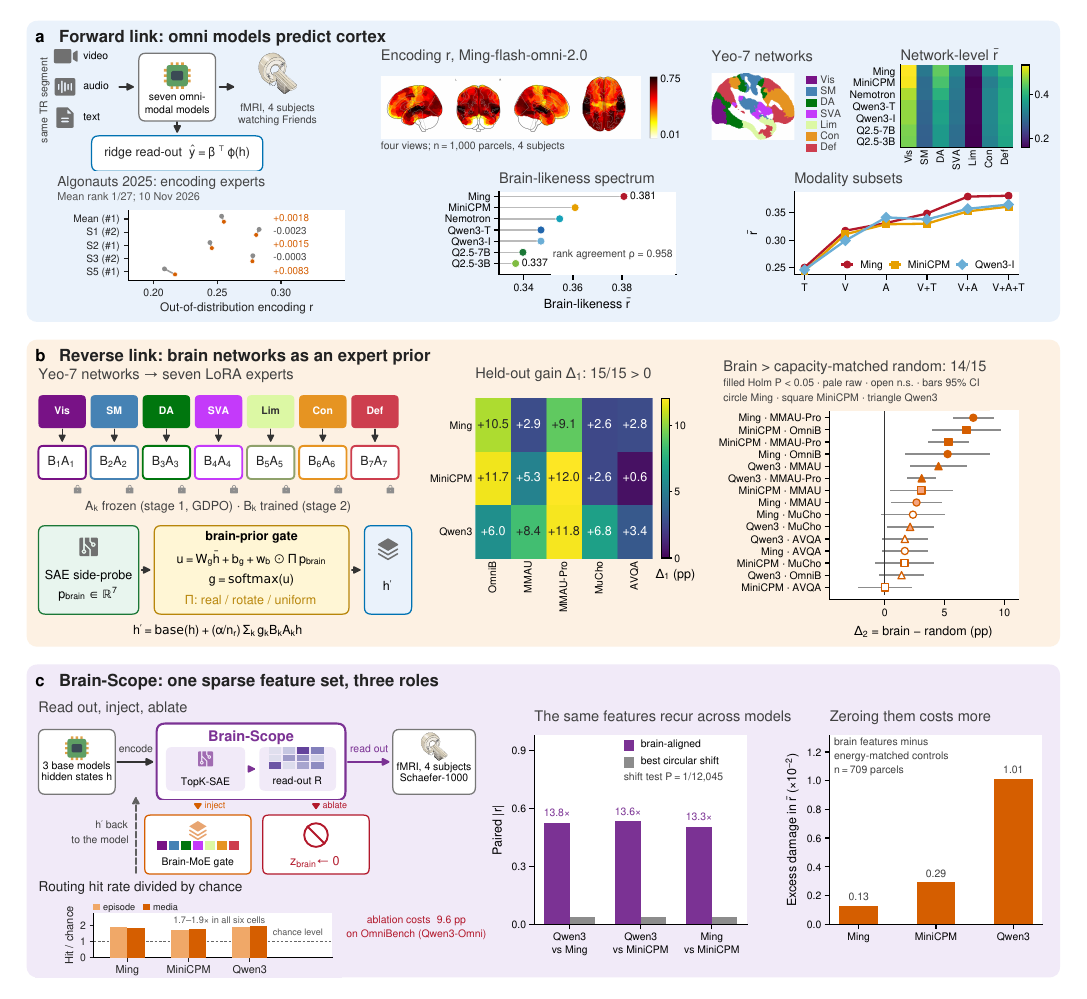}
\caption{\textbf{The Platonic brain bridge: forward link, reverse link and one set of sparse autoencoders.} \textbf{a}, Forward link (model to brain). Ridge read-outs map hidden states of seven omni models to 1,000 cortical parcels in four participants watching \textit{Friends}. Shown are encoding $r$ for Ming-flash-omni-2.0 (four views); the five Algonauts 2025 leaderboard columns on 10 November 2026 (orange, our entry; grey, the entry ranked second by mean score); a sagittal section of the Yeo seven networks (Yeo-7) with its colour key; network-level $\bar r$ per model; the brain-likeness spectrum (rank agreement $\rho = 0.958$); and $\bar r$ of three bases under six input subsets. \textbf{b}, Reverse link (brain to model). The Yeo-7 networks map onto seven low-rank adaptation (LoRA) experts $\mathbf{B}_k\mathbf{A}_k$ (lock, $\mathbf{A}_k$ frozen after stage one); a read-only side-probe supplies the brain prior $\pbrain$ to the gate, and the gated experts add residually to the frozen base. Heat map, held-out gain $\Delta_1$ over that base (percentage points, pp; three bases $\times$ five benchmarks); forest plot, $\Delta_2$ (brain minus capacity-matched random experts) with 95\% McNemar intervals (filled, Holm-corrected $P<0.05$; pale, uncorrected $P<0.05$ only; open, $P\ge0.05$; shape, base). \textbf{c}, Brain-Scope in three roles: read out, inject $\pbrain$, ablate ($z_{\mathrm{brain}}\leftarrow 0$). Bars, from left: the routing hit rate, the share of held-out questions whose top-1 routed expert is the annotated network, divided by the independence baseline (dashed line at 1; two splits per base, $P<2\times10^{-4}$); paired $|r|$ of matched brain-aligned features against the best circular shift ($P = 1/12{,}045$); and excess damage in $\bar r$ from zeroing them against activation-matched controls ($n = 709$ parcels), which also costs 9.6 pp of OmniBench accuracy on Qwen3-Omni.}
\label{fig:ov}
\end{figure*}
We address these questions with the Platonic brain bridge hypothesis: the correspondence between multimodal large language models and the human brain is usable in both directions. From model to brain, omni models converge on structured representations that map onto cortex. From brain to model, the functional-network organization of the brain can be engineered as an architectural prior, a structural preference fixed before task training: here, one expert per cortical network, each pretrained only on the questions of its network. Such a prior should raise accuracy on held-out questions of tasks unrelated to the recordings, and its gains should localize to specific components and grow with headroom, the accuracy that a base has yet to gain on a benchmark. We inject an organizational principle, the Yeo seven networks (Yeo-7), a standard division of human cortex\cite{Yeo2011}, rather than fitting model weights to brain recordings. The pretrained content is confined to a frozen feature subspace and the brain prior reaches the expert gate through a read-only path, so each control can switch off, shuffle or randomize one of them while the rest of the model stays unchanged; validation uses benchmarks unrelated to the recordings.

The study makes three contributions. (1) Brain-MoE, which injects the seven-network organization as an architectural prior, improves accuracy on held-out questions in all 15 base-model and benchmark pairs and exceeds capacity-matched random experts in 14 of them (Fig.~\ref{fig:ov}b). (2) Brain-AVQA supplies the network-labelled questions that training this map needs; to our knowledge, it is the first audiovisual question-answering benchmark in which each question is labelled with the cortical network most engaged by its source window, so that network priors are testable per question. (3) Brain-Scope, a set of sparse autoencoders with readouts fitted to fMRI, computes the brain prior that steers the gate; removing its brain-aligned features impairs brain prediction on all three bases and benchmark accuracy on one (Fig.~\ref{fig:ov}c). Brain-likeness analyses of seven omni models, stable across participants, structured along modality and depth and differing along a single axis, motivate the design of Brain-MoE, and our mixture of encoding experts built on omni-model features ranks first by participant-mean score on the post-challenge Algonauts 2025 leaderboard (Fig.~\ref{fig:ov}a). Brain correspondence thus becomes a design tool, not only a measurement.
\section*{Results}

Of the three questions raised in the Introduction, the Results centre on usability: can brain-network supervision form expert subspaces that are more effective and more transferable than capacity-matched alternatives, and thereby improve the task adaptation of multimodal large language models? We first describe how Brain-MoE turns that supervision into frozen experts and how the control arms isolate its contribution. We then test transfer to unseen questions and unrelated public benchmarks, locate the source of the gain, and use Brain-Scope to show how the brain prior is read out, used and tested by intervention. Finally, brain-likeness analyses of seven omni models explain the design choices, and an encoding system built on the same network-indexed principle is the leading entry on a public benchmark.

\subsection*{Brain-MoE turns brain-network supervision into frozen expert subspaces}

\begin{figure*}[!tbp]
\centering
\includegraphics[width=\textwidth]{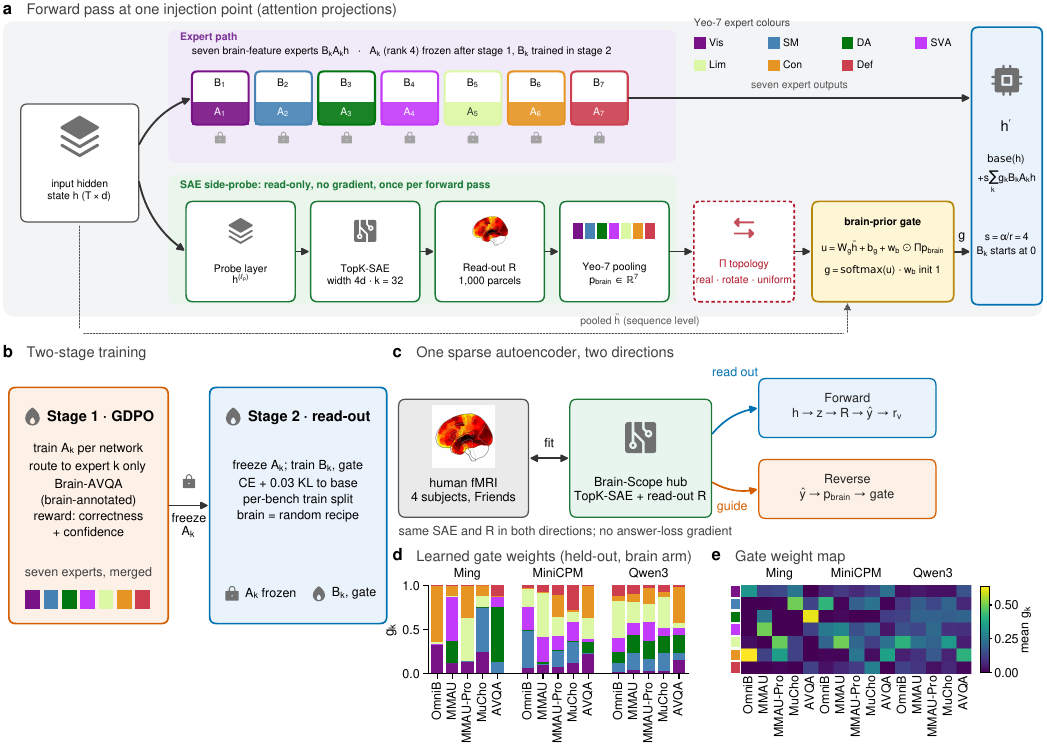}
\caption{\textbf{Forward data flow of Brain-MoE, two-stage training, the Brain-Scope loop and learned gating weights.} \textbf{a}, Forward computation at one injection point (attention projections of eight layers, zero-based indices 20 to 27 for MiniCPM-o 4.5 and Ming-flash-omni-2.0 and 28 to 35 for Qwen3-Omni). The hidden state $\mathbf{h}$ takes two paths: the expert path (purple) with seven brain-feature experts $\mathbf{B}_k\mathbf{A}_k\mathbf{h}$ (Yeo-7 colours; $\mathbf{A}_k$ frozen after stage one, lock icon; $\mathbf{B}_k$ trained in stage two), and the read-only sparse-autoencoder bypass (green), which passes the probe-layer state through the TopK sparse autoencoder, the brain readout $\mathbf{R}$ and Yeo-7 pooling to give $\pbrain$, which enters the gate with one learnable gain per network (red dashed box: the controls real, rotate and uniform, which act on the routing targets in the Brain-AVQA topology control; Methods); for MiniCPM-o 4.5 the readout acts on the probe-layer state directly; the expert outputs are weighted by $\mathbf{g}$ and added residually to the frozen base projection. \textbf{b}, Two-stage training. Stage one trains $\mathbf{A}_k$ and $\mathbf{B}_k$ network by network with GDPO-based policy-gradient training on Brain-AVQA and merges the seven experts; stage two freezes $\mathbf{A}_k$ (lock) and trains $\mathbf{B}_k$ and the gate (flame) on the training split of each benchmark; brain and random arms share architecture and stage-two steps and data, and random experts skip stage one. \textbf{c}, The two directions of one sparse autoencoder: Brain-Scope (TopK-SAE and readout $\mathbf{R}$, fitted from human fMRI) reads out brain responses in the forward chain and aggregates predicted responses into $\pbrain$ in the reverse chain, with identical parameters and no answer-loss gradient in either direction. \textbf{d}, Gating weights learned by the brain arm on held-out questions of five public benchmarks: mean weight $g_k$ of the seven experts in 15 model and benchmark combinations (stacks normalized per cell; no error bars). \textbf{e}, The same gating weights as a heat map, with the seven networks as rows and the 15 cells as columns.}
\label{fig:moe}
\end{figure*}

Brain-MoE (brain mixture of experts) is a detachable module that gives a frozen omni model seven experts, one for each Yeo-7 network (Fig.~\ref{fig:moe}a). We used three bases: Ming-flash-omni-2.0 (about 100B parameters)\cite{Ming2026}, MiniCPM-o 4.5 (9B)\cite{MiniCPMo2026} and the Instruct variant of Qwen3-Omni (30B)\cite{Qwen3Omni2025}, abbreviated below as Ming, MiniCPM and Qwen3. Each expert is a low-rank adaptation (LoRA) module\cite{Hu2022LoRA} with a down-projection $\mathbf{A}_k$ (rank 4) and an up-projection $\mathbf{B}_k$, placed in parallel with the attention projections of the layers after mid-depth, and Yeo-7 network $k$ maps to expert $k$ by a fixed one-to-one rule. At each injection point a gate operates once per input on the token-averaged hidden state $\bar{\mathbf{h}}$, and its scores for the seven experts sum a term computed from the hidden state and a term computed from a brain prior, the predicted response of each network to the current input,
\begin{subequations}\label{eq:1}
\begin{align}
\mathbf{u}&=\mathbf{W}_g\bar{\mathbf{h}}+\mathbf{b}_g+\mathbf{w}_b\odot\pbrain(\mathbf{h}),\label{eq:1a}\\
\mathbf{g}&=\mathrm{softmax}(\mathbf{u}),\label{eq:1b}\\
\mathbf{h}'&=\mathrm{base}(\mathbf{h})+\frac{\alpha}{n_r}\sum_{k=1}^{7}g_k\,\mathbf{B}_k\mathbf{A}_k\mathbf{h},\label{eq:1c}
\end{align}
\end{subequations}
where $\mathbf{W}_g$ and $\mathbf{b}_g$ are the trainable weights and bias of the gate, $\pbrain\in\mathbb{R}^{7}$ is computed read-only by Brain-Scope, a sparse autoencoder (SAE) with a brain readout head described below, $\mathbf{w}_b$ is a learnable gain per network, the scale is $\alpha/n_r = 4$ (rank $n_r = 4$, $\alpha = 16$) and $\mathrm{base}(\cdot)$ is frozen. The gate weights $\mathbf{g}\in\mathbb{R}^{7}$ are non-negative and sum to one, and the output of expert $k$ is added to the frozen base with weight $g_k$. Read-only means that no gradient flows back into the autoencoder, the readout head or the base along this path.

The experts acquire their content from brain-network supervision in two stages (Fig.~\ref{fig:moe}b). Brain-AVQA supplies that supervision: each 8-s window of the sitcom \textit{Friends}, watched by four participants of the CNeuroMod dataset as released for Algonauts 2025\cite{CNeuroMod2023,Gifford2025}, was labelled with the Yeo-7 network most engaged in their fMRI (a group-level label; Methods); Qwen3.5-Omni-Plus\cite{Qwen35_2026} generated a four-choice question targeting the capability of that network, and Gemini 3.5 Flash\cite{Gemini35_2026} audited each question on audio and video jointly. The benchmark contains 2,362 questions from 81 half-episode segments, 2,096 (88.7\%) requiring joint audiovisual answering. In stage one, expert $k$ is trained only on the questions of network $k$, by policy-gradient reinforcement learning with the reward-decoupled advantage of Group reward-Decoupled Normalization Policy Optimization (GDPO)\cite{Liu2026GDPO}. Stage two adapts the module to a target benchmark: it freezes every $\mathbf{A}_k$ and trains only $\mathbf{B}_k$ and the gate with cross-entropy plus a Kullback-Leibler (KL) constraint to the base. Brain data therefore enter the model as supervision that decides which subspace each expert spans; once learned, the subspace is frozen, so downstream training can use it but cannot overwrite it.

This design lets us ask what brain-network supervision adds beyond the capacity and routing of a standard LoRA mixture of experts. A cell is one base model paired with one benchmark; an arm is one version of the system evaluated in that cell. Every cell was run with three arms: the frozen base; capacity-matched random experts, which share the architecture, parameter count and stage-two training of the brain arm but start from randomly initialized down-projections and a shuffled prior; and brain experts with the real prior. $\Delta_1=\acc(\mathrm{MoE})-\acc(\mathrm{base})$ measures the gain of the module over the base, and $\Delta_2=\acc(\mathrm{brain})-\acc(\mathrm{random})$ measures the part of that gain that is not produced by added parameters, the mixture architecture and stage-two training alone; both are tested with paired exact McNemar tests. A subset of cells adds the remaining arms of a $2\times2$ factorial design that separates expert content from brain prior. Any advantage of brain over random experts therefore reflects stage-one pretraining under brain-network supervision rather than the added capacity of the module.

\subsection*{Brain-supervised experts transfer to unseen questions and unrelated benchmarks}

\begin{figure*}[!tbp]
\centering
\includegraphics[width=\textwidth]{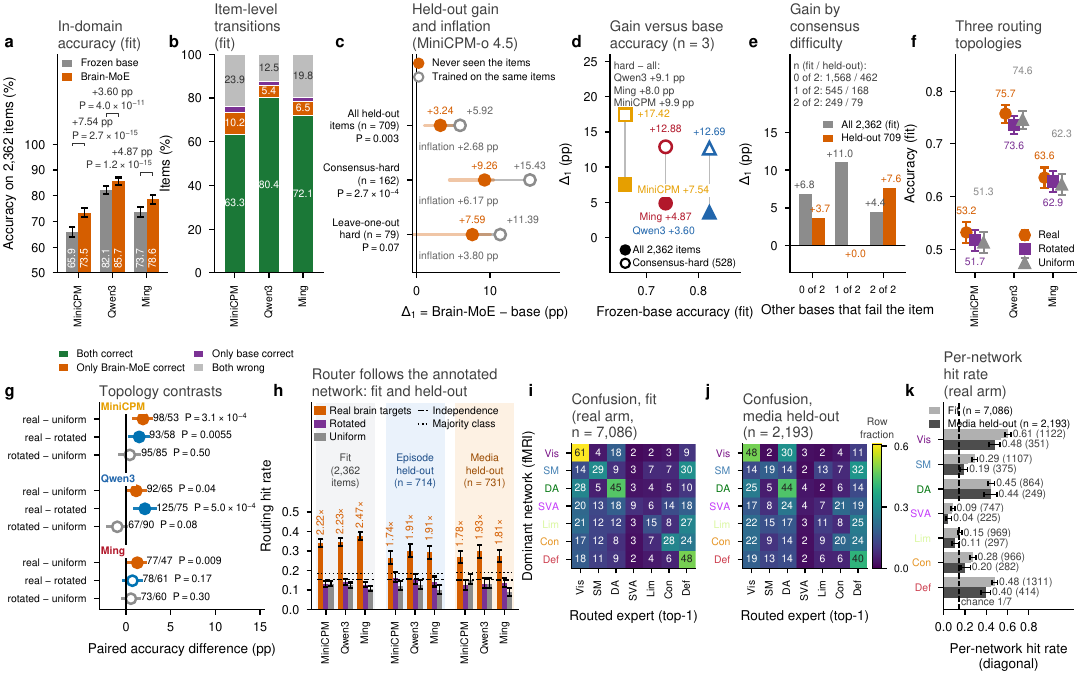}
\caption{\textbf{Brain-MoE improves accuracy and its routing depends on real brain topology.} \textbf{a}, Accuracy of the frozen base (grey) and Brain-MoE (orange) on all 2,362 Brain-AVQA questions in the in-domain fitting regime; error bars, Wilson 95\% confidence intervals (CIs); $\Delta_1$ and paired exact McNemar $P$ in parentheses. \textbf{b}, Per-question outcome transitions per base. \textbf{c}, $\Delta_1$ of MiniCPM-o 4.5 on the 709 held-out questions under three question selections (all, consensus hard and leave-one-out hard): filled, experts and routing network trained without these questions; open, trained on all 2,362 questions. \textbf{d}, In-domain $\Delta_1$ against frozen-base accuracy (filled, all questions; open, consensus hard questions, $n = 528$). \textbf{e}, $\Delta_1$ of MiniCPM-o 4.5 stratified by leave-one-out difficulty, the number of the other two frozen bases that miss the question; grey, all questions ($n = 1{,}568$, 545 and 249); orange, held-out questions ($n = 462$, 168 and 79). \textbf{f}, Accuracy of three bases in the in-domain topology control under real (orange), rotated (purple) and uniform (grey) routing topologies; error bars, Wilson 95\% CIs; compared only between arms. \textbf{g}, Nine paired differences with 95\% paired CIs; filled, $P<0.05$. \textbf{h}, Routing hit rate (top-1 expert equals annotated dominant network) under three topologies for the in-domain topology control (grey background) and the held-out splits by half-episode ($n = 714$, blue; labelled episode in the panel) and by whole episode ($n = 731$, orange; labelled media); error bars, Wilson 95\% CIs; dashed line, independence baseline; dotted line, majority-class rate; multiplier labels, real-arm hit rate relative to that baseline (held out, 1.74 to 1.93$\times$). \textbf{i},\textbf{j}, Routing confusion matrices of the real arm (rows, annotated network; columns, top-1 expert; row percentages) for the in-domain topology control (\textbf{i}) and held-out whole episodes (\textbf{j}), pooled over the three bases ($n = 7{,}086$ and 2,193 question-base pairs). \textbf{k}, Per-network hit rates from the diagonals of \textbf{i},\textbf{j}; error bars, Wilson 95\% CIs; dashed line, 1/7 chance.}
\label{fig:avqa}
\end{figure*}

On MiniCPM, the experts first improved accuracy on Brain-AVQA questions held out from training. The questions were split, stratified by network and difficulty, into 1,653 for training and 709 held out. On the 709 held-out questions, Brain-MoE exceeded the frozen base by +3.24 percentage points (pp) (paired exact McNemar test, two-sided $P = 0.0032$, $n = 709$), with larger gains on questions the frozen bases answered poorly; models trained on the same questions gained more, which measures the training effect (Fig.~\ref{fig:avqa}c to e). In the in-domain fitting regime, in which the module is trained and evaluated on all 2,362 questions, Brain-MoE raised accuracy on all three bases, by +3.60, +4.87 and +7.54 pp on Qwen3, Ming and MiniCPM, and the gain rose as base accuracy fell (Fig.~\ref{fig:avqa}a,b,d). These questions share their source with the stage-one supervision, so they show that the module works but not that its benefit extends beyond the brain-labelled material.

\begin{figure*}[!tbp]
\centering
\includegraphics[width=\textwidth]{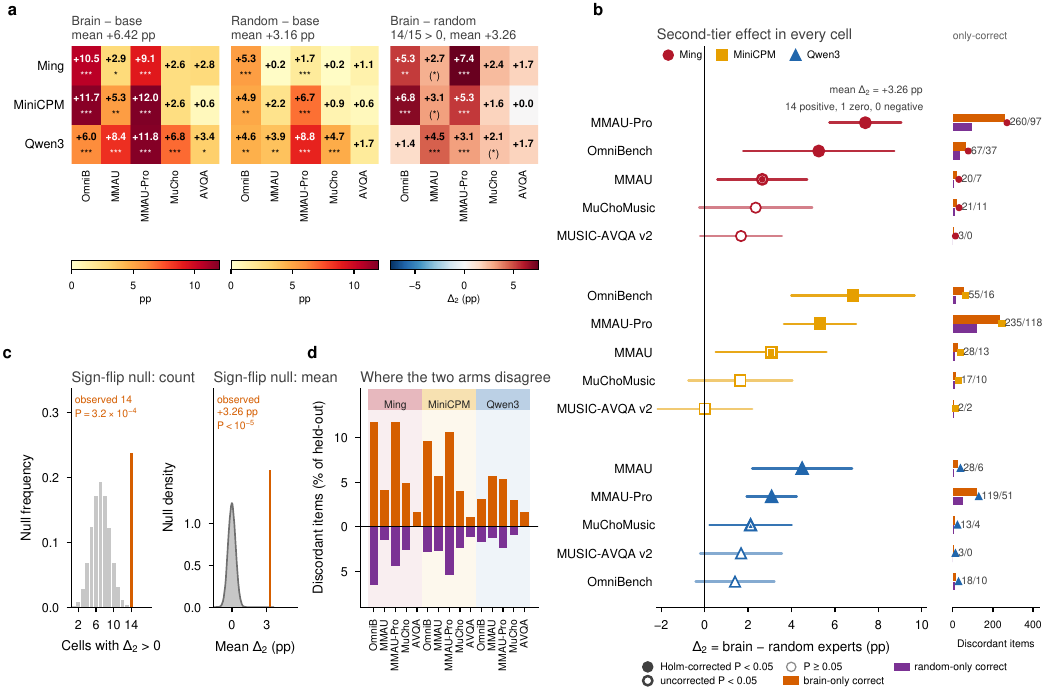}
\caption{\textbf{Two levels of evidence on three heterogeneous bases and five public benchmarks.} \textbf{a}, Heat maps of held-out accuracy gains in pp: left, gain $\Delta_1$ of the brain-pretrained expert arm over the frozen base; middle, gain of the capacity-matched random expert arm over the same base (shared 0 to 12 scale); right, their difference $\Delta_2$ (zero-centred diverging scale). Stars, per-question paired exact McNemar tests (left and middle, arm versus base; right, brain versus random; *$P<0.05$, **$P<0.01$, ***$P<0.001$; bracketed stars, $P<0.05$ that fails Holm correction over 15 cells). Columns, OmniBench, MMAU, MMAU-Pro, MuChoMusic, MUSIC-AVQA v2.0 (held-out $n$ = 571, 490, 2,203, 425, 179); rows, Ming-flash-omni-2.0, MiniCPM-o 4.5, Qwen3-Omni. \textbf{b}, Forest plot of $\Delta_2$ in the 15 cells grouped by base (dark red circles, Ming-flash-omni-2.0; gold squares, MiniCPM-o 4.5; blue triangles, Qwen3-Omni); bars, 95\% confidence intervals from discordant pairs; filled, Holm-corrected $P<0.05$; open with an inner dot, uncorrected $P<0.05$ only; open, $P \ge 0.05$; the sub-axis on the right gives the numbers of questions answered correctly only by the brain arm (orange) and only by the random arm (purple). \textbf{c}, Null distributions of the question-level joint sign-flip permutation ($N_{\mathrm{perm}} = 100{,}000$; one sign per question shared across the three bases): left, number of cells with $\Delta_2 > 0$ (observed 14 marked); right, kernel density of the mean $\Delta_2$ over 15 cells (observed value marked). \textbf{d}, Fraction of held-out questions per cell answered correctly only by the brain arm (orange, upward) and only by the random arm (purple, downward); background shading groups bases.}
\label{fig:matrix}
\end{figure*}

The main test is therefore transfer to tasks that the brain data never touched. We evaluated the same experts on OmniBench\cite{OmniBench2024}, MMAU test-mini\cite{MMAU2024}, a multiple-choice subset of MMAU-Pro\cite{MMAUPro2025}, MuChoMusic\cite{MuChoMusic2024} and a four-choice adaptation of MUSIC-AVQA v2.0\cite{MusicAVQA2022,MusicAVQA2_2024}, each split strictly by media key with identical held-out sets across cells. Brain-pretrained experts improved held-out accuracy in all 15 base and benchmark cells and exceeded random experts in 14 (Fig.~\ref{fig:ov}b; Fig.~\ref{fig:matrix}). $\Delta_1$, which includes the effect of stage-two training on each benchmark, ranged from +0.56 to +11.98 pp (mean +6.42 pp; Fig.~\ref{fig:matrix}a); mean $\Delta_2$, the contrast with identically trained random experts, was +3.26 pp, and the brain arm equalled the random arm in the remaining cell (Fig.~\ref{fig:matrix}b). Because the three bases answer the same questions, the primary test was a question-level joint sign-flip permutation (100,000 draws): the observed 14 positive cells gave right-tail $P = 3.2\times10^{-4}$, and no permutation reached the observed mean ($P = 1/100{,}001$; Fig.~\ref{fig:matrix}c). Per-cell McNemar tests gave $P<0.05$ in nine cells, six remaining below 0.05 after Holm correction (raw $P = 2.5\times10^{-18}$ to 0.0042; Fig.~\ref{fig:matrix}a,b,d). The gain held across seeds: on MiniCPM, three stage-two seeds on four benchmarks left $\Delta_2$ positive in all 12 seed-benchmark runs (+1.65 to +6.83 pp), and a second stage-one seed gave $\Delta_2 = +6.13$ pp on OmniBench against +6.83 pp with the first. Subspaces learned from brain-labelled audiovisual questions therefore help a frozen model adapt to audio, music and audiovisual reasoning tasks unrelated to the brain recordings used for~supervision.

\subsection*{The added value resides in the expert subspace and scales with headroom}

\begin{figure*}[!tbp]
\centering
\includegraphics[width=\textwidth]{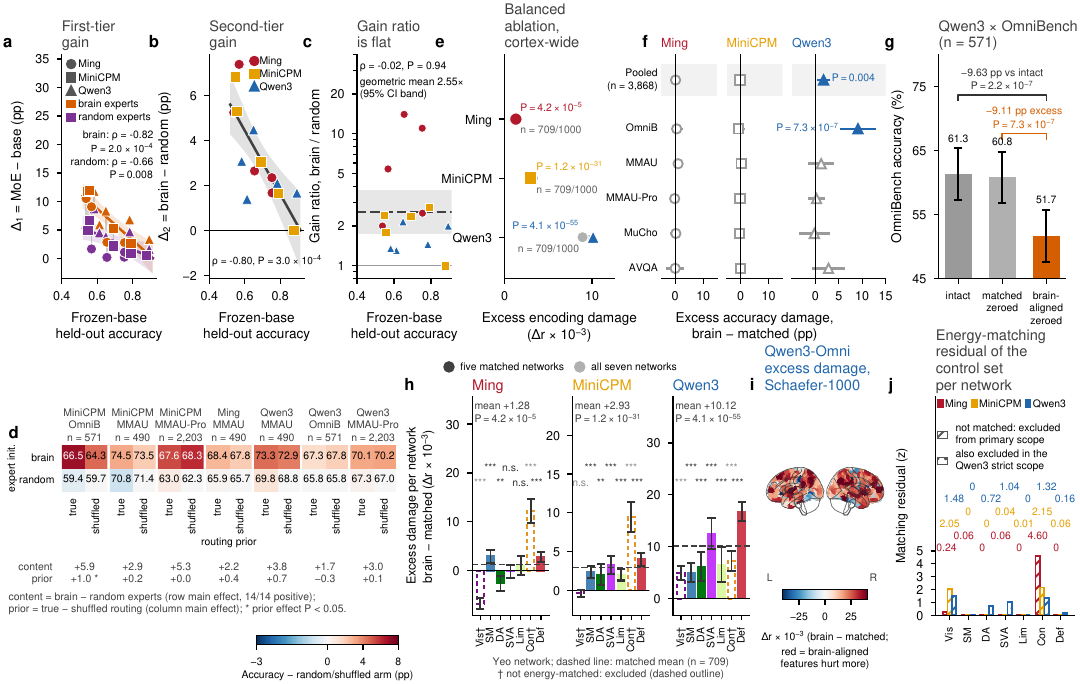}
\caption{\textbf{Gain scales with base headroom and resides in expert content, and zeroing brain-aligned features impairs encoding on three bases and accuracy on Qwen3-Omni.} \textbf{a}, First-level effect $\Delta_1$ against frozen-base held-out accuracy for the brain arm (orange) and the random arm (purple), 15 cells each (circle Ming, square MiniCPM, triangle Qwen3); in \textbf{a} and \textbf{b}, lines and shading are least-squares fits with 95\% confidence bands. \textbf{b}, Second-level effect $\Delta_2$ against base accuracy; Spearman $\rho = -0.80$, two-sided $P = 3.0\times10^{-4}$, $n = 15$ cells. \textbf{c}, Ratio of brain-arm to random-arm gain (log axis); dashed line and shading, geometric mean over the 15 cells and its 95\% CI (normal approximation on the log ratios). \textbf{d}, Held-out accuracy (\%) of seven complete $2\times2$ factorial cells; below each panel, mean content and prior main effects (pp); *, prior effect within brain experts (+2.28 pp), paired exact McNemar $P = 0.0072$. \textbf{e}, Whole-cortex excess encoding damage of the matched ablation ($\Delta r\times10^{-3}$): coloured, primary criterion (five matched networks, $n = 709$ parcels); grey, reference criterion (all seven networks). \textbf{f}, Excess downstream damage of the same zeroing (pp), pooled ($n = 3{,}868$) and per benchmark; filled, exact McNemar $P<0.05$; bars, 95\% CIs from discordant pairs. \textbf{g}, Accuracy of Qwen3-Omni on OmniBench ($n = 571$) intact, with the matched control zeroed and with the 98-feature brain set zeroed; error bars, 95\% binomial CIs; brackets, accuracy change relative to intact (black) and to the matched control (orange, $-9.11$ pp, an excess damage of +9.11 pp). \textbf{h}, Per-network excess encoding damage on the three bases; error bars, 95\% CIs across parcels; stars, within-network Wilcoxon signed-rank tests (*$P<0.05$, **$P<0.01$, ***$P<0.001$; n.s., $P\ge0.05$); dashed open bars ($\dagger$), networks that failed activation matching. \textbf{i}, Parcel-wise excess damage for Qwen3-Omni (lateral glass brains). \textbf{j}, Matching audit: matching residual of the control set of each network (hatched, excluded from the primary criterion).}
\label{fig:head}
\end{figure*}

The gain over random experts comes from what the brain experts learned in stage one. A further arm (brain experts, shuffled prior) splits $\Delta_2$ into two pathways by a rule fixed in advance. The content pathway is the gain of brain over random experts when both use a shuffled prior; the routing pathway is the further gain from restoring the real prior to the brain experts. The content pathway was positive in all five cells with this arm (+1.93 to +7.31 pp). Seven complete $2\times2$ cells, four of them among these five, which cross brain or random experts with real or shuffled prior, show the same split (Fig.~\ref{fig:head}d): content effects were positive in all 14 contrasts (+1.40 to +7.18 pp). Relative to a LoRA mixture of the same size and stage-two training, the independent value of brain-network supervision on these benchmarks therefore lies in the expert subspace that it produces.

The size of that value depended on headroom. Base accuracy ranged from 0.5377 to 0.8939 across the 15 cells, and both effect sizes decreased with it: Spearman $\rho = -0.82$ for $\Delta_1$ ($P = 2.0\times10^{-4}$, $n = 15$) and $-0.80$ for $\Delta_2$ ($P = 3.0\times10^{-4}$) (Fig.~\ref{fig:head}a,b). For $\Delta_1$ such a relation could arise simply because a weak base has more headroom (1 minus base accuracy); $\Delta_2$ contains no base term, so its correlation indicates that the advantage of brain-pretrained over random experts also falls as base accuracy rises. The ratio of brain-arm to random-arm gain, by contrast, was unrelated to base accuracy ($\rho = -0.02$, $P = 0.94$; Fig.~\ref{fig:head}c), and regression to the mean, base confounding, question count and model scale did not account for the pattern (Supplementary Note~1). Brain-supervised experts thus gained a geometric mean of 2.55 times the gain of random experts, a ratio with no detectable relation to base accuracy, and they help most in the cells with most room to improve.

\subsection*{Brain-Scope makes the brain prior readable, usable and testable}

\begin{figure*}[!tbp]
\centering
\includegraphics[width=\textwidth]{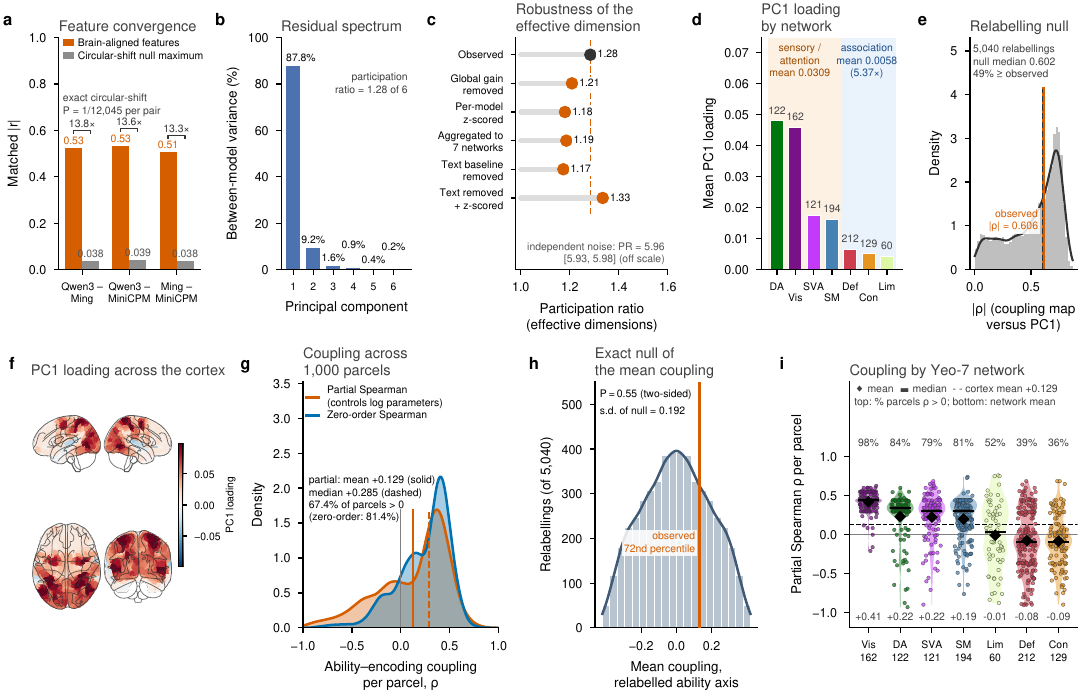}
\caption{\textbf{Brain-aligned sparse features converge across models, whereas inter-model differences concentrate on one low-dimensional shared axis.} \textbf{a}, Mean activation-series correlation $|r|$ after one-to-one Hungarian matching of the brain-aligned sparse features of each model (185, 184 and 190 for Qwen3-Omni, Ming-flash-omni-2.0 and MiniCPM-o 4.5) over 13,383 shared stimulus grid points for three pairs of independently trained models (orange) against the largest matched correlation reached by any circular shift of one series (grey, the permutation null maximum); ratios in parentheses; exact circular-shift permutation $P = 1/12{,}045$ per pair. \textbf{b}, Variance spectrum of the inter-model residual matrix, the seven Schaefer-1000 encoding maps with the shared map removed; participation ratio 1.28 out of 6 possible dimensions. \textbf{c}, Robustness of the effective dimension to five treatments addressing four alternative explanations (orange points); dashed line, observed value; independent noise at the measured residual amplitude (2,000 simulations) lies outside the axis range and is annotated as text. \textbf{d}, Mean loading of the first residual principal component on the Yeo-7 networks (bar colours, network colours; parcel counts above bars); sensory and attention networks (orange background) against association networks (blue background). \textbf{e}, Relabelling null (all 7! = 5,040 label permutations) of $|$Spearman $\rho|$ between the capability-brain coupling map and PC1; orange line, observed value; dashed line, null median. \textbf{f}, Cortical distribution of PC1 loadings (glass-brain projection, four views; diverging colour scale centred at zero). \textbf{g}, Distribution over 1,000 parcels of the coupling between the capability composite and parcel-wise brain-likeness: orange, partial correlation controlling parameter count; blue, zero-order; vertical lines, mean and median. \textbf{h}, Exact permutation null of the mean coupling under all 5,040 model relabellings; orange line, observed value (two-sided $P = 0.55$). \textbf{i}, The same parcel-wise partial correlations grouped by Yeo-7 network. Each violin is the distribution over the parcels of one network, with individual parcels as jittered points; the black diamond marks the mean, the horizontal bar the median and the dashed line the cortex-wide mean. The numbers above each violin give the fraction of parcels with a positive correlation, the numbers below give the network mean, and the parcel count of each network is given on the $x$ axis.}
\label{fig:feat}
\end{figure*}

The expert subspace is one of the two components of Brain-MoE; the other is the brain prior that steers the gate. Brain-Scope is the interface through which this prior enters the model (Fig.~\ref{fig:moe}c), and it can be examined at three points. First, the autoencoders that supply the prior contain brain-aligned features that recur across models (for MiniCPM the prior readout bypasses the autoencoder; Methods). On intermediate hidden states of each model we trained TopK SAEs\cite{Gao2025SAE}, which express each hidden state with a few active features (width four times the hidden width, 32 active per token), with linear parcel-wise readout heads fitted on fMRI; the name follows Qwen-Scope\cite{Deng2026}. The brain-aligned features of each pair of frozen models, each set selected by alignment with the same fMRI network responses and matched one-to-one by the Hungarian algorithm\cite{Kuhn1955} over 13,383 shared time points, had mean matched correlations of 0.5257, 0.5344 and 0.5068, 13.3 to 13.8 times the largest value reached in an exhaustive circular-shift permutation (exact $P = 1/12{,}045$; Fig.~\ref{fig:feat}a). The autoencoders that compute the prior therefore contain brain-aligned features that independently trained bases share, consistent with one network partition serving three different bases.

Second, routing trained on the measured network profile sends questions preferentially to the expert of their annotated network, also on unseen material. Two further splits held out half-episodes ($n = 714$) and whole episodes ($n = 731$). The routing hit rate, the fraction of questions on which the top-1 expert chosen by the routing network (Methods) matches the annotated network, was 0.2647 to 0.2997 in all six combinations of three bases and two splits, 1.74 to 1.93 times the independence baseline (permutation of routed labels across questions, 5,000 permutations per cell, none reaching the observed value, $P<2\times10^{-4}$), whereas rotated and uniform controls stayed at 0.088 to 0.164 (Fig.~\ref{fig:avqa}h to k); hit rates on unseen whole episodes and unseen half-episodes did not differ (0.2795 versus 0.2857, pooled over the three bases; two-sample permutation test, $P = 0.659$). In the in-domain topology control ($n = 2{,}362$, arms compared per question), real routing exceeded uniform routing on all three bases by +1.14 to +1.91 pp ($P = 3.1\times10^{-4}$ to 0.038) and rotated routing by +0.72 to +2.12 pp ($P = 0.0055$, $5.0\times10^{-4}$ and 0.17 on MiniCPM, Qwen3 and Ming; Fig.~\ref{fig:avqa}f,g), and on MiniCPM the real map was more accurate than all 20 random maps (empirical $P = 0.048$, the smallest value 20 draws can give). Brain-derived routing thus carries network identity to the experts in a form that generalizes to unseen episodes.

Third, intervening on brain-aligned features of the same autoencoders links the prior to what the models compute. The most aligned features of each Yeo-7 network (Ming 111, MiniCPM 115, Qwen3 98) were set to zero and compared with an equal-size, non-overlapping, activation-matched control set; excess damage, measured with the frozen brain readout head, is the damage from zeroing the brain features minus that from zeroing the control set. Under the primary criterion, five networks whose control passed matching ($n = 709$ parcels), excess encoding damage was positive on all three bases: largest on Qwen3 (+0.01012; across-parcel Wilcoxon signed-rank test, two-sided $P = 4.1\times10^{-55}$), with Ming at +0.00128 ($P = 4.2\times10^{-5}$) and MiniCPM at +0.00293 ($P = 1.2\times10^{-31}$) (Fig.~\ref{fig:head}e,h to j); the sign held under the reference and strict criteria. On Qwen3, removing the 98 features also lowered OmniBench accuracy from 0.6130 to 0.5166 ($-9.63$ pp from unrounded accuracies), an excess damage of +9.11 pp over the matched control (paired exact test, $P = 7.3\times10^{-7}$, $n = 571$) and +1.73 pp pooled over five benchmarks ($n = 3{,}868$, $P = 0.0041$) (Fig.~\ref{fig:head}f,g), whereas Ming and MiniCPM showed near-zero downstream excess damage. On Qwen3, zeroing the features that predict cortex is therefore more harmful to benchmark accuracy than zeroing the control features.

\subsection*{Brain-likeness analyses ground the design choices}

\begin{figure*}[!tbp]
\centering
\includegraphics[width=\textwidth]{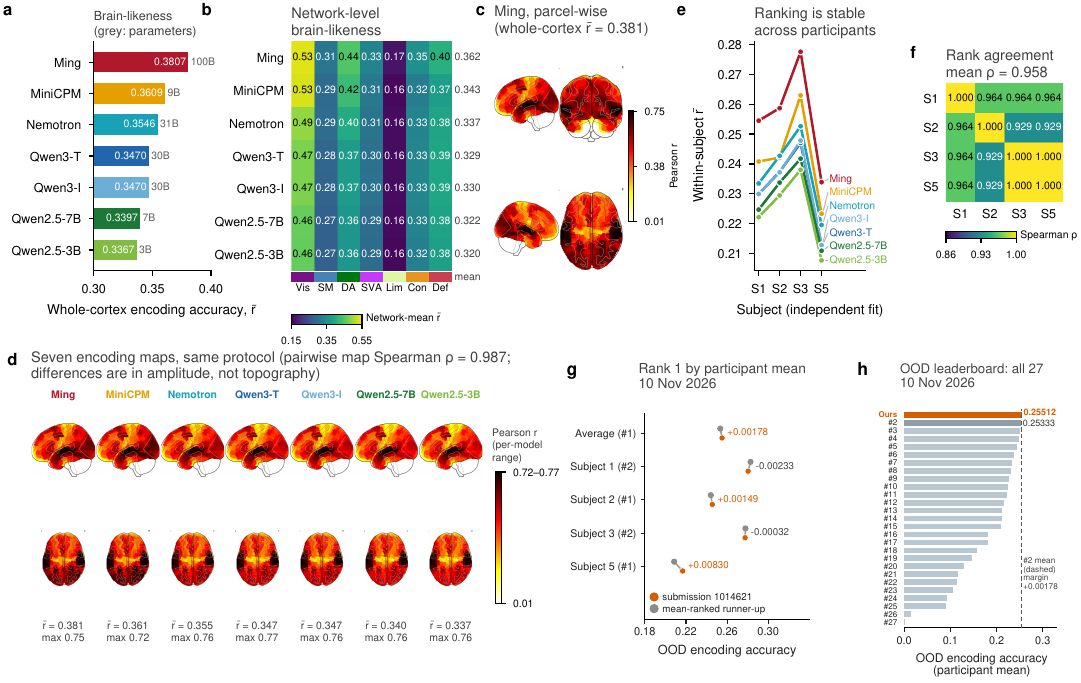}
\caption{\textbf{Brain-likeness spectrum of seven omni models, its stability across participants and benchmark performance.} \textbf{a}, Whole-cortex brain-likeness $\bar r$ (validation Pearson correlation between predicted and participant-averaged responses, mean over 1,000 Schaefer parcels) of seven models ($n = 7$) under one encoding protocol; nominal parameter counts in grey. \textbf{b}, Network-mean $\bar r$ of the seven models (rows, sorted by $\bar r$) within the Yeo-7 networks (columns; Vis, visual; SM, somatomotor; DA, dorsal attention; SVA, salience/ventral attention; Lim, limbic; Con, control; Def, default); the mean column is the cross-network mean. \textbf{c}, Parcel-wise encoding accuracy of Ming-flash-omni-2.0 as a glass brain (four views; colour scale from the minimum to the maximum parcel $r$ of this model). \textbf{d}, Parcel-wise encoding maps of the seven models (left sagittal above axial view; each map coloured over its own range, with $\bar r$ and maximum below); mean pairwise Spearman $\rho = 0.987$ ($n = 21$ pairs). \textbf{e}, Brain-likeness of the seven models under independent fits for each of the four participants ($n = 7$ models $\times$ 4 participants), one line per model; each readout is fitted to the responses of one participant instead of the participant average used in \textbf{a}, so values are lower than in \textbf{a}. \textbf{f}, Spearman correlation matrix of model rankings between participant pairs ($n = 7$ models); off-diagonal mean $\rho = 0.958$. \textbf{g}, Score of submission 1014621 (orange) and submission 1034419, ranked second by mean score (grey), in the five leaderboard columns (participant mean and four participants) of the public post-challenge Algonauts 2025 out-of-distribution leaderboard, with signed differences annotated ($n = 27$ entries). \textbf{h}, Participant-mean score of all 27 entries; submission 1014621 in orange, dashed line at the entry ranked second. Panels g and h report the 10 November 2026 leaderboard snapshot.}
\label{fig:spec}
\end{figure*}

The results above test the prior; its design, in turn, rests on brain-likeness analyses of seven omni models under one encoding protocol, the model-to-brain direction of the hypothesis. For each model, a ridge encoding model predicts the response of each of the 1,000 cortical parcels of the Schaefer-1000 atlas\cite{Schaefer2018} from hidden states given the video, audio and aligned transcripts watched by the participants; brain-likeness is the validation Pearson correlation with the participant-averaged response, averaged over parcels (equation~(\ref{eq:4})). Besides the three bases, the panel included Nemotron-3-Nano-Omni (31B)\cite{Nemotron2026}, Qwen3-Omni Thinking (30B) and Qwen2.5-Omni 7B and 3B\cite{Qwen25Omni2025}. The bases resemble one another closely enough to share one design: brain-likeness lay on a narrow spectrum from 0.337 to 0.381 whose ranking agreed across the four participants when the readout was refitted for each of them (mean pairwise Spearman $\rho = 0.958$, $n = 7$ models; Fig.~\ref{fig:ov}a; Fig.~\ref{fig:spec}a, e, f), and once the shared map was removed, the first principal component held 87.8\% of the remaining differences between models (participation ratio 1.28 out of 6; Fig.~\ref{fig:feat}b to d). The models thus differ mainly in how strongly they express one shared pattern, which is compatible with injecting the same seven-network partition into all of them.

\begin{figure*}[!tp]
\centering
\includegraphics[width=\textwidth]{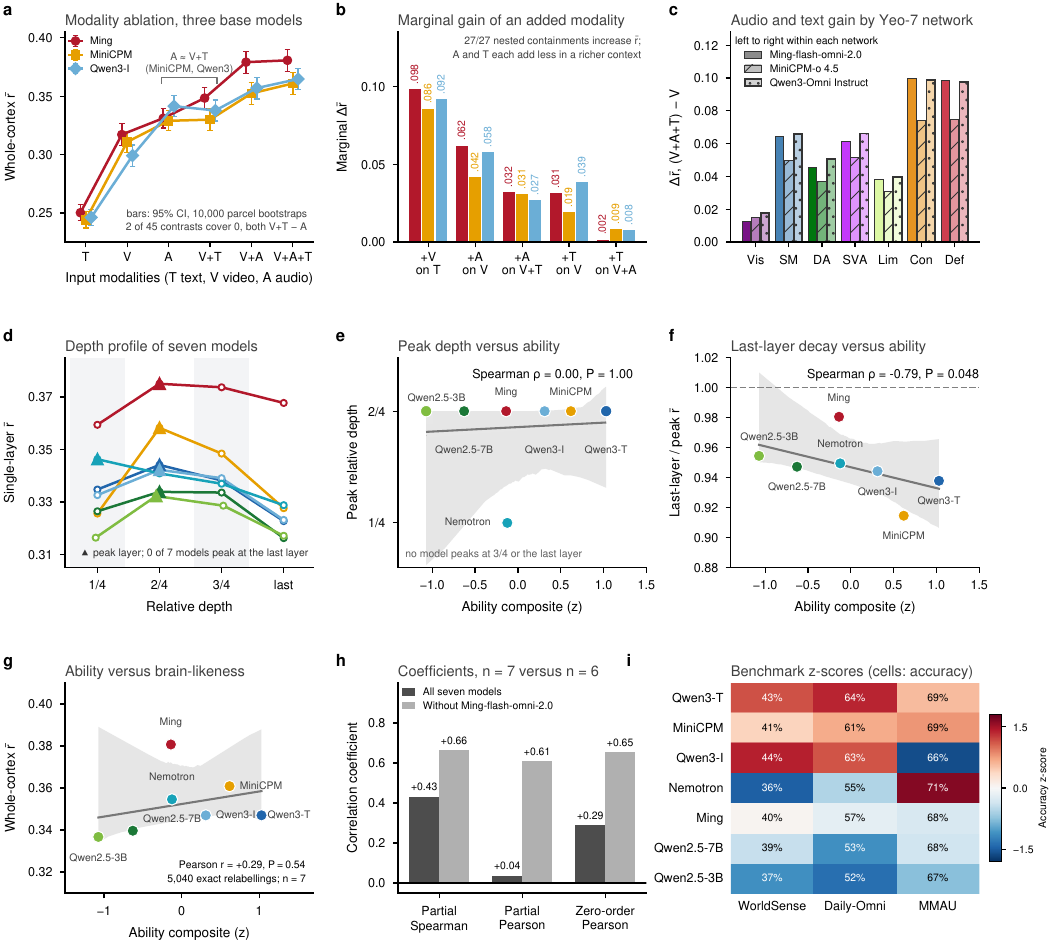}
\caption{\textbf{Modality structure, depth structure and the capability-brain-likeness relation.} \textbf{a}, Whole-cortex brain-likeness of three base models (Ming-flash-omni-2.0, circles; MiniCPM-o 4.5, squares; Qwen3-Omni, diamonds) under six input-modality subsets (T text, V video, A audio and combinations); all three follow T $<$ V $<$ \{A, V+T\} $<$ V+A $<$ V+A+T; error bars, 95\% confidence intervals (paired parcel bootstrap, 10,000 resamples, $n = 1{,}000$ parcels); A and V+T overlap in MiniCPM-o 4.5 and Qwen3-Omni. \textbf{b}, Marginal gain of one added channel between nested subsets; audio and subtitles each add less in a richer context, and all 27 nested comparisons raise brain-likeness. \textbf{c}, Increment of full-modality over video-only input decomposed by Yeo-7 network (network colours; within each network, left to right, Ming, MiniCPM, Qwen3). \textbf{d}, Single-layer brain-likeness of seven models at four relative depths (1/4, 2/4, 3/4, last); triangles mark the peak of each model; 0/7 models peak at the last layer. \textbf{e}, Peak relative depth against a capability composite (mean $z$-score over WorldSense, Daily-Omni and MMAU); Spearman $\rho = 0.00$, exhaustive permutation $P = 1.00$, $n = 7$. \textbf{f}, Last-to-peak brain-likeness ratio against the capability composite; $\rho = -0.786$, $P = 0.048$, $n = 7$; no correlation on this seven-model panel remains significant after Holm correction over the family of six (smallest corrected $P = 0.29$); the partial Pearson coefficient in \textbf{h} is outside that family. \textbf{g}, Capability composite against whole-cortex $\bar r$; zero-order Pearson $r = +0.29$, exhaustive label permutation (5,040) $P = 0.54$. \textbf{h}, Partial Spearman, partial Pearson (controlling $\log_{10}$ parameters) and zero-order Pearson coefficients on all seven models (dark grey) and on six models without Ming-flash-omni-2.0 (light grey). \textbf{i}, Composition of the capability composite: accuracy $z$-scores of the seven models on the three benchmarks (raw accuracies in cells). In \textbf{e} to \textbf{g}, grey lines are least-squares fits and shading is the bootstrap 95\% confidence band (2,000 resamples); point colours follow \textbf{d}.}
\label{fig:mod}
\end{figure*}

The same analyses fix where the experts and the probe are placed. No model peaked at its last layer, and the best single layer lay at a quarter to half of the depth (Fig.~\ref{fig:mod}d), consistent with earlier reports\cite{LopezCardona2025}, so the probe reads the half-depth layer and the experts sit just after it. Qwen3-Omni Thinking and Instruct, which differ only in post-training, were indistinguishable (difference $2\times10^{-5}$), consistent with a report that instruction tuning did not consistently improve brain alignment in the tested models\cite{Gao2025}; brain-likeness was thus insensitive to this difference in post-training. On the three bases, brain-likeness rose with every added input channel, from 0.250, 0.244 and 0.246 for text alone to 0.381, 0.361 and 0.365 for all three (Ming, MiniCPM and Qwen3), and the increment of full-modality over video-only input was concentrated in the control and default networks (Fig.~\ref{fig:mod}a to c), which motivates supervising experts with audiovisual rather than text-only questions. On this panel neither the shared axis nor brain-likeness tracked a capability composite of three public benchmarks\cite{WorldSense2025,DailyOmni2025,MMAU2024} (Fig.~\ref{fig:mod}e to i; Fig.~\ref{fig:feat}e to i), so these analyses guide the design of the prior rather than predict which model is most~capable.

\subsection*{Network-indexed encoding experts lead a public brain-encoding benchmark}

The organizing principle of Brain-MoE, experts indexed by cortical networks, also applies in the model-to-brain direction, where a public benchmark scores the resulting predictions. In the Algonauts 2025 challenge\cite{Gifford2025}, entrants predict fMRI responses to films outside the training distribution, and the organizers score the predictions against withheld responses. Our encoding system applies the same indexing: brain predictions read out from different omni models act as experts, and their mixture is organized, for each participant, by hemisphere and Yeo-7 network, with mixture weights fitted separately for each group. The system is separate from the LoRA experts above and from the fixed seven-model protocol. In a development test on labelled held-out films, this network-organized mixing exceeded an equal-weight pool of the same experts in all 16 participant-film folds (participant means +0.0061 to +0.0112 in $r$; Supplementary Table 16); the test compares network-wise with uniform weighting, not with other groupings. The submitted system, a larger mixture that uses the same network organization (Methods), ranked first by participant-mean correlation among 27 entries on the public post-challenge Algonauts 2025 out-of-distribution leaderboard (10 November 2026 snapshot; Fig.~\ref{fig:spec}g,h): $r = 0.25512$ for submission 1014621, 0.00178 above the second entry and 0.00383 above the third, with first place for sub-02 and sub-05 and second place for sub-01 and sub-03. Indexing experts by cortical networks is thus applicable in both directions, although the leaderboard rank reflects the whole system rather than this component alone: it organizes what a model learns from the brain and how models are combined to predict~it.

\section*{Discussion}

This study asked whether brain-network supervision can form expert subspaces that improve the task adaptation of multimodal large language models, and the answer on three bases and five public benchmarks is yes. Each contribution corresponds to one result. Brain-MoE turns the seven-network organization of cortex into frozen expert subspaces that transfer: held-out accuracy rose in all 15 base-benchmark pairs, brain experts exceeded capacity-matched random experts in 14, and the gain, located in the expert content, was on average about two and a half times that of random experts, with no detectable trend across base accuracy. Brain-AVQA makes this supervision possible by labelling every question with the network most engaged, at group level, by its source window. Brain-Scope makes the prior readable, usable and testable: its autoencoders hold brain-aligned features that recur across models, brain-derived routing sends questions preferentially to the matching expert on unseen episodes and, on Qwen3, removing its brain-aligned features lowers benchmark accuracy. Brain-likeness analyses of seven omni models supplied the design evidence, and our mixture of encoding experts built on omni-model features ranked first by mean score on the post-challenge Algonauts 2025 leaderboard. Together, these results show that brain-network supervision is a practical route to task adaptation, and they give the brain-to-model direction of the Platonic brain bridge hypothesis a concrete, testable form (Supplementary Table~10).

The controls also show what brain-network supervision adds to existing adaptation methods (Supplementary Table~7). LoRA and mixture-of-experts capacity alone, matched in the random arm, recovered only part of the gain. Brain-tuning and steering\cite{Moussa2025,Policzer2025,Xiao2026,Merlin2026} change parameters or activations with brain-derived targets, whereas here the brain contributes an organizational label on training questions and a frozen fMRI-fitted readout for routing, and the learned subspace is frozen before downstream training. Earlier organizational priors used hand-defined cognitive domains, as in MiCRo\cite{AlKhamissi2026}, topographic constraints, as in Topoformer and Topo-Omni\cite{Binhuraib2025,AlKhamissi2026b}, or a Yeo-7 expert prior for brain-to-image reconstruction, as in FPED\cite{Ren2026}; routing costs\cite{Cook2025} and untrained convolutional architectures\cite{Kazemian2025} show that architecture can shape brain correspondence. Here the partition comes from measured cortex and is tested as the expert subspace of an omni generative model on public capability benchmarks. Brain-Scope extends sparse-autoencoder analyses of brain encoding and vision transformers\cite{Lepori2026,Li2026ViSAE,Guo2026a} by linking encoding, routing and capability ablation within one set of sparse autoencoders, and the encoding analyses build on the Algonauts 2025~encoders\cite{dAscoli2025,Gokce2026,Scotti2025,Abdollahi2025,Ying2026BrainLoRA}.

Future work should settle three open questions. First, random experts received no stage-one training, so the comparison shows value beyond matched capacity but cannot yet separate the network partition from auxiliary pretraining; pretraining experts on the same questions under a non-brain grouping, random or cognitive-domain, would make that separation. Second, brain-derived routing selected the matching expert on unseen episodes and helped in the in-domain topology control, but on public benchmarks it added little beyond the expert content in most cells; larger benchmarks and more seeds are needed to decide whether it improves task accuracy on its own. Third, the evidence covers three omni bases, five adapted benchmark subsets, one main training seed per cell and the seven-network resolution of the Yeo atlas. Testing more model families, text-only and vision-language models, finer parcellations and other atlases would show how far the prior generalizes. Beyond these tests, the detachable module could in principle be added to the policies of embodied agents without retraining the base, and extended to channels such as proprioception and touch.

On the present evidence, the macroscopic functional organization of the human brain is a usable architectural prior for multimodal large language models: supervising experts with it yields transferable subspaces whose benefit is largest in the cells with most headroom. Brain correspondence can thus serve as a design tool as well as a~measurement.

\section*{Methods}

The Methods follow the order of the Results. We first describe Brain-MoE, with its training, control arms and evaluation benchmarks, and then Brain-Scope, which supplies the brain prior. The brain-likeness analyses that motivated the design and the Algonauts 2025 evaluation come next. The fMRI data, ethics, the use of large language models, computational resources, the statistical procedures shared by all analyses and code availability close the section.

\subsection*{Brain-MoE architecture}
Brain-MoE places seven low-rank experts in parallel with the attention projections of eight consecutive layers of the frozen base (zero-based indices 20 to 27 for MiniCPM-o 4.5 and Ming-flash-omni-2.0, and 28 to 35 for Qwen3-Omni), with expert $k$ corresponding one-to-one to network $k$ of the Yeo seven. The injection points are the q and v projections for MiniCPM-o 4.5 and Qwen3-Omni (two per layer) and the fused QKV projection for Ming-flash-omni-2.0 (one per layer). Each expert is a low-rank adaptation module $\mathbf{B}_k\mathbf{A}_k$ of rank $n_r = 4$, where the down-projection $\mathbf{A}_k\in\mathbb{R}^{n_r\times d}$ ($d$ being the hidden width of the model) is the brain-feature extractor and the up-projection $\mathbf{B}_k\in\mathbb{R}^{d_o\times n_r}$ is the output projection, $d_o$ being the output dimension of the projection; the scale is $\alpha/n_r = 4$ ($\alpha = 16$). Attention projections determine how the model weights its input, the closest analogue of attentional allocation across brain networks. The band starts three to five blocks after the half-depth block of each base (36, 32 and 48 layers for MiniCPM-o 4.5, Ming-flash-omni-2.0 and Qwen3-Omni), because the depth analysis (below) shows brain-likeness peaking at half depth in all three, so the injection points lie just past that~peak.

The brain prior $\pbrain(\mathbf{h})$ is computed by a read-only bypass probe and does not participate in backpropagation. Its autoencoder and readout head are fitted on fMRI once and then frozen, so the prior carries brain-fitted parameters but is predicted from the stimulus alone: no recording is used in stage two or at inference. The hidden state $\mathbf{h}^{(\ell_p)}$ of the probe layer $\ell_p$ (the single layer whose hidden states the autoencoder reads) is encoded token by token by the frozen sparse autoencoder (equation~(\ref{eq:5}), below; $\mathbf{z}$ denotes its sparse code), averaged over the sequence into $\bar{\mathbf{z}}$, mapped by the brain readout head to predicted responses of the 1,000 cortical parcels of the Schaefer-1000 atlas (equation~(\ref{eq:6}), below; weights $\mathbf{R}$, bias $\mathbf{b}_R$), and aggregated by the Yeo-7 mask matrix $\mathbf{Y}\in\mathbb{R}^{7\times1000}$ ($Y_{k,v}=\mathbf{1}[v\in\mathcal{N}_k]/|\mathcal{N}_k|$, where $\mathcal{N}_k$ is the set of parcels of network $k$) into the mean predicted response of each network, then standardized across networks:
\begin{equation}
\begin{aligned}
\mathbf{p}_7&=\mathbf{Y}\big(\mathbf{R}\bar{\mathbf{z}}+\mathbf{b}_R\big),\\
\pbrain(\mathbf{h})&=\frac{\mathbf{p}_7-\mathrm{mean}(\mathbf{p}_7)}{\mathrm{std}(\mathbf{p}_7)+10^{-6}}.
\end{aligned}
\label{eq:9}
\end{equation}
The probe layers are blocks 18 (MiniCPM-o 4.5), 24 (Qwen3-Omni) and 16 (Ming-flash-omni-2.0), counting transformer blocks from one, the numbering also used in the depth analyses, whereas injection layers are given as zero-based indices. On each base the probe is the half-depth block; in stage two the hook of Qwen3-Omni and Ming-flash-omni-2.0 is attached one block later (blocks 25 and 17). In every case the probe precedes every injection point, so no expert output has entered the hidden state it reads. For MiniCPM-o 4.5 the readout head takes the token-averaged hidden state of the probe block directly, without the sparse autoencoder. The prior is computed inside the same forward pass, from the probe-layer hidden state of the item being answered, by a hook that propagates no gradient: nothing is backpropagated into the sparse autoencoder, the readout head or the frozen base along this path, and the only trained parameters later in the same forward pass are the per-network gain $\mathbf{w}_b$ and the gate. It is computed once per forward pass and shared by all injection points.

At each injection point the hidden state $\mathbf{h}\in\mathbb{R}^{L\times d}$ ($L$ tokens) is averaged over the sequence to $\bar{\mathbf{h}}$; the gate softmax uses temperature $T_g = 1$ with all seven experts participating and no truncation to a subset of experts, and the expert sum is scaled by $\alpha/n_r$. The gate linear layer $\mathbf{W}_g$ is initialized from a normal distribution with standard deviation $10^{-3}$ and the bias $\mathbf{b}_g$ at zero, so the initial gate is approximately uniform; $\mathbf{B}_k$ is zero-initialized when the experts are created, so the initial output equals the base. Each injection point has its own gate ($\mathbf{W}_g$, $\mathbf{b}_g$) and prior gain $\mathbf{w}_b$. The per-network learnable prior gain $\mathbf{w}_b\in\mathbb{R}^{7}$ is initialized at 1 and optimized in the same group as the gate; as $w_{b,j}\to0$ the gate reduces to an ordinary mixture of experts driven by the hidden state alone. This is the gated forward pass of equation~(\ref{eq:1}). In stage one, described next, the gates are not used and the expert mixture is a fixed one-hot vector.

\subsection*{Two-stage training}
The experts acquire their brain-derived content in stage one and are adapted to each benchmark in stage two. Stage one trained experts network by network on the Brain-AVQA benchmark, described below, by policy-gradient reinforcement learning with the reward-decoupled advantage of GDPO\cite{Liu2026GDPO}: each reward is normalized within the group of sampled answers, the normalized rewards are summed with weights and the sum is normalized again. The policy loss takes the REINFORCE form\cite{Williams1992}, without the clipped importance-ratio objective. For network $k$, the training set was restricted to questions annotated with dominant network $k$, the trainable parameters were $\mathbf{A}_k$, $\mathbf{B}_k$ and an auxiliary routing network (a two-layer perceptron with 512 hidden units that maps cached frozen-base features of the window, taken at the probe block for MiniCPM-o 4.5 and Qwen3-Omni and at block 24 for Ming-flash-omni-2.0, to seven logits $\mathbf{v}_q$), and the expert mixture was fixed to the one-hot vector $\mathbf{e}_k$, so that all residual correction flowed through expert $k$ without cross-talk. Questions were processed one per update. For question $q$, the forward pass gives the vector of option log-likelihoods $\boldsymbol{\ell}_q$ (normalized over the four answer letters), from which $G = 4$ answers $\mathrm{ans}_o$ ($o = 1,\dots,G$) are sampled at temperature $T_{\mathrm{samp}} = 1.2$ and compared with the correct option $\mathrm{ans}^\star_q$, each receiving a two-dimensional reward (correctness, confidence of that answer):
\begin{equation}
\begin{aligned}
\boldsymbol\psi_o&=\big[\mathbf{1}(\mathrm{ans}_o=\mathrm{ans}^\star_q),\ \mathrm{softmax}(\boldsymbol{\ell}_q)[\mathrm{ans}_o]\big],\\
\hat\psi_{o,j}&=\frac{\psi_{o,j}-\mathrm{mean}_{o'}\psi_{o',j}}{\mathrm{std}_{o'}\psi_{o',j}+\epsilon_0},\\
\tilde A_o&=\textstyle\sum_{j=1}^{2}\omega_j\hat\psi_{o,j},\\
\hat A_o&=\mathrm{clip}\Big(\frac{\tilde A_o-\mathrm{mean}_{o'}\tilde A_{o'}}{\mathrm{std}_{o'}\tilde A_{o'}+\epsilon_0},-c_{\mathrm{clip}},c_{\mathrm{clip}}\Big),
\end{aligned}
\label{eq:s1reward}
\end{equation}
where $\epsilon_0 = 10^{-6}$, the clipping threshold is $c_{\mathrm{clip}} = 5.0$ and the reward weights are $\boldsymbol\omega=(1.0,0.3)$. The policy loss is advantage-weighted REINFORCE\cite{Williams1992} with a supervised anchor and a routing-alignment term:
\begin{equation}
\begin{aligned}
\mathcal{L}_{\mathrm{stage1}}={}&w_{\mathrm{hard}}\Big(c_{\mathrm{gdpo}}\big[-\tfrac{1}{G}\textstyle\sum_o\boldsymbol{\ell}_q[\mathrm{ans}_o]\hat A_o\big]\\
&\quad+c_{\mathrm{sft}}\big[-\boldsymbol{\ell}_q[\mathrm{ans}^\star_q]\big]\Big)\\
&+c_{\mathrm{route}}\,\mathrm{KL}\big(\mathbf{g}^\star_q\,\|\,\mathrm{softmax}(\mathbf{v}_q)\big),
\end{aligned}
\label{eq:s1loss}
\end{equation}
where $\mathbf{g}^\star_q$ is the softmax of the seven measured network $z$-scores of the window of question $q$, $c_{\mathrm{gdpo}} = 1.0$, $c_{\mathrm{sft}} = 0.5$, $c_{\mathrm{route}} = 0.2$ (the routing term is averaged over the seven components, an effective weight of 0.2/7), and the hard-question weight is $w_{\mathrm{hard}} = 2.0$ for the 227 questions that carry a preliminary difficulty flag (1.0 otherwise). Each network expert was trained for one epoch with AdamW (learning rate $10^{-4}$, gradient-norm clipping at 1.0), after one full-batch update of the routing network (learning rate $2\times10^{-4}$). The experts trained for the seven networks were merged by slot into one checkpoint ($\mathbf{A}_k$ and $\mathbf{B}_k$ of expert $k$ taken from the run on network $k$) whose down-projections $\{\mathbf{A}_k\}$ are the frozen brain-feature subspace reused in all subsequent experiments. A preliminary comparison of three training modes, differing only in the parameter set entering the optimizer (frozen $\mathbf{A}$ with trained $\mathbf{B}$ and gate; gate only; $\mathbf{A}$ and $\mathbf{B}$ trained together), in which only the frozen-A mode let brain experts outperform random experts, motivated the frozen-A design; that comparison served as a qualitative design basis and is not reported as a result.

Stage two froze all $\mathbf{A}_k$ and trained only $\{\mathbf{B}_k\}$ and the gate parameters $\{\mathbf{W}_g,\mathbf{b}_g,\mathbf{w}_b\}$ with cross-entropy plus a KL divergence to the base distribution:
\begin{equation}
\begin{aligned}
\mathcal{L}_{\mathrm{stage2}}&=-\boldsymbol{\ell}_q[\mathrm{ans}^\star_q]+c_{\mathrm{KL}}\,\mathrm{KL}\big(\boldsymbol\varpi_{\mathrm{base}}\,\|\,\boldsymbol\varpi_{\mathrm{MoE}}\big),\\
c_{\mathrm{KL}}&=0.03,
\end{aligned}
\label{eq:10}
\end{equation}
where $\boldsymbol{\ell}_q$ and $\mathrm{ans}^\star_q$ are defined as in equation~(\ref{eq:s1reward}), and $\boldsymbol\varpi_{\mathrm{base}}$ and $\boldsymbol\varpi_{\mathrm{MoE}}$ are the softmax distributions over options for the same question with experts off and on. We used an up-projection learning rate of $7\times10^{-5}$, a gate learning rate of $3\times10^{-4}$, the AdamW optimizer\cite{Loshchilov2019}, one question per forward pass with gradient accumulation over 4 questions, gradient-norm clipping at 1.0 and three epochs of training. At the start of stage two the brain arm loads $\mathbf{A}_k$ and $\mathbf{B}_k$ from the merged stage-one checkpoint, whereas the random arm keeps $\mathbf{A}_k$ at its random initialization and $\mathbf{B}_k$ at zero; in both arms the gates and $\mathbf{w}_b$ are newly initialized and the stage-one routing network is not used. The control arms test whether the resulting gain comes from brain pretraining rather than from the added~parameters.

\subsection*{Control arms and factorial design}
Each (base $\times$ benchmark) cell evaluates three arms: base (all experts off), random experts (down-projections randomly initialized with the same shape and not trained in stage one, the network-to-expert map of the prior shuffled, stage-two architecture and training identical) and brain experts (stage-one $\mathbf{A}_k$ with the real network-to-expert correspondence). The two effect sizes are $\Delta_1=\acc(\mathrm{MoE})-\acc(\mathrm{base})$ and $\Delta_2=\acc(\mathrm{brain})-\acc(\mathrm{random})$, both in percentage points. Supplementary Table 14 lists both effect sizes and the per-cell $P$ values on the five public benchmarks.

Topology controls on Brain-AVQA apply a data-independent operator $\Pi$ to the seven-network target $\mathbf{g}^\star$ on which the routing network is trained: real, the target is unchanged; rotated, its seven components are shifted cyclically by two positions, so each expert receives the target of another network and the multiset of components is unchanged; uniform, every component is set to 1/7, which removes the information. The shuffled prior of the factorial design is a recorded random permutation of the components of $\pbrain$; expert parameters, steps and compute are identical across operators. The routing network is re-initialized before routing training under every operator, and only the expert LoRA parameters are carried over from stage one. The random-map control draws 20 permutations of the seven networks at random (seed 20260905) from the 5,040 possible ones, after removing the identity and a cyclic shift, and repeats the routing training under each. The empirical $P$ of the real map is $(1+m)/21$, where $m$ is the number of random maps at least as good as it; with 20 draws its smallest value is 0.048. The in-domain topology control assembles stage-one expert checkpoints and runs four full-batch updates of the routing network under each operator, without stage-two training, ; its absolute accuracies are compared only between arms under per-question pairing with topology as the sole variable. Supplementary Table 12 lists the three-arm accuracies and paired comparisons, and Supplementary Note 1 reports the retraining under 20 random maps.

The four-arm factorial design crosses expert initialization (brain/random) with brain prior (real/shuffled) as independent factors; the arms BT (brain experts, real prior), BS (brain experts, shuffled prior), RT (random experts, real prior) and RS (random experts, shuffled prior) have held-out accuracies obeying the saturated two-factor model
\begin{equation}
\acc(x_{\mathrm{c}},x_{\mathrm{p}})=\mu+\theta_{\mathrm{c}}\,x_{\mathrm{c}}+\theta_{\mathrm{p}}\,x_{\mathrm{p}}+\gamma\,x_{\mathrm{c}}x_{\mathrm{p}},
\label{eq:11}
\end{equation}
with $x_{\mathrm{c}}$ indicating brain-pretrained experts and $x_{\mathrm{p}}$ a real prior, whose four parameters are determined uniquely: $\mu=\acc_{\mathrm{RS}}$, content effect under the shuffled prior $\theta_{\mathrm{c}}=\acc_{\mathrm{BS}}-\acc_{\mathrm{RS}}$, prior effect for random experts $\theta_{\mathrm{p}}=\acc_{\mathrm{RT}}-\acc_{\mathrm{RS}}$ and interaction $\gamma=(\acc_{\mathrm{BT}}-\acc_{\mathrm{BS}})-(\acc_{\mathrm{RT}}-\acc_{\mathrm{RS}})$; the main effects shown in Fig.~\ref{fig:head}d average over the two levels of the other factor ($\theta_{\mathrm{c}}+\gamma/2$ and $\theta_{\mathrm{p}}+\gamma/2$). These arms were run in seven cells on three public benchmarks (OmniBench, MMAU and MMAU-Pro; Supplementary Table 15). In an eighth cell (Ming $\times$ OmniBench) the four runs used two different media-key splits; it is listed separately in that table on the 474 questions held out in both splits and is not included in Fig.~\ref{fig:head}d. Supplementary Table 5 gives the three-arm decomposition into content and routing pathways, and Supplementary Table 15 the four-arm accuracies and factorial~effects.

\subsection*{Brain-AVQA benchmark}
Brain-AVQA annotates each audiovisual question, beyond video, question, options and answer, with the Yeo-7 network most strongly engaged, in the fMRI of the participants during passive viewing, by the 8-s window from which the question was generated; the label belongs to the stimulus window, not to an answering process. Construction has five steps. First, each half-episode of the \textit{Friends} stimuli watched by the four participants is cut into non-overlapping 8-s windows on a fixed grid, and each window is assigned the five or six fMRI samples that it covers (repetition time TR, 1.49 s; Data and participants, below), shifted by a haemodynamic lag of three TRs (about 4.5 s). For each participant the response is averaged over these TRs and over the parcels of each Yeo-7 network; the seven network means are averaged over the participants with data for that half-episode (four, or three for three half-episodes), each network is $z$-scored across the windows of the same half-episode, and the network with the largest $z$-score becomes the dominant-network label of the window. The labels use a lag of three TRs whereas the encoding analyses use two; label reliability is nearly identical at lags of two, three and four TRs (Supplementary Table 8). Second, a multimodal generative model (Qwen3.5-Omni-Plus\cite{Qwen35_2026}) generates a four-choice question from the audiovisual content of the window, with a prompt requiring that its solution depend on the capability associated with the dominant network. Third, an audit model independent of the question writer (Gemini 3.5 Flash) performs joint audiovisual verification of each question, retaining only those whose answer is uniquely correct, derivable from the window content and free of textual shortcuts. Fourth, questions and options occasionally generated in Chinese are translated into English question by question by Qwen3-Max\cite{Qwen3_2025}, a proprietary API model of the Qwen3 series (temperature 0, five questions per batch), with option order and answer index unchanged, and translated questions pass through the third-step audit again. Fifth, a reproducible pseudo-random option permutation is applied to every question so that correct options are balanced across the four positions (591, 591, 590 and 590 questions), removing positional~priors.

The final benchmark contains 2,362 questions from 81 half-episode segments of seasons 1 to 6; the dominant-network distribution is default 437, visual 374, somatomotor 369, limbic 323, control 322, dorsal attention 288 and salience/ventral attention 249; 2,096 questions require joint audiovisual answering, 259 visual only and 7 auditory only. Consensus hard questions are answered correctly by at most one of the three frozen bases: 528 of the 2,362 questions, the 342 missed by two bases plus the 186 missed by all three. The strata missed by zero to three bases hold 1,228, 606, 342 and 186 questions; 99 and 63 of the last two fall in the held-out split, giving 162 held-out consensus hard questions. Leave-one-out hard questions are missed by both of the other two frozen bases, a per-base set of 249 for MiniCPM-o 4.5, 271 for Ming-flash-omni-2.0 and 380 for Qwen3-Omni; a base never enters its own difficulty label. Both sets are used in the hard-question analyses.

The held-out split is at question level: stratified by dominant network and a preliminary difficulty flag (227 flagged questions), 30\% of questions in each stratum were drawn at random as held out (seed 20260612), giving 1,653 training and 709 held-out questions. The held-out evaluation was run on MiniCPM-o 4.5 only. Held-out questions never appear in training, and stratified sampling leaves held-out and training sets without detectable differences in base accuracy (three bases, Mann-Whitney $U$ tests $P = 0.44$, 0.47, 0.53) or consensus-difficulty composition ($\chi^2$ test $P = 0.525$).

The dominant-network label is a group-level property. Its reliability was estimated by labelling two disjoint participant pairs independently: the two halves of the participant sample agreed on 0.353 of windows averaged over the three possible splits (Cohen $\kappa = 0.243$, $n = 2{,}362$) against a chance level of 0.143, and the bootstrap interval of the observed value does not overlap the permutation null (Supplementary Table 8). Agreement reaches 0.759 when the label from any three participants is compared with the four-participant label and falls to 0.285 for single participant pairs, and all four participants agree on 6.39\% of windows, so the label does not support a window-by-window anatomical reading. The dominant-network distribution is close to uniform because of the network-level standardization, which compresses a largest-to-smallest count ratio of 10.1 to 1.76 (Supplementary Table 9). Labels re-derived from the fMRI reproduced the released values with $r = 0.9988$ and 97.6\% identical dominant networks ($n = 2{,}362$; Supplementary Tables 8 and 9).

Brain-AVQA analyses use the stage-one experts without stage-two training: the merged experts are held fixed, the routing network is trained by four full-batch updates on the training questions, and each question is answered through the expert with the largest routing output (top-1). For the held-out result, experts and routing network were trained on the 1,653 training questions only. In the in-domain fitting regime and the in-domain topology control they were trained and evaluated on all 2,362 questions; topology-control accuracies are compared only between arms. Generalization beyond the training benchmark is tested on public~benchmarks.

\subsection*{Public benchmarks and held-out splits}
Downstream generalization was evaluated using multiple-choice tasks drawn or adapted from five public benchmarks unrelated to the brain data: OmniBench\cite{OmniBench2024}, MMAU test-mini\cite{MMAU2024}, an eligible multiple-choice subset of MMAU-Pro\cite{MMAUPro2025}, MuChoMusic\cite{MuChoMusic2024} and a four-choice adaptation of MUSIC-AVQA v2.0\cite{MusicAVQA2022,MusicAVQA2_2024}, covering general omni reasoning, audio and music understanding and joint audiovisual question answering. MMAU-Pro includes open-ended and instruction-following questions as well as multiple-choice questions with up to ten options; our loader retains questions with two to four options and a valid answer. For entries listing multiple audio files, the loader uses only the first file; this adaptation does not test the native multi-audio protocol. MUSIC-AVQA v2.0 originally uses answer classification rather than per-question multiple-choice options. Our loader constructs four-choice items by sampling three distractors from the answer pool of the same question type and excludes items without three eligible distractors, in addition to applying media-availability checks. The reported results therefore apply to these adapted subsets, not the complete native benchmark protocols. Question counts, task-format eligibility, media exclusions and held-out sizes are given in Supplementary Table 13. Each benchmark is split strictly by media key, so all questions of one audio or video clip fall wholly on the training or the held-out side. The three bases and two arms of a benchmark share identical held-out sets. Each (base $\times$ benchmark $\times$ arm) cell is an independent stage-two training run (training fraction 0.5, 3 epochs, random seed 0) with held-out evaluation. The gains in these cells enter the headroom analyses. A supplementary evaluation on MiniCPM-o 4.5 (brain arm against the frozen base; Supplementary Note 1) used AVQA\cite{Yang2022AVQA}: media availability, two to four answer options and a valid answer mapping defined an eligible subset of 2,957 questions. Ten questions affected by video-decoder failures were excluded before the media-key split into 1,474 training and 1,473 held-out questions.

\subsection*{Headroom and routing analyses}
Two further analyses describe the dependence of the gain on headroom and on routing; they are not used to establish the gain. The relation between headroom and effect size was measured by Spearman rank correlation between the headroom of each cell $\eta_{\mathrm{headroom}}=1-\acc(\mathrm{base})$ and the two effect sizes ($n = 15$). Headroom and held-out size are collinear at benchmark level (Spearman $\rho = +0.90$, $n = 5$), so effect size and evidence strength ($-\log_{10}P$) were regressed jointly by least squares on $z$-standardized headroom and $\log_{10}n$, reporting standardized coefficients. Regression to the mean was bounded by simulating 5,000 draws with a constant true gain and independent sampling of the two arms; base confounding was tested by shuffling benchmark labels within each base (20,000 Monte Carlo draws); robustness was checked by leaving out each benchmark or each base in turn and by comparing linear and quadratic fits in base accuracy. Supplementary Note 1 reports the results of these checks.

Transfer of routing structure to held-out episodes was measured by the routing hit rate on held-out questions, the fraction whose highest-weighted expert matches the annotated dominant~network. The baseline is the independence expectation $\sum_c P(\mathrm{route}=c)P(\mathrm{dominant}=c)$, with significance from permuting routing labels across questions (both marginals preserved, 5,000 permutations). Routing analyses use two splits independent of the primary held-out set (question-level, 709 questions): a split grouped by half-episode (25 of 81 half-episodes held out, $n = 714$; the other half of an episode may be in training) and a split grouped by whole episode (15 of 44 episodes held out, $n = 731$), evaluated separately. Experts and routing network for the routing hit-rate analysis were trained on the training side of the corresponding split, and the top-1 expert of a question is the largest output of the routing~network.

\subsection*{Brain-Scope: sparse autoencoders and brain readout}
Brain-Scope is the component shared by both halves of the study: its features are read out to fMRI in the forward direction and supply the brain prior in the reverse direction. For each base used in brain-to-model experiments, MiniCPM-o 4.5, Qwen3-Omni-30B-A3B-Instruct (the Instruct variant of Qwen3-Omni) and Ming-flash-omni-2.0, a TopK sparse autoencoder was trained on the hidden states $\mathbf{h}\in\mathbb{R}^{d}$ of the probe layer (layers 18, 24 and 16, respectively). The autoencoder and its loss are
\begin{equation}
\begin{aligned}
\mathbf{z}&=\mathrm{TopK}_{k_{\mathrm{act}}}\big(\mathbf{W}_e\mathbf{h}+\mathbf{b}_e\big)\in\mathbb{R}^{W},\\
\hat{\mathbf{h}}&=\mathbf{W}_d\mathbf{z}+\mathbf{b}_d,\\
\mathcal{L}_{\mathrm{SAE}}&=\mathbb{E}_{\mathbf{h}}\big\lVert\mathbf{h}-\hat{\mathbf{h}}\big\rVert_2^2+c_{\mathrm{aux}}\mathcal{L}_{\mathrm{aux}},
\end{aligned}
\label{eq:5}
\end{equation}
where $\mathrm{TopK}_{k_{\mathrm{act}}}$ keeps only the $k_{\mathrm{act}}$ largest pre-activations, the width is $W = 4d$ (16,384 for Ming-flash-omni-2.0 and MiniCPM-o 4.5, 8,192 for Qwen3-Omni), $k_{\mathrm{act}} = 32$, and decoder columns are normalized to unit norm; the auxiliary loss $\mathcal{L}_{\mathrm{aux}}$ reconstructs the residual with the 512 longest-inactive features to avoid dead features, with weight $c_{\mathrm{aux}} = 1/32$. Each autoencoder was trained on token hidden states sampled at random from the TR shards of the stimuli available for each base (179,422 TRs of \textit{Friends} and the Movie10 films of the same CNeuroMod release for Ming-flash-omni-2.0 and Qwen3-Omni, 13,383 \textit{Friends} TRs for MiniCPM-o 4.5), with 5\% of stimulus files held out for validation (up to 200,000 training and 20,000 validation tokens per model). Training used AdamW\cite{Loshchilov2019} with per-dimension standardized inputs and rows above the 99.9th percentile of L2 norm dropped, learning rate $3\times10^{-4}$, batch size 256, 30 epochs, weight decay $10^{-4}$ and random seed 42.

The brain readout head $\mathbf{R}\in\mathbb{R}^{1000\times W}$ is a linear ridge readout from sparse features, mean-pooled over the tokens of each TR and standardized per feature, to the responses of the 1,000 parcels averaged over the four participants. It was fitted in closed form on all stimuli with fMRI except \textit{Friends} season 6, which served as validation set; a single $\lambda_R$ shared by all parcels was chosen from seven values between $10^{3}$ and $10^{6}$ by the mean parcel correlation on the validation set:
\begin{equation}
\begin{aligned}
\hat{\mathbf{y}}_{\tau+2}&=\mathbf{R}\,\mathbf{z}_\tau+\mathbf{b}_R,\\
\hat{\mathbf{R}}&=\arg\min_{\mathbf{R}}\sum_{\tau\in\mathcal{T}_{\mathrm{train}}}\big\lVert\bar{\mathbf{y}}_{\tau+2}-\mathbf{R}\mathbf{z}_\tau-\mathbf{b}_R\big\rVert_2^2\\
&\quad+\lambda_R\lVert\mathbf{R}\rVert_F^2,
\end{aligned}
\label{eq:6}
\end{equation}
where $\bar{\mathbf{y}}_\tau$ is the participant-averaged response of the 1,000 parcels at TR $\tau$ (Data and participants, below), $\mathcal{T}_{\mathrm{train}}$ is the set of training TRs and responses lag features by two TRs. Feature activation series were pooled over the tokens of each TR by mean, maximum or firing rate (the fraction of tokens on which the feature is active). The alignment $a_{j,k}$ of feature $j$ with network $k$ is the Pearson correlation between its pooled activation series and the mean fMRI response of network $k$, clipped at zero. For each network, features whose alignment was significant at a Benjamini-Hochberg false discovery rate of 0.05 were ranked by selectivity margin, $a_{j,k}$ minus the median alignment of feature $j$ with the other six networks, and the top 50 entries per network were retained. These entries were pooled and deduplicated by feature to form the brain-aligned set $\mathcal{F}_{\mathrm{brain}}(M)$ (185, 184 and 190 features for Qwen3-Omni, Ming-flash-omni-2.0 and MiniCPM-o 4.5). Control sets $\mathcal{F}_{\mathrm{ctrl}}(M)$ for cross-model matching are 30 random draws of the same size from the remaining features, with the same mix of pooling schemes as the brain-aligned set (seed 20260726). The alignment under mean pooling is used for feature selection in the matched ablation. The network-level aggregation $\pbrain(\mathbf{h})\in\mathbb{R}^{7}$ averages $\mathbf{R}\mathbf{z}$ within the Yeo seven networks and standardizes across networks; it is the brain prior of the Brain-MoE gate (equation~(\ref{eq:9})).

\subsection*{Cross-model feature matching and exact permutation}
To test whether independently trained models express the correspondence through the same features, for a model pair $(M_1,M_2)$, activation series $\mathbf{z}^{(1)}_j,\mathbf{z}^{(2)}_{j'}\in\mathbb{R}^{N_t}$ of their brain-aligned features were taken on the stimulus grid shared by both models ($N_t = 13{,}383$ TR segments from 81 \textit{Friends} half-episode segments), similarity was the absolute series correlation, and the Hungarian algorithm found the optimal one-to-one matching $\pi^\star$ within the set $\mathcal{B}$ of one-to-one correspondences from $\mathcal{F}_{\mathrm{brain}}(M_1)$ to $\mathcal{F}_{\mathrm{brain}}(M_2)$:
\begin{equation}
\begin{aligned}
\pi^\star&=\arg\max_{\pi\in\mathcal{B}}\ \sum_{j\in\mathcal{F}_{\mathrm{brain}}(M_1)}\Big|\mathrm{corr}\big(\mathbf{z}^{(1)}_j,\ \mathbf{z}^{(2)}_{\pi(j)}\big)\Big|,\\
\bar c_{\mathrm{obs}}&=\frac{1}{|\mathcal{F}_{\mathrm{brain}}|}\sum_{j}\Big|\mathrm{corr}\big(\mathbf{z}^{(1)}_j,\mathbf{z}^{(2)}_{\pi^\star(j)}\big)\Big|.
\end{aligned}
\label{eq:7}
\end{equation}
The control statistic applies the same algorithm to each of the 30 random control sets $\mathcal{F}_{\mathrm{ctrl}}$ and reports their mean; no draw reached the observed value in any pair. Because activation series are autocorrelated, significance uses circular-shift permutation, shifting all series of one model as a block by $s$ grid points and re-solving the matching at every shift:
\begin{equation}
\begin{aligned}
\bar c(s)&=\max_{\pi\in\mathcal{B}}\frac{1}{|\mathcal{F}_{\mathrm{brain}}|}\sum_j\Big|\mathrm{corr}\big(\mathbf{z}^{(1)}_j,\ \mathrm{shift}_s\,\mathbf{z}^{(2)}_{\pi(j)}\big)\Big|,\\
P&=\frac{1+\big|\{s:\bar c(s)\ge\bar c_{\mathrm{obs}}\}\big|}{N_s},
\end{aligned}
\label{eq:8}
\end{equation}
where all integer shifts between 5\% and 95\% of the grid length (670 to 12,713 grid points; 12,044 shifts) were enumerated, so that $N_s = 12{,}045$ including the identity; the excluded band removes shifts that leave the two series nearly aligned. The three model pairs form one family corrected by the Holm method. Supplementary Table 3 lists the matched correlations, null maxima and $P$ values of the three pairs. Matching identifies shared features; the matched ablation, described next, tests whether such features matter for encoding and~accuracy.

\subsection*{Matched ablation}
The matched ablation tests whether zeroing brain-aligned Brain-Scope features impairs brain encoding and benchmark accuracy. From the encoding $\mathbf{z}$ of the frozen sparse autoencoder at the probe layer, the 24 features with the highest alignment $a_{j,k}$ under mean pooling are taken for each Yeo-7 network $k$, and the deduplicated union over seven networks forms the brain set $S_{\mathrm{brain}}$ (Ming-flash-omni-2.0 111, MiniCPM-o 4.5 115, Qwen3-Omni 98; distinct from the brain-aligned sets used for cross-model matching). Activation statistics of a feature (firing rate and mean activation) were computed on the TR-averaged hidden states of 2,000 sampled \textit{Friends} TRs; the control set $S_{\mathrm{matched}}$ is chosen network by network from features outside $S_{\mathrm{brain}}$ whose alignment (maximum over networks) lies between the 40th and 60th percentiles. Matching is greedy nearest neighbour on $z$-scored log firing rate and log mean activation magnitude, with equal size and zero overlap. The intervention zeroes the selected features, reconstructs the hidden state through the decoder and continues the forward pass; zeroing feature $j$ subtracts $z_{j,t}\mathbf{W}_d[:,j]$ from the hidden state. Matching used the TR-averaged statistics; at token level, where the zeroing acts, the control features carried less activation than the brain~features.

Feature selection uses the alignment $a_{j,k}$, whereas damage is measured with the frozen brain readout head of equation~(\ref{eq:6}) (for MiniCPM-o 4.5, the readout on token-averaged hidden states), applied to hidden states re-extracted with and without zeroing on two \textit{Friends} season-6 half-episodes (up to 120 TRs each). With $r_v$ the validation correlation of parcel $v$ on intact states and $r_v(S)$ that after zeroing set $S$, the measures are parcel-wise encoding damage $D_v(S)=r_v-r_v(S)$ and excess damage $\mathrm{Excess}_v=D_v(S_{\mathrm{brain}})-D_v(S_{\mathrm{matched}})$, tested by across-parcel Wilcoxon signed-rank tests. Matching quality is the within-network ratio of mean activation of brain to control arm; matching failed for the visual and control networks on all three bases (ratios 1.57 to 7.04), which are therefore excluded from the primary criterion ($n = 709$ parcels); the reference criterion includes all seven networks ($n = 1{,}000$), and the strict criterion for Qwen3-Omni further excludes the dorsal attention and salience/ventral attention networks with larger matching residuals ($n = 466$). The downstream capability ablation uses the same zeroing on the brain set and on its one-to-one paired control features (98 per arm on Qwen3-Omni), replacing the measure with answer accuracy on held-out questions of the five public benchmarks, $\mathrm{Excess}^{\mathrm{cap}}=\acc(S_{\mathrm{matched}})-\acc(S_{\mathrm{brain}})$, assessed by per-question paired exact tests (pooled over five benchmarks, $n = 3{,}868$, and separately per benchmark). Supplementary Table 6 reports the excess damage under the primary, reference and strict criteria.

\subsection*{Model panel and feature extraction}
The remaining analyses measure brain-likeness and supplied the design evidence for Brain-MoE. The panel comprises the seven omni models listed in Supplementary Table 1, run with frozen weights. The nominal 31B count for Nemotron-3-Nano-Omni follows its official model card. The Qwen2.5-Omni 3B checkpoint was released after the technical report cited for that family.

For each TR segment (video, audio and subtitles input together) of model $M$, we extracted hidden states $\mathbf{h}^{(\ell)}_t\in\mathbb{R}^{d}$ at four equally spaced relative depths (1/4, 2/4, 3/4 and last layer), averaged them over the tokens covered by that TR, reduced each layer independently to 512 dimensions by principal component analysis (PCA) and concatenated:
\begin{equation}
\begin{aligned}
\tilde{\mathbf{h}}_\tau=\Big[&\mathrm{PCA}_{512}\big(\mathrm{pool}_{t\in\tau}\,\mathbf{h}^{(\ell_1)}_t\big);\ \dots;\\
&\mathrm{PCA}_{512}\big(\mathrm{pool}_{t\in\tau}\,\mathbf{h}^{(\ell_4)}_t\big)\Big]\in\mathbb{R}^{2048}.
\end{aligned}
\label{eq:2}
\end{equation}
PCA was fitted on the training set (seasons 1 to 5 of \textit{Friends}) and applied as a projection on the validation set (season 6; Data and participants, below). These 2,048-dimensional features feed the encoding model.

\subsection*{Brain encoding model}
Features $\tilde{\mathbf{h}}_\tau$ were paired with the parcel responses averaged over the four participants, $\bar{\mathbf{y}}_{\tau+2}$, at a fixed lag of two TRs. For each parcel $v$, one linear readout was fitted to this participant-averaged response by ridge regression,
\begin{equation}
\begin{aligned}
\hat{\boldsymbol\beta}_{v}=\arg\min_{\boldsymbol\beta}&\sum_{\tau\in\mathcal{T}_{\mathrm{train}}}\big(\bar y_{v,\tau+2}-\boldsymbol\beta^{\top}\phi(\tilde{\mathbf{h}}_\tau)\big)^2\\
&+\lambda\lVert\boldsymbol\beta\rVert_2^2,
\end{aligned}
\label{eq:3}
\end{equation}
where $\phi(\cdot)$ is a standardization operator fitted on the training set and the regularization strength $\lambda$, one value shared by all parcels, is chosen by leave-one-out cross-validation among 10 log-spaced candidates between $10^{2}$ and $10^{6}$. The encoding accuracy of parcel $v$ on the validation set is the Pearson correlation $r_{v}$ between predicted and participant-averaged responses, and whole-cortex brain-likeness is
\begin{equation}
\bar r(M)=\frac{1}{1000}\sum_{v=1}^{1000}\mathrm{corr}\big(\bar y_{v},\hat y_{v}\big),
\label{eq:4}
\end{equation}
the mean over 1,000 parcels. For the participant-stability analysis the same features and protocol were refitted to the responses of each participant separately. Every model-to-brain comparison in this study uses this~score.

\subsection*{Modality and depth analyses}
Modality analyses were run on Ming-flash-omni-2.0, MiniCPM-o 4.5 and Qwen3-Omni, the three bases of the brain-to-model experiments. For a modality subset $\mathcal{S}\subseteq\{V,A,T\}$ (video, audio, subtitles), hidden states were re-extracted with only the channels in $\mathcal{S}$ and processed by equations~(\ref{eq:2}) and (\ref{eq:3}), giving brain-likeness $\bar r_{\mathcal{S}}$ for six of the seven non-empty subsets (all but A+T). Overlap between channels was judged by comparing the increment of V+A over V with the increment of A over T, and saturation by $\bar r_{VAT}-\bar r_{VA}\approx0$. Uncertainty on every subset value and nested contrast came from a paired bootstrap over the 1,000 parcels with 10,000 resamples; because parcels are spatially correlated and one validation split was used, these intervals are optimistic (too narrow).

In the modality ablation (Supplementary Table 2), training rows vary by condition for Ming-flash-omni-2.0 (113,412 to 114,003) and are constant for the other two models, whereas the validation split is identical in every cell ($n = 23{,}324$). Qwen3-Omni modality features came from a separate extraction pass at layers 8, 16, 24 and 48 in float32. Its whole-cortex value is 0.365 against 0.347 for the depth-matched pass used in the seven-model spectrum, so absolute values are comparable across the three models only up to that offset. All six Qwen3-Omni conditions come from that one pass, so contrasts within the model are exact. A control on a separate all-layer Qwen3-Omni feature store, in which the quartile layer set (12, 24, 36 and 48) was replaced by layers 8, 16, 24 and 48 under the same protocol, gave a cost of the layer set of 0.00309 in whole-cortex brain-likeness (0.23216 to 0.22907; 1.3\% relative). The pooled features of that pass consist of the per-unit mean and standard deviation of hidden states over the tokens of each TR. One standard-deviation dimension (that of hidden unit 940) accounts for 90 to 96\% of the feature variance in the audio-only and text-only conditions at layers 8, 16 and 24, reducing their effective dimension to about one (participation ratio 1.2 to 1.3). It accounts for 22 to 34\% in the other conditions and at most 1\% in the other two models. That dimension can only lower the audio-only and text-only values: the audio share of the full score is therefore a lower bound, and any contrast that subtracts the text-only value is an upper bound. No control without that channel is available for the Qwen3-Omni features.

Depth analyses fitted separate encoding models at four relative depths for each of the seven models (7 $\times$ 4 = 28 fits), recording the peak relative depth and the last-to-peak ratio. Relations between peak position and model size or capability composite were measured by Spearman rank correlation with exact $P$ values from all 5,040 permutations at $n = 7$. The capability composite is the mean $z$-score over three public capability benchmarks, WorldSense\cite{WorldSense2025}, Daily-Omni\cite{DailyOmni2025} and MMAU\cite{MMAU2024}, used only as an external measure of model capability. Supplementary Table 11 lists the single-layer brain-likeness of the seven models at the four depths. Modality and depth analyses identify where the correspondence is strongest; the effective dimension of the differences between models is quantified~next.

\subsection*{Quantifying effective dimension}
To quantify how the models differ once the shared map is removed, the Schaefer-1000 encoding maps of the seven models (participant-averaged) are stacked into $\mathbf{M}\in\mathbb{R}^{7\times1000}$. The shared map $\bar{\mathbf{m}}\in\mathbb{R}^{1000}$ is the mean over models, and the residual matrix $\mathbf{E}=\mathbf{M}-\mathbf{1}\bar{\mathbf{m}}^{\top}$ is centred across models, so its rank is at most 6. Eigendecomposition of $\mathbf{E}\mathbf{E}^{\top}$ gives eigenvalues $\xi_1\ge\dots\ge\xi_6$, and the effective dimension of inter-model differences is the participation ratio,
\begin{equation}
\mathrm{PR}(\mathbf{E})=\frac{\left(\sum_i\xi_i\right)^2}{\sum_i\xi_i^2},\qquad 1\le\mathrm{PR}\le K-1,
\label{eq:12}
\end{equation}
where $K = 7$ is the number of models, PR = 1 means all inter-model differences lie along one direction and PR = 6 means that differences spread equally over six directions.

Four alternative explanations were tested by recomputing PR after each treatment: a global scalar gain per model (regressed out as $\hat\zeta_i=\mathbf{m}_i^{\top}\bar{\mathbf{m}}/\bar{\mathbf{m}}^{\top}\bar{\mathbf{m}}$, with $\mathbf{m}_i$ row $i$ of $\mathbf{M}$, or removed by per-model $z$-standardization; measured gains 0.960 to 1.073), measurement noise (2,000 centred $7\times1000$ Gaussian matrices at the measured residual standard deviation 0.0213), parcel resolution (aggregation to seven networks, random subsets of 100, 250 or 500 parcels, or split halves) and a signal-to-noise map (text-only encoding map residualized from the map of each model). Under the five treatments PR stayed between 1.17 and 1.33 (Supplementary Table 4). The geometric corollary was tested by correlating the parcel-wise capability-brain coupling map with the first residual principal component against the median over 5,040 model-label permutations ($|\rho| = 0.606$ against a null median of 0.602). A low effective dimension permits the seven-network partition to be injected as a whole.

\subsection*{Algonauts 2025 out-of-distribution evaluation}
This leaderboard evaluation uses the official out-of-distribution set of the Algonauts 2025 challenge: training responses comprise about 55 h of \textit{Friends} and about 10 h of Movie10 films, and evaluation stimuli are about 2 h of six films absent from the released training responses\cite{Gifford2025,Scotti2025}. True responses are withheld; the organizers average parcel-wise Pearson correlations over parcels, films and participants\cite{Gifford2025}. The public leaderboard reports five columns (four participants and their mean). Figures report 27 ranked entries captured on 10 November 2026 at 08:23 UTC; the raw response and numerical source data are supplied with this~\mbox{submission}.

In the encoding system, an expert is the brain prediction obtained by a participant-specific, parcel-wise ridge readout from the internal layers of one omni model, and the mixture of experts is organized by hemisphere and Yeo-7 network (14 groups per participant). This organization was validated without any out-of-distribution response. With experts from four model families fitted on \textit{Friends}, each of four Movie10 films was held out in turn, all settings were chosen on the other three films, and the held-out film was scored by the parcel-mean Pearson correlation. The network-organized pool was compared with the equal-weight average over families and with a single expert (Supplementary Table 16). Because these experts had been explored in earlier development, the comparison is development evidence rather than an independent test, and the pool tested here is smaller than the submitted system.

The current leaderboard result is submission 1014621, uploaded on 28 October 2026, and is separate from the fixed seven-model brain-likeness protocol. This extension starts from the archived predictions of submission 972781 and retains nine successive residual-refinement stages. Documented candidate banks include Qwen3-Omni Instruct and Thinking\cite{Qwen3Omni2025}, Qwen2.5-Omni 3B and 7B\cite{Qwen25Omni2025}, MiniCPM-o 4.5\cite{MiniCPMo2026}, Ming-flash-omni-2.0, Nemotron-3-Nano-Omni\cite{Nemotron2026}, Gemma 4\cite{Gemma4_2026}, TLive-Omni\cite{TLiveOmni2026}, MiMo-V2.5\cite{MiMo2026} and Inkling-Small\cite{Inkling2026}, including internal audio and vision branches of MiniCPM. For newly generated banks, PCA or hybrid PCA/partial least squares (PLS) bases were fitted on \textit{Friends} data, ridge penalties were selected using whole-Movie10-title validation, and deployment readouts were refitted on the labelled \textit{Friends} and Movie10 data, with delays of 2, 3, 4 and 5 TRs.

The final refinement combined a cortical-network-reweighted residual, a stage-specific network-routed residual and an additional encoding-pool residual with weights 0.52, 0.28 and 0.20, respectively. Network reweighting used the seven Yeo networks; the final combined residual was centred and orthogonalized parcel-wise against the preceding prediction and residual on scored rows, then normalized. For the final prediction, 24 recorded signed mixture coefficients, one per participant and out-of-distribution film, combined the preceding prediction with this residual; residual additions at the first and last five rows of each clip were tapered. Each parcel was standardized using the pooled scored rows of both film halves before conversion to float32. Returned participant-by-film out-of-distribution scores informed these mixture coefficients, so this is an adaptive post-challenge benchmark result rather than an untouched out-of-distribution generalization test. A local replay using the recorded coefficients and prediction-combination code reproduced all 48 final prediction arrays exactly from the archived stage-8 prediction and stage-9 residual. Supplementary Data 1 provides this code, the coefficients and verification reports; the required archived prediction and residual arrays are not included. This reproducibility record starts from the identified archived anchor; it does not reconstruct every earlier feature-producing ancestor of that anchor. In the current snapshot, submission 1014621 ranked first by participant-mean correlation ($r = 0.25512$), ahead of submission 1034419 ($r = 0.25333$) by 0.00178 from unrounded scores; its participant ranks were 2, 1, 2 and 1 for sub-01, sub-02, sub-03 and sub-05,~respectively.

\subsection*{Data and participants}
All analyses above draw on one fMRI dataset. Brain imaging data came from the public naturalistic-viewing dataset of the CNeuroMod project\cite{Gifford2025,CNeuroMod2023}. Four participants (sub-01, sub-02, sub-03, sub-05; three men and one woman, all right-handed, within the 31-to-47-year age range of the cohort at recruitment in 2018, native or bilingual French speakers with advanced English comprehension) watched seasons 1 to 6 of the sitcom \textit{Friends} in a 3 T scanner (TR = 1.49 s). Acquisition used a Siemens Prisma Fit scanner with a 64-channel head/neck coil and an accelerated simultaneous multi-slice gradient-echo echo-planar sequence (slice acceleration 4, TE = 37 ms, flip angle 52 degrees, 2 mm isotropic voxels, 60 slices). Whole-brain functional images were preprocessed by the data providers with fMRIPrep 20.2.5, with slice-timing correction disabled, and released projected onto the Schaefer-1000 cortical parcellation\cite{Schaefer2018} in MNI152NLin2009cAsym space, with each parcel labelled by its Yeo-7 network membership (visual, somatomotor, dorsal attention, salience/ventral attention, limbic, control, default)\cite{Yeo2011}. The providers standardized and smoothed each run (5-mm kernel) and regressed out 24 head-motion parameters, the global signal and the mean white-matter and cerebrospinal-fluid signals with high-pass filtering, without scrubbing. The cortical response at each TR is a vector $\mathbf{y}_{s,\tau}\in\mathbb{R}^{1000}$, where $s$ indexes participant and $\tau$ indexes TR. Seasons 1 to 5 formed the training set and season 6 the validation set; the split is at episode level, so the two sets share no continuous narrative segment. The stimulus segment at each TR consists of video frames, audio waveform and subtitles; modality analyses mask one or two input channels with all other processing unchanged.

\subsection*{Ethics}
The CNeuroMod project was approved by the institutional research ethics board of the CIUSSS du Centre-Sud-de-l'\^Ile-de-Montr\'eal and most recently renewed by the Comit\'e d'\'ethique de la recherche Vieillissement et neuroimagerie (CER-VN), project number CER VN 18-19-22. Participants gave written informed consent to participate and, separately, to share their data. The present study used only the publicly released, de-identified data and collected no new human data.

\subsection*{Use of large language models}
Large language models were used as research instruments in the construction of Brain-AVQA, as documented above: Qwen3.5-Omni-Plus\cite{Qwen35_2026} generated four-choice questions from the audiovisual content of each window; Qwen3-Max, a proprietary API model of the Qwen3 series, translated occasional Chinese items into English (temperature 0, five questions per batch); and Gemini 3.5 Flash, independent of the generator, audited every question jointly on audio and video and again after translation, retaining only questions with a unique correct answer derivable from the window without textual shortcuts. The authors reviewed the audited benchmark, including per-network examples and difficulty labels, and take full responsibility for its content. Large language models also assisted language editing of the manuscript, bibliographic verification and reference-related editorial corrections; the authors reviewed all such output and take full responsibility for the text. No large language model generated the core research reasoning or analyses.

\subsection*{Computational resources and software}
All experiments ran on a single machine with 8 NVIDIA RTX A6000 GPUs (48 GB; CUDA 12.9). Stage-two training per cell took 0.3 to 1.2 h for MiniCPM-o 4.5 (one GPU), 2.9 to 15.4 h for Qwen3-Omni (two GPUs) and 1.2 to 6.5 h for Ming-flash-omni-2.0 (five to eight GPUs); stage-one network-wise training took about 0.08 h per network on MiniCPM-o 4.5. Each (base $\times$ benchmark $\times$ arm) cell uses a single training seed, and the two arms share split and seed. Seeds: training and split, 0 in both stages; row-level held-out split, 20260612; media and episode grouping, 20260813; answer-position permutation, 20260812; stage-two repeats, 1 and 2; stage-one repeat, 1; random-map draw, 20260905. Encoding models and statistics were run in Python 3.10.20 with PyTorch 2.8.0, NumPy 1.26.4, SciPy 1.15.3 and scikit-learn 1.7.2. Brain-MoE training used one environment per base, with Transformers 4.57.1 and PyTorch 2.5.1 for Ming-flash-omni-2.0, Transformers 4.51.0 and PyTorch 2.8.0 for MiniCPM-o 4.5, and Transformers 4.57.6, PyTorch 2.3.1 (CUDA 11.8 build) and Python 3.11.15 for Qwen3-Omni.

\subsection*{Statistics and reproducibility}
No statistical method was used to predetermine sample size; the $n$ of every analysis is stated in the Results and figure legends. Data exclusions: benchmark questions were excluded only by the task-format eligibility rules described above (two to four options for MMAU-Pro, three eligible distractors for MUSIC-AVQA), when the option count or answer field could not be parsed, or when the media file was unavailable or could not be decoded (Supplementary Table 13 and AVQA details above), and matched-ablation networks whose control arm failed activation matching were excluded from the primary criterion. Replication: each cell of the main matrix was trained with a single seed, repeated runs reproduced results bit-for-bit under the same seed, and the paired design controls common randomness; on MiniCPM-o 4.5 stage two was repeated with two further seeds on four benchmarks and stage one with one further seed (Supplementary Note 1); independent replication across three bases and five benchmarks is the unit of generalization. Randomization: question-level and media-level splits were drawn by seeded pseudo-random sampling; no other allocation was randomized, because the study consists of computational experiments in which no participant or sample is assigned to a condition. Blinding: not applicable, because all evaluations are automated and no human rater was involved; the investigators were not blind to arm identity during analysis, because arm identity cannot be masked: the arms differ in model configuration rather than in a label.

Multiplicity: the correlations computed on the seven-model panel are reported raw and after Holm correction across that family, which holds six coefficients: peak relative depth against parameter count and against the capability composite, the last-to-peak ratio against the capability composite, the partial Spearman and the zero-order Pearson coefficient between the capability composite and whole-cortex brain-likeness, and the model-relabelling test of the mean parcel-wise coupling (raw $P = 0.571$, 1.000, 0.048, 0.38, 0.54 and 0.55). The smallest raw value, $P = 0.048$ for the last-to-peak ratio, becomes $P = 0.29$ after correction, so no correlation on this panel is significant once the family is accounted for. The seven models are not seven independent families, because Qwen3-Omni Thinking and Instruct differ only in post-training and Qwen2.5-Omni 3B and 7B are two sizes of one family, so the effective number of independent units is smaller than the nominal $n = 7$.

Comparisons of two arms on the same questions use per-question paired exact McNemar tests: with $n_{10}$ and $n_{01}$ the discordant counts in the two directions, under the null $n_{10}\sim\mathrm{Binomial}(n_{10}+n_{01},1/2)$ and
\begin{equation}
\begin{aligned}
P&=\min\{1,\ 2P_{\mathrm{tail}}\},\\
P_{\mathrm{tail}}&=\sum_{i\le\min(n_{10},n_{01})}\tbinom{n_{10}+n_{01}}{i}\big(\tfrac12\big)^{n_{10}+n_{01}},
\end{aligned}
\label{eq:13}
\end{equation}
two-sided unless stated otherwise, with a significance level of 0.05 for all tests. The $n_{\mathrm{f}}$ tests of a family are corrected by the Holm step-down method: with $P$ values in ascending order, $P^{\mathrm{Holm}}_{(i)}=\max_{j\le i}\min\{1,(n_{\mathrm{f}}-j+1)P_{(j)}\}$; whenever the array size changes, corrected values are recomputed from raw $P$ values. When the permutation space is enumerable, all permutations are used (5,040 for seven models, 12,045 circular shifts) and the lower bound of $P$ is the reciprocal of the number of permutations; otherwise Monte Carlo sampling with $N_{\mathrm{perm}}$ permutations gives $(1+\#\{\mathrm{extreme}\})/(1+N_{\mathrm{perm}})$.

Because the three bases evaluate the same questions, pooled inference uses a question-level joint sign-flip permutation instead of Stouffer or Fisher combination: with $\delta_{m,b,q}\in\{-1,0,+1\}$ the paired difference of question $q$ on base $m$ and benchmark $b$, each permutation $u$ draws one sign $\varepsilon^{(u)}_q\in\{-1,+1\}$ per question, shared across the three bases,
\begin{equation}
\begin{aligned}
Q^{(u)}_{\mathrm{count}}&=\sum_{(m,b)}\mathbf{1}\Big[\sum_q\varepsilon^{(u)}_q \delta_{m,b,q}>0\Big],\\
Q^{(u)}_{\mathrm{mean}}&=\frac{1}{15}\sum_{(m,b)}\frac{1}{n_b}\sum_q\varepsilon^{(u)}_q \delta_{m,b,q},
\end{aligned}
\label{eq:14}
\end{equation}
with $N_{\mathrm{perm}} = 100{,}000$ and random seed 0.

Effect sizes are reported at cell level because averaging across benchmarks reverses the sign of the correlation between base accuracy and gain (cell-level Spearman $\rho = -0.82$, base-level +1.00). Comparisons across parcels use Wilcoxon signed-rank tests (large-sample normal approximation); rank correlations use Spearman coefficients with mid-ranks for ties, with exact $P$ values (all 5,040 permutations) for the seven-model panel and $P$ values from the $t$ approximation for the 15 cells. Accuracies and hit rates carry Wilson score intervals (Fig.~\ref{fig:avqa}a,f,h,k) or, in Fig.~\ref{fig:head}g, normal-approximation binomial intervals; paired differences carry Clopper-Pearson intervals on the discordant-pair proportion (Fig.~\ref{fig:avqa}c,g) or normal-approximation intervals (Fig.~\ref{fig:matrix}b and Fig.~\ref{fig:head}f); intervals across parcels are the mean $\pm$ 1.96 standard errors (Fig.~\ref{fig:head}h) or a paired bootstrap over the 1,000 parcels (Fig.~\ref{fig:mod}a). The tests used are exact, permutation-based or rank-based and do not assume normally distributed data; the $\chi^2$ test compares categorical counts. All analyses were performed in Python with PyTorch, NumPy, SciPy and scikit-learn (versions under Computational resources and software).

\subsection*{Code availability}
An open-source release of the brain encoding pipeline, Brain-Scope training, Brain-MoE two-stage training, control arms and factorial design, matched ablation, Brain-AVQA construction and audit, and statistical analysis and plotting code is available at Zenodo (DOI: \mbox{10.5281/zenodo.22326682})\cite{BrainAVQAdata}, in the same deposit as the data. The plotting code for the updated leaderboard panels, final-stage replay code and local verification record are provided in Supplementary Data 1. The data and annotations of that deposit are released under a CC BY 4.0 licence, which is also the licence recorded for the deposit as a whole, whereas the code is released under the MIT licence, as stated in the licence section of the deposit description and in the README of the~record.

\section*{Data availability}
CNeuroMod naturalistic-viewing fMRI and Algonauts 2025 challenge data were obtained under their original licences from the public Algonauts 2025 repository and the CNeuroMod data portal\cite{Gifford2025,CNeuroMod2023} (Methods). OmniBench\cite{OmniBench2024}, MMAU\cite{MMAU2024}, MMAU-Pro\cite{MMAUPro2025}, MuChoMusic\cite{MuChoMusic2024}, MUSIC-AVQA v2.0\cite{MusicAVQA2_2024}, AVQA\cite{Yang2022AVQA}, WorldSense\cite{WorldSense2025} and Daily-Omni\cite{DailyOmni2025} are public benchmarks available from the sources given in the cited publications. The Brain-AVQA benchmark constructed here (2,362 questions with dominant-network annotations, consensus difficulty labels, three held-out splits, audit records and generation scripts) is available under a CC BY 4.0 licence at Zenodo (DOI: \mbox{10.5281/zenodo.22326682})\cite{BrainAVQAdata}; the benchmark distributes only the episode and time coordinates of 8-s windows and slice checksums, and video stimuli are obtained from these official repositories and sliced locally. The same Zenodo record contains parcel-wise encoding maps, Brain-Scope checkpoints and readout heads, per-question predictions and figure files from release 1.0.2 in vector PDF and derived SVG/PNG formats. The archived prediction and residual arrays of the leaderboard submission are available from the corresponding author on reasonable request. The weights of the seven omni models are publicly available.

\FloatBarrier
\setlength{\bibsep}{0pt}

\backmatter

\bibliography{refs}

\begin{thebibliography}{10}
\expandafter\ifx\csname url\endcsname\relax
  \def\url#1{\burl{#1}}\fi
\expandafter\ifx\csname urlprefix\endcsname\relax\def\urlprefix{URL }\fi
\providecommand{\bibinfo}[2]{#2}
\providecommand{\eprint}[2][]{\url{#2}}
\providecommand{\doi}[1]{\url{https://doi.org/#1}}
\bibcommenthead

\bibitem{Huh2024}
\bibinfo{author}{Huh, M.}, \bibinfo{author}{Cheung, B.}, \bibinfo{author}{Wang,
  T.} \& \bibinfo{author}{Isola, P.} \emph{\bibinfo{title}{Position: the
  {Platonic} representation hypothesis}}.
\newblock \emph{\bibinfo{booktitle}{Proceedings of the 41st International
  Conference on Machine Learning}}, Vol. \bibinfo{volume}{235},
  \bibinfo{pages}{20617--20642} (\bibinfo{publisher}{PMLR},
  \bibinfo{year}{2024}).

\bibitem{Gurnee2026}
\bibinfo{author}{Gurnee, W.} \emph{et~al.}
\newblock \bibinfo{title}{Verbalizable representations form a global workspace
  in language models} (\bibinfo{year}{2026}).
\newblock \bibinfo{note}{Preprint at \url{https://arxiv.org/abs/2607.15495}}.

\bibitem{Yamins2014}
\bibinfo{author}{Yamins, D. L.~K.} \emph{et~al.}
\newblock \bibinfo{title}{Performance-optimized hierarchical models predict
  neural responses in higher visual cortex}.
\newblock \emph{\bibinfo{journal}{Proc. Natl Acad. Sci. USA}}
  \textbf{\bibinfo{volume}{111}}, \bibinfo{pages}{8619--8624}
  (\bibinfo{year}{2014}).

\bibitem{Schrimpf2021}
\bibinfo{author}{Schrimpf, M.} \emph{et~al.}
\newblock \bibinfo{title}{The neural architecture of language: integrative
  modeling converges on predictive processing}.
\newblock \emph{\bibinfo{journal}{Proc. Natl Acad. Sci. USA}}
  \textbf{\bibinfo{volume}{118}}, \bibinfo{pages}{e2105646118}
  (\bibinfo{year}{2021}).

\bibitem{Goldstein2022}
\bibinfo{author}{Goldstein, A.} \emph{et~al.}
\newblock \bibinfo{title}{Shared computational principles for language
  processing in humans and deep language models}.
\newblock \emph{\bibinfo{journal}{Nat. Neurosci.}}
  \textbf{\bibinfo{volume}{25}}, \bibinfo{pages}{369--380}
  (\bibinfo{year}{2022}).

\bibitem{Mischler2024}
\bibinfo{author}{Mischler, G.}, \bibinfo{author}{Li, Y.~A.},
  \bibinfo{author}{Bickel, S.}, \bibinfo{author}{Mehta, A.~D.} \&
  \bibinfo{author}{Mesgarani, N.}
\newblock \bibinfo{title}{Contextual feature extraction hierarchies converge in
  large language models and the brain}.
\newblock \emph{\bibinfo{journal}{Nat. Mach. Intell.}}
  \textbf{\bibinfo{volume}{6}}, \bibinfo{pages}{1467--1477}
  (\bibinfo{year}{2024}).

\bibitem{LopezCardona2025}
\bibinfo{author}{Lopez-Cardona, A.}, \bibinfo{author}{Idesis, S.},
  \bibinfo{author}{Masias-Bruns, M.}, \bibinfo{author}{Abadal, S.} \&
  \bibinfo{author}{Arapakis, I.} \emph{\bibinfo{title}{Brain--language model
  alignment: insights into the {Platonic} hypothesis and intermediate-layer
  advantage}}.
\newblock \emph{\bibinfo{booktitle}{Proceedings of UniReps: the Third Edition
  of the Workshop on Unifying Representations in Neural Models}}, Vol.
  \bibinfo{volume}{322} of \emph{\bibinfo{series}{Proceedings of Machine
  Learning Research}}, \bibinfo{pages}{81--101} (\bibinfo{publisher}{PMLR},
  \bibinfo{year}{2026}).
\newblock
  \urlprefix\url{https://proceedings.mlr.press/v322/lopez-cardona26a.html}.

\bibitem{Doerig2025}
\bibinfo{author}{Doerig, A.} \emph{et~al.}
\newblock \bibinfo{title}{High-level visual representations in the human brain
  are aligned with large language models}.
\newblock \emph{\bibinfo{journal}{Nat. Mach. Intell.}}
  \textbf{\bibinfo{volume}{7}}, \bibinfo{pages}{1220--1234}
  (\bibinfo{year}{2025}).

\bibitem{Du2025}
\bibinfo{author}{Du, C.} \emph{et~al.}
\newblock \bibinfo{title}{Human-like object concept representations emerge
  naturally in multimodal large language models}.
\newblock \emph{\bibinfo{journal}{Nat. Mach. Intell.}}
  \textbf{\bibinfo{volume}{7}}, \bibinfo{pages}{860--875}
  (\bibinfo{year}{2025}).

\bibitem{Gifford2025}
\bibinfo{author}{Gifford, A.~T.} \emph{et~al.}
\newblock \bibinfo{title}{The {Algonauts} {Project} 2025 {Challenge}: how the
  human brain makes sense of multimodal movies} (\bibinfo{year}{2025}).
\newblock \bibinfo{note}{Preprint at \url{https://arxiv.org/abs/2501.00504}}.

\bibitem{dAscoli2025}
\bibinfo{author}{d'Ascoli, S.}, \bibinfo{author}{Rapin, J.},
  \bibinfo{author}{Benchetrit, Y.}, \bibinfo{author}{Banville, H.} \&
  \bibinfo{author}{King, J.-R.} \emph{\bibinfo{title}{{TRIBE}: {TRImodal} brain
  encoder for whole-brain {fMRI} response prediction}}.
\newblock \emph{\bibinfo{booktitle}{International Conference on Learning
  Representations (ICLR)}}, \bibinfo{pages}{152742--152759}
  (\bibinfo{year}{2026}).
\newblock
  \urlprefix\url{https://proceedings.iclr.cc/paper_files/paper/2026/hash/f7e8bc4c853e3e58bc487e213c79c587-Abstract-Conference.html}.

\bibitem{Gokce2026}
\bibinfo{author}{Gokce, A.}, \bibinfo{author}{AlKhamissi, B.} \&
  \bibinfo{author}{Schrimpf, M.}
\newblock \bibinfo{title}{{MIRAGE}: adaptive multimodal gating for whole-brain
  {fMRI} encoding} (\bibinfo{year}{2026}).
\newblock \bibinfo{note}{Preprint at \url{https://arxiv.org/abs/2605.29850}}.

\bibitem{Scotti2025}
\bibinfo{author}{Scotti, P.~S.} \& \bibinfo{author}{Tripathy, M.}
\newblock \bibinfo{title}{Insights from the {Algonauts} 2025 winners}
  (\bibinfo{year}{2025}).
\newblock \bibinfo{note}{Preprint at \url{https://arxiv.org/abs/2508.10784}}.

\bibitem{Abdollahi2025}
\bibinfo{author}{Abdollahi, H.}, \bibinfo{author}{Majoumerd, A. H.~M.},
  \bibinfo{author}{Baboukani, A. H.~B.}, \bibinfo{author}{Suratgar, A.~A.} \&
  \bibinfo{author}{Menhaj, M.~B.}
\newblock \bibinfo{title}{Probing multimodal fusion in the brain: the dominance
  of audiovisual streams in naturalistic encoding} (\bibinfo{year}{2025}).
\newblock \bibinfo{note}{Preprint at \url{https://arxiv.org/abs/2507.19052}}.

\bibitem{Ying2026BrainLoRA}
\bibinfo{author}{Ying, Z.~J.} \emph{et~al.}
\newblock \bibinfo{title}{Scaling parameter-efficient fine-tuning for brain
  encoding} (\bibinfo{year}{2026}).
\newblock \bibinfo{note}{Preprint at \url{https://arxiv.org/abs/2610.26802}}.

\bibitem{Merlin2026}
\bibinfo{author}{Merlin, G.} \& \bibinfo{author}{Toneva, M.}
  \emph{\bibinfo{title}{When language models lose their mind: the consequences
  of brain misalignment}}.
\newblock \emph{\bibinfo{booktitle}{International Conference on Learning
  Representations (ICLR)}}, \bibinfo{pages}{83181--83235}
  (\bibinfo{year}{2026}).
\newblock
  \urlprefix\url{https://proceedings.iclr.cc/paper_files/paper/2026/hash/864b96035c34651e12202f8cd1d5e945-Abstract-Conference.html}.

\bibitem{Li2019}
\bibinfo{author}{Li, Z.} \emph{et~al.} \emph{\bibinfo{title}{Learning from
  brains how to regularize machines}}.
\newblock \emph{\bibinfo{booktitle}{Advances in Neural Information Processing
  Systems (NeurIPS)}}, Vol.~\bibinfo{volume}{32}, \bibinfo{pages}{9497--9507}
  (\bibinfo{publisher}{Curran Associates, Inc.}, \bibinfo{year}{2019}).
\newblock
  \urlprefix\url{https://proceedings.neurips.cc/paper_files/paper/2019/file/70117ee3c0b15a2950f1e82a215e812b-Paper.pdf}.

\bibitem{Moussa2025}
\bibinfo{author}{Moussa, O.} \& \bibinfo{author}{Toneva, M.}
  \emph{\bibinfo{title}{Brain-tuning improves generalizability and efficiency
  of brain alignment in speech models}}.
\newblock \emph{\bibinfo{booktitle}{Advances in Neural Information Processing
  Systems (NeurIPS)}}, Vol.~\bibinfo{volume}{38},
  \bibinfo{pages}{135575--135600} (\bibinfo{publisher}{Curran Associates,
  Inc.}, \bibinfo{year}{2025}).

\bibitem{Policzer2025}
\bibinfo{author}{Policzer, N.}, \bibinfo{author}{Braunstein, C.} \&
  \bibinfo{author}{Toneva, M.} \emph{\bibinfo{title}{The one where they
  brain-tune for social cognition: multi-modal brain-tuning on {Friends}}}.
\newblock \emph{\bibinfo{booktitle}{NeurIPS 2025 Workshop on Interpreting
  Cognition in Deep Learning Models}} (\bibinfo{year}{2025}).
\newblock \urlprefix\url{https://openreview.net/forum?id=5kyIDVYrUv}.

\bibitem{Xiao2026}
\bibinfo{author}{Xiao, M.}, \bibinfo{author}{Du, K.} \& \bibinfo{author}{Lin,
  Z.}
\newblock \bibinfo{title}{Beyond representational alignment with brain-guided
  language models for robust reasoning}.
\newblock \emph{\bibinfo{journal}{Nat. Mach. Intell.}}
  \textbf{\bibinfo{volume}{8}}, \bibinfo{pages}{1275--1289}
  (\bibinfo{year}{2026}).

\bibitem{Toneva2019}
\bibinfo{author}{Toneva, M.} \& \bibinfo{author}{Wehbe, L.}
  \emph{\bibinfo{title}{Interpreting and improving natural-language processing
  (in machines) with natural language-processing (in the brain)}}.
\newblock \emph{\bibinfo{booktitle}{Advances in Neural Information Processing
  Systems (NeurIPS)}}, Vol.~\bibinfo{volume}{32}, \bibinfo{pages}{14887--14897}
  (\bibinfo{publisher}{Curran Associates, Inc.}, \bibinfo{year}{2019}).
\newblock
  \urlprefix\url{https://proceedings.neurips.cc/paper_files/paper/2019/file/749a8e6c231831ef7756db230b4359c8-Paper.pdf}.

\bibitem{AlKhamissi2026}
\bibinfo{author}{AlKhamissi, B.} \emph{et~al.} \emph{\bibinfo{title}{Mixture of
  cognitive reasoners: modular reasoning with brain-like specialization}}.
\newblock \emph{\bibinfo{booktitle}{International Conference on Learning
  Representations (ICLR)}}, \bibinfo{pages}{15683--15722}
  (\bibinfo{year}{2026}).
\newblock
  \urlprefix\url{https://proceedings.iclr.cc/paper_files/paper/2026/hash/1a596f504fc914cabfbfdd78da071930-Abstract-Conference.html}.

\bibitem{Ren2026}
\bibinfo{author}{Ren, Y.}, \bibinfo{author}{Shi, P.}, \bibinfo{author}{Ma, Z.},
  \bibinfo{author}{He, X.} \& \bibinfo{author}{Li, X.}
\newblock \bibinfo{title}{{FPED}: a functional-network prior-guided
  mixture-of-experts framework for interpretable brain decoding}
  (\bibinfo{year}{2026}).
\newblock \bibinfo{note}{Preprint at \url{https://arxiv.org/abs/2605.19279}}.

\bibitem{Cook2025}
\bibinfo{author}{Cook, J.}, \bibinfo{author}{Akarca, D.},
  \bibinfo{author}{Costa, R.~P.} \& \bibinfo{author}{Achterberg, J.}
  \emph{\bibinfo{title}{Brain-like processing pathways form in models with
  heterogeneous experts}}.
\newblock \emph{\bibinfo{booktitle}{Advances in Neural Information Processing
  Systems (NeurIPS)}}, Vol.~\bibinfo{volume}{38}, \bibinfo{pages}{93092--93129}
  (\bibinfo{publisher}{Curran Associates, Inc.}, \bibinfo{year}{2025}).

\bibitem{Binhuraib2025}
\bibinfo{author}{Binhuraib, T.}, \bibinfo{author}{Tuckute, G.} \&
  \bibinfo{author}{Blauch, N.~M.} \emph{\bibinfo{title}{Topoformer: brain-like
  topographic organization in {Transformer} language models through spatial
  querying and reweighting}}.
\newblock \emph{\bibinfo{booktitle}{ICLR 2024 Workshop on Representational
  Alignment (Re-Align)}} (\bibinfo{year}{2024}).
\newblock \urlprefix\url{https://openreview.net/forum?id=3pLMzgoZSA}.

\bibitem{AlKhamissi2026b}
\bibinfo{author}{AlKhamissi, B.} \emph{et~al.}
\newblock \bibinfo{title}{Discovering functionally selective brain regions with
  a deep topographic multimodal model} (\bibinfo{year}{2026}).
\newblock \bibinfo{note}{Preprint at \url{https://arxiv.org/abs/2606.09770}}.

\bibitem{Koepke2026}
\bibinfo{author}{Koepke, A.~S.}, \bibinfo{author}{Zverev, D.},
  \bibinfo{author}{Ginosar, S.} \& \bibinfo{author}{Efros, A.~A.}
\newblock \bibinfo{title}{Back into {Plato}'s cave: examining cross-modal
  representational convergence at scale} (\bibinfo{year}{2026}).
\newblock \bibinfo{note}{Preprint at \url{https://arxiv.org/abs/2604.18572}}.

\bibitem{Jia2026}
\bibinfo{author}{Jia, X.}
\newblock \bibinfo{title}{Do language models align with brains? {Prediction}
  scores are not enough} (\bibinfo{year}{2026}).
\newblock \bibinfo{note}{Preprint at \url{https://arxiv.org/abs/2605.14025}}.

\bibitem{Yeo2011}
\bibinfo{author}{Yeo, B. T.~T.} \emph{et~al.}
\newblock \bibinfo{title}{The organization of the human cerebral cortex
  estimated by intrinsic functional connectivity}.
\newblock \emph{\bibinfo{journal}{J. Neurophysiol.}}
  \textbf{\bibinfo{volume}{106}}, \bibinfo{pages}{1125--1165}
  (\bibinfo{year}{2011}).

\bibitem{Ming2026}
\bibinfo{author}{{Inclusion AI}}.
\newblock \bibinfo{title}{Ming-flash-omni-2.0} (\bibinfo{year}{2026}).
\newblock \bibinfo{note}{Model card at
  \url{https://huggingface.co/inclusionAI/Ming-flash-omni-2.0} (accessed 11
  November 2026)}.

\bibitem{MiniCPMo2026}
\bibinfo{author}{Cui, J.} \emph{et~al.}
\newblock \bibinfo{title}{{MiniCPM-o} 4.5: towards real-time full-duplex
  omni-modal interaction} (\bibinfo{year}{2026}).
\newblock \bibinfo{note}{Preprint at \url{https://arxiv.org/abs/2604.27393}}.

\bibitem{Qwen3Omni2025}
\bibinfo{author}{Xu, J.} \emph{et~al.}
\newblock \bibinfo{title}{{Qwen3-Omni} technical report}
  (\bibinfo{year}{2025}).
\newblock \bibinfo{note}{Preprint at \url{https://arxiv.org/abs/2509.17765}}.

\bibitem{Hu2022LoRA}
\bibinfo{author}{Hu, E.~J.} \emph{et~al.} \emph{\bibinfo{title}{{LoRA}:
  low-rank adaptation of large language models}}.
\newblock \emph{\bibinfo{booktitle}{International Conference on Learning
  Representations (ICLR)}} (\bibinfo{year}{2022}).
\newblock \urlprefix\url{https://openreview.net/forum?id=nZeVKeeFYf9}.

\bibitem{CNeuroMod2023}
\bibinfo{author}{Boyle, J.~A.} \emph{et~al.} \emph{\bibinfo{title}{The
  {Courtois} {NeuroMod} project: quality assessment of the initial data release
  (2020)}}.
\newblock \emph{\bibinfo{booktitle}{Conference on Cognitive Computational
  Neuroscience (CCN)}}, \bibinfo{pages}{389--392}
  (\bibinfo{publisher}{Cognitive Computational Neuroscience},
  \bibinfo{address}{Oxford, UK}, \bibinfo{year}{2023}).

\bibitem{Qwen35_2026}
\bibinfo{author}{{Qwen Team}}.
\newblock \bibinfo{title}{{Qwen3.5-Omni} technical report}
  (\bibinfo{year}{2026}).
\newblock \bibinfo{note}{Preprint at \url{https://arxiv.org/abs/2604.15804}}.

\bibitem{Gemini35_2026}
\bibinfo{author}{{Google DeepMind}}.
\newblock \bibinfo{title}{{Gemini} 3.5 {Flash} model card}
  (\bibinfo{year}{2026}).
\newblock
  \bibinfo{note}{\url{https://deepmind.google/models/model-cards/gemini-3-5-flash/}
  (accessed 30 August 2026)}.

\bibitem{Liu2026GDPO}
\bibinfo{author}{Liu, S.-Y.} \emph{et~al.} \emph{\bibinfo{title}{{GDPO}: group
  reward-decoupled normalization policy optimization for multi-reward {RL}
  optimization}}.
\newblock \emph{\bibinfo{booktitle}{Proceedings of the 43rd International
  Conference on Machine Learning}}, Vol. \bibinfo{volume}{306},
  \bibinfo{pages}{75978--76000} (\bibinfo{publisher}{PMLR},
  \bibinfo{year}{2026}).
\newblock \urlprefix\url{https://proceedings.mlr.press/v306/liu26aa.html}.

\bibitem{OmniBench2024}
\bibinfo{author}{Li, Y.} \emph{et~al.} \emph{\bibinfo{title}{{OmniBench}:
  towards the future of universal omni-language models}}.
\newblock \emph{\bibinfo{booktitle}{Advances in Neural Information Processing
  Systems (NeurIPS)}}, Vol.~\bibinfo{volume}{38}, \bibinfo{pages}{34822--34851}
  (\bibinfo{publisher}{Curran Associates, Inc.}, \bibinfo{year}{2025}).
\newblock
  \urlprefix\url{https://proceedings.neurips.cc/paper_files/paper/2025/file/2c7c4a12a9dcf7dace9896d08154c705-Paper-Datasets_and_Benchmarks_Track.pdf}.

\bibitem{MMAU2024}
\bibinfo{author}{Sakshi, S.} \emph{et~al.} \emph{\bibinfo{title}{{MMAU}: a
  massive multi-task audio understanding and reasoning benchmark}}.
\newblock \emph{\bibinfo{booktitle}{International Conference on Learning
  Representations (ICLR)}}, \bibinfo{pages}{84929--84964}
  (\bibinfo{year}{2025}).
\newblock
  \urlprefix\url{https://proceedings.iclr.cc/paper_files/paper/2025/file/d36f208919582785db965fe648b9fe59-Paper-Conference.pdf}.

\bibitem{MMAUPro2025}
\bibinfo{author}{Kumar, S.} \emph{et~al.} \emph{\bibinfo{title}{{MMAU-Pro}: a
  challenging and comprehensive benchmark for holistic evaluation of audio
  general intelligence}}.
\newblock \emph{\bibinfo{booktitle}{Proceedings of the AAAI Conference on
  Artificial Intelligence}}, Vol.~\bibinfo{volume}{40},
  \bibinfo{pages}{22688--22697} (\bibinfo{publisher}{AAAI Press},
  \bibinfo{year}{2026}).

\bibitem{MuChoMusic2024}
\bibinfo{author}{Weck, B.} \emph{et~al.} \emph{\bibinfo{title}{{MuChoMusic}:
  evaluating music understanding in multimodal audio-language models}}.
\newblock \emph{\bibinfo{booktitle}{Proc. 25th International Society for Music
  Information Retrieval Conference (ISMIR)}}, \bibinfo{pages}{825--833}
  (\bibinfo{publisher}{ISMIR}, \bibinfo{year}{2024}).

\bibitem{MusicAVQA2022}
\bibinfo{author}{Li, G.} \emph{et~al.} \emph{\bibinfo{title}{Learning to answer
  questions in dynamic audio-visual scenarios}}.
\newblock \emph{\bibinfo{booktitle}{Proc. IEEE/CVF Conference on Computer
  Vision and Pattern Recognition (CVPR)}}, \bibinfo{pages}{19086--19096}
  (\bibinfo{publisher}{IEEE}, \bibinfo{year}{2022}).

\bibitem{MusicAVQA2_2024}
\bibinfo{author}{Liu, X.}, \bibinfo{author}{Dong, Z.} \&
  \bibinfo{author}{Zhang, P.} \emph{\bibinfo{title}{Tackling data bias in
  {MUSIC-AVQA}: crafting a balanced dataset for unbiased question-answering}}.
\newblock \emph{\bibinfo{booktitle}{Proc. IEEE/CVF Winter Conference on
  Applications of Computer Vision (WACV)}}, \bibinfo{pages}{4466--4475}
  (\bibinfo{publisher}{IEEE}, \bibinfo{year}{2024}).

\bibitem{Gao2025SAE}
\bibinfo{author}{Gao, L.} \emph{et~al.} \emph{\bibinfo{title}{Scaling and
  evaluating sparse autoencoders}}.
\newblock \emph{\bibinfo{booktitle}{International Conference on Learning
  Representations (ICLR)}}, \bibinfo{pages}{26721--26754}
  (\bibinfo{year}{2025}).
\newblock
  \urlprefix\url{https://proceedings.iclr.cc/paper_files/paper/2025/hash/42ef3308c230942d223c411adf182c88-Abstract-Conference.html}.

\bibitem{Deng2026}
\bibinfo{author}{Deng, B.} \emph{et~al.}
\newblock \bibinfo{title}{{Qwen-Scope}: turning sparse features into
  development tools for large language models} (\bibinfo{year}{2026}).
\newblock \bibinfo{note}{Preprint at \url{https://arxiv.org/abs/2605.11887}}.

\bibitem{Kuhn1955}
\bibinfo{author}{Kuhn, H.~W.}
\newblock \bibinfo{title}{The {Hungarian} method for the assignment problem}.
\newblock \emph{\bibinfo{journal}{Nav. Res. Logist. Q.}}
  \textbf{\bibinfo{volume}{2}}, \bibinfo{pages}{83--97} (\bibinfo{year}{1955}).

\bibitem{Schaefer2018}
\bibinfo{author}{Schaefer, A.} \emph{et~al.}
\newblock \bibinfo{title}{Local-global parcellation of the human cerebral
  cortex from intrinsic functional connectivity {MRI}}.
\newblock \emph{\bibinfo{journal}{Cereb. Cortex}}
  \textbf{\bibinfo{volume}{28}}, \bibinfo{pages}{3095--3114}
  (\bibinfo{year}{2018}).

\bibitem{Nemotron2026}
\bibinfo{author}{{NVIDIA}} \emph{et~al.}
\newblock \bibinfo{title}{Nemotron 3 {Nano} {Omni}: efficient and open
  multimodal intelligence} (\bibinfo{year}{2026}).
\newblock \bibinfo{note}{Preprint at \url{https://arxiv.org/abs/2604.24954}}.

\bibitem{Qwen25Omni2025}
\bibinfo{author}{Xu, J.} \emph{et~al.}
\newblock \bibinfo{title}{{Qwen2.5-Omni} technical report}
  (\bibinfo{year}{2025}).
\newblock \bibinfo{note}{Preprint at \url{https://arxiv.org/abs/2503.20215}}.

\bibitem{Gao2025}
\bibinfo{author}{Gao, C.} \emph{et~al.}
\newblock \bibinfo{title}{Increasing alignment of large language models with
  language processing in the human brain}.
\newblock \emph{\bibinfo{journal}{Nat. Comput. Sci.}}
  \textbf{\bibinfo{volume}{5}}, \bibinfo{pages}{1080--1090}
  (\bibinfo{year}{2025}).

\bibitem{WorldSense2025}
\bibinfo{author}{Hong, J.} \emph{et~al.} \emph{\bibinfo{title}{{WorldSense}:
  evaluating real-world omnimodal understanding for multimodal {LLMs}}}.
\newblock \emph{\bibinfo{booktitle}{International Conference on Learning
  Representations (ICLR)}}, \bibinfo{pages}{52423--52443}
  (\bibinfo{year}{2026}).
\newblock
  \urlprefix\url{https://proceedings.iclr.cc/paper_files/paper/2026/file/55dc32df563a35bf406717fb52c118d5-Paper-Conference.pdf}.

\bibitem{DailyOmni2025}
\bibinfo{author}{Zhou, Z.}, \bibinfo{author}{Wang, R.}, \bibinfo{author}{Wu,
  Z.} \& \bibinfo{author}{Jiang, Y.-G.}
\newblock \bibinfo{title}{{Daily-Omni}: towards audio-visual reasoning with
  temporal alignment across modalities} (\bibinfo{year}{2025}).
\newblock \bibinfo{note}{Preprint at \url{https://arxiv.org/abs/2505.17862}}.

\bibitem{Kazemian2025}
\bibinfo{author}{Kazemian, A.}, \bibinfo{author}{Elmoznino, E.} \&
  \bibinfo{author}{Bonner, M.~F.}
\newblock \bibinfo{title}{Convolutional architectures are cortex-aligned de
  novo}.
\newblock \emph{\bibinfo{journal}{Nat. Mach. Intell.}}
  \textbf{\bibinfo{volume}{7}}, \bibinfo{pages}{1834--1844}
  (\bibinfo{year}{2025}).

\bibitem{Lepori2026}
\bibinfo{author}{Lepori, M.~A.}, \bibinfo{author}{Kay, K.} \&
  \bibinfo{author}{Tuckute, G.}
\newblock \bibinfo{title}{Interpreting brain responses to language with sparse
  features from language models} (\bibinfo{year}{2026}).
\newblock \bibinfo{note}{Preprint at \url{https://arxiv.org/abs/2606.06857}}.

\bibitem{Li2026ViSAE}
\bibinfo{author}{Li, T.}, \bibinfo{author}{Chen, Y.}, \bibinfo{author}{Ma, M.}
  \& \bibinfo{author}{Peng, X.} \emph{\bibinfo{title}{Inside the visual mind:
  neuroscience-motivated concept circuits for interpreting and steering vision
  transformers}}.
\newblock \emph{\bibinfo{booktitle}{Proceedings of the 43rd International
  Conference on Machine Learning}}, Vol. \bibinfo{volume}{306},
  \bibinfo{pages}{66270--66288} (\bibinfo{publisher}{PMLR},
  \bibinfo{year}{2026}).
\newblock \urlprefix\url{https://proceedings.mlr.press/v306/li26s.html}.

\bibitem{Guo2026a}
\bibinfo{author}{Guo, D.}, \bibinfo{author}{Wu, J.} \& \bibinfo{author}{Yiu,
  S.~M.}
\newblock \bibinfo{title}{Sparse autoencoders map brain-{LLM} alignment onto
  cortical semantic topography} (\bibinfo{year}{2026}).
\newblock \bibinfo{note}{Preprint at \url{https://arxiv.org/abs/2605.23035}}.

\bibitem{Williams1992}
\bibinfo{author}{Williams, R.~J.}
\newblock \bibinfo{title}{Simple statistical gradient-following algorithms for
  connectionist reinforcement learning}.
\newblock \emph{\bibinfo{journal}{Mach. Learn.}} \textbf{\bibinfo{volume}{8}},
  \bibinfo{pages}{229--256} (\bibinfo{year}{1992}).

\bibitem{Loshchilov2019}
\bibinfo{author}{Loshchilov, I.} \& \bibinfo{author}{Hutter, F.}
  \emph{\bibinfo{title}{Decoupled weight decay regularization}}.
\newblock \emph{\bibinfo{booktitle}{International Conference on Learning
  Representations (ICLR)}} (\bibinfo{year}{2019}).
\newblock \urlprefix\url{https://openreview.net/forum?id=Bkg6RiCqY7}.

\bibitem{Qwen3_2025}
\bibinfo{author}{{Qwen Team}}.
\newblock \bibinfo{title}{{Qwen3-Max}: just scale it} (\bibinfo{year}{2025}).
\newblock \bibinfo{note}{Qwen blog, 24 September 2025,
  \url{https://qwen.ai/blog?id=qwen3-max} (accessed 5 September 2026)}.

\bibitem{Yang2022AVQA}
\bibinfo{author}{Yang, P.} \emph{et~al.} \emph{\bibinfo{title}{{AVQA}: a
  dataset for audio-visual question answering on videos}}.
\newblock \emph{\bibinfo{booktitle}{Proc. 30th ACM International Conference on
  Multimedia}}, \bibinfo{pages}{3480--3491} (\bibinfo{publisher}{ACM},
  \bibinfo{year}{2022}).

\bibitem{Gemma4_2026}
\bibinfo{author}{{Gemma Team}} \emph{et~al.}
\newblock \bibinfo{title}{Gemma 4 technical report} (\bibinfo{year}{2026}).
\newblock \bibinfo{note}{Preprint at \url{https://arxiv.org/abs/2607.02770}}.

\bibitem{TLiveOmni2026}
\bibinfo{author}{Hu, Y.} \emph{et~al.}
\newblock \bibinfo{title}{{TLive-Omni}: an omni-modal understanding model for
  e-commerce live streaming} (\bibinfo{year}{2026}).
\newblock \bibinfo{note}{Preprint at \url{https://arxiv.org/abs/2608.20958}}.

\bibitem{MiMo2026}
\bibinfo{author}{{Xiaomi MiMo Team}}.
\newblock \bibinfo{title}{{MiMo-V2.5} model card} (\bibinfo{year}{2026}).
\newblock \bibinfo{note}{\url{https://huggingface.co/XiaomiMiMo/MiMo-V2.5}
  (accessed 5 September 2026)}.

\bibitem{Inkling2026}
\bibinfo{author}{{Thinking Machines Lab}}.
\newblock \bibinfo{title}{{Inkling-Small} model card} (\bibinfo{year}{2026}).
\newblock \bibinfo{note}{Model card at
  \url{https://thinkingmachines.ai/model-card/inkling-small/} (accessed 11
  November 2026)}.

\bibitem{BrainAVQAdata}
\bibinfo{author}{Zhang, P.}, \bibinfo{author}{Tian, B.}, \bibinfo{author}{Li,
  X.} \& \bibinfo{author}{Liu, L.}
\newblock \bibinfo{title}{Brain-{AVQA} and the {Platonic} brain bridge: data,
  model checkpoints and code}.
\newblock \bibinfo{howpublished}{Zenodo} (\bibinfo{year}{2026}).
\newblock \urlprefix\url{https://doi.org/10.5281/zenodo.22326682}.
\newblock \bibinfo{note}{Version 1.0.2}.

\end{thebibliography}


\begin{thebibliography}{10}
\expandafter\ifx\csname url\endcsname\relax
  \def\url#1{\burl{#1}}\fi
\expandafter\ifx\csname urlprefix\endcsname\relax\def\urlprefix{URL }\fi
\providecommand{\bibinfo}[2]{#2}
\providecommand{\eprint}[2][]{\url{#2}}
\providecommand{\doi}[1]{\url{https://doi.org/#1}}
\bibcommenthead

\bibitem{Yang2022AVQA}
\bibinfo{author}{Yang, P.} \emph{et~al.} \emph{\bibinfo{title}{{AVQA}: a
  dataset for audio-visual question answering on videos}}.
\newblock \emph{\bibinfo{booktitle}{Proc. 30th ACM International Conference on
  Multimedia}}, \bibinfo{pages}{3480--3491} (\bibinfo{publisher}{ACM},
  \bibinfo{year}{2022}).

\bibitem{dAscoli2025}
\bibinfo{author}{d'Ascoli, S.}, \bibinfo{author}{Rapin, J.},
  \bibinfo{author}{Benchetrit, Y.}, \bibinfo{author}{Banville, H.} \&
  \bibinfo{author}{King, J.-R.} \emph{\bibinfo{title}{{TRIBE}: {TRImodal} brain
  encoder for whole-brain {fMRI} response prediction}}.
\newblock \emph{\bibinfo{booktitle}{International Conference on Learning
  Representations (ICLR)}}, \bibinfo{pages}{152742--152759}
  (\bibinfo{year}{2026}).
\newblock
  \urlprefix\url{https://proceedings.iclr.cc/paper_files/paper/2026/hash/f7e8bc4c853e3e58bc487e213c79c587-Abstract-Conference.html}.

\bibitem{dAscoli2026}
\bibinfo{author}{d'Ascoli, S.} \emph{et~al.}
\newblock \bibinfo{title}{A foundation model of vision, audition, and language
  for in-silico neuroscience} (\bibinfo{year}{2026}).
\newblock \bibinfo{note}{Preprint at \url{https://arxiv.org/abs/2605.04326}}.

\bibitem{Gokce2026}
\bibinfo{author}{Gokce, A.}, \bibinfo{author}{AlKhamissi, B.} \&
  \bibinfo{author}{Schrimpf, M.}
\newblock \bibinfo{title}{{MIRAGE}: adaptive multimodal gating for whole-brain
  {fMRI} encoding} (\bibinfo{year}{2026}).
\newblock \bibinfo{note}{Preprint at \url{https://arxiv.org/abs/2605.29850}}.

\bibitem{Binhuraib2025}
\bibinfo{author}{Binhuraib, T.}, \bibinfo{author}{Tuckute, G.} \&
  \bibinfo{author}{Blauch, N.~M.} \emph{\bibinfo{title}{Topoformer: brain-like
  topographic organization in {Transformer} language models through spatial
  querying and reweighting}}.
\newblock \emph{\bibinfo{booktitle}{ICLR 2024 Workshop on Representational
  Alignment (Re-Align)}} (\bibinfo{year}{2024}).
\newblock \urlprefix\url{https://openreview.net/forum?id=3pLMzgoZSA}.

\bibitem{AlKhamissi2026b}
\bibinfo{author}{AlKhamissi, B.} \emph{et~al.}
\newblock \bibinfo{title}{Discovering functionally selective brain regions with
  a deep topographic multimodal model} (\bibinfo{year}{2026}).
\newblock \bibinfo{note}{Preprint at \url{https://arxiv.org/abs/2606.09770}}.

\bibitem{Xiao2026}
\bibinfo{author}{Xiao, M.}, \bibinfo{author}{Du, K.} \& \bibinfo{author}{Lin,
  Z.}
\newblock \bibinfo{title}{Beyond representational alignment with brain-guided
  language models for robust reasoning}.
\newblock \emph{\bibinfo{journal}{Nat. Mach. Intell.}}
  \textbf{\bibinfo{volume}{8}}, \bibinfo{pages}{1275--1289}
  (\bibinfo{year}{2026}).

\bibitem{Moussa2025}
\bibinfo{author}{Moussa, O.} \& \bibinfo{author}{Toneva, M.}
  \emph{\bibinfo{title}{Brain-tuning improves generalizability and efficiency
  of brain alignment in speech models}}.
\newblock \emph{\bibinfo{booktitle}{Advances in Neural Information Processing
  Systems (NeurIPS)}}, Vol.~\bibinfo{volume}{38},
  \bibinfo{pages}{135575--135600} (\bibinfo{publisher}{Curran Associates,
  Inc.}, \bibinfo{year}{2025}).

\bibitem{Policzer2025}
\bibinfo{author}{Policzer, N.}, \bibinfo{author}{Braunstein, C.} \&
  \bibinfo{author}{Toneva, M.} \emph{\bibinfo{title}{The one where they
  brain-tune for social cognition: multi-modal brain-tuning on {Friends}}}.
\newblock \emph{\bibinfo{booktitle}{NeurIPS 2025 Workshop on Interpreting
  Cognition in Deep Learning Models}} (\bibinfo{year}{2025}).
\newblock \urlprefix\url{https://openreview.net/forum?id=5kyIDVYrUv}.

\bibitem{AlKhamissi2026}
\bibinfo{author}{AlKhamissi, B.} \emph{et~al.} \emph{\bibinfo{title}{Mixture of
  cognitive reasoners: modular reasoning with brain-like specialization}}.
\newblock \emph{\bibinfo{booktitle}{International Conference on Learning
  Representations (ICLR)}}, \bibinfo{pages}{15683--15722}
  (\bibinfo{year}{2026}).
\newblock
  \urlprefix\url{https://proceedings.iclr.cc/paper_files/paper/2026/hash/1a596f504fc914cabfbfdd78da071930-Abstract-Conference.html}.

\bibitem{Ren2026}
\bibinfo{author}{Ren, Y.}, \bibinfo{author}{Shi, P.}, \bibinfo{author}{Ma, Z.},
  \bibinfo{author}{He, X.} \& \bibinfo{author}{Li, X.}
\newblock \bibinfo{title}{{FPED}: a functional-network prior-guided
  mixture-of-experts framework for interpretable brain decoding}
  (\bibinfo{year}{2026}).
\newblock \bibinfo{note}{Preprint at \url{https://arxiv.org/abs/2605.19279}}.

\bibitem{Wei2025}
\bibinfo{author}{Wei, Y.} \emph{et~al.} \emph{\bibinfo{title}{{MoRE-Brain}:
  routed mixture of experts for interpretable and generalizable cross-subject
  {fMRI} visual decoding}}.
\newblock \emph{\bibinfo{booktitle}{Advances in Neural Information Processing
  Systems (NeurIPS)}}, Vol.~\bibinfo{volume}{38}, \bibinfo{pages}{62254--62286}
  (\bibinfo{publisher}{Curran Associates, Inc.}, \bibinfo{year}{2025}).

\bibitem{Yu2026}
\bibinfo{author}{Yu, J.} \emph{et~al.} \emph{\bibinfo{title}{Local intrinsic
  dimension of representations predicts alignment and generalization in {AI}
  models and human brain}}.
\newblock \emph{\bibinfo{booktitle}{Proceedings of the 43rd International
  Conference on Machine Learning}}, Vol. \bibinfo{volume}{306},
  \bibinfo{pages}{150763--150792} (\bibinfo{publisher}{PMLR},
  \bibinfo{year}{2026}).
\newblock \urlprefix\url{https://proceedings.mlr.press/v306/yu26ag.html}.

\bibitem{Cheng2026}
\bibinfo{author}{Cheng, E.}, \bibinfo{author}{Vaidya, A.~R.} \&
  \bibinfo{author}{Antonello, R.~J.} \emph{\bibinfo{title}{Abstraction induces
  the brain alignment of language and speech models}}.
\newblock \emph{\bibinfo{booktitle}{Proceedings of the 43rd International
  Conference on Machine Learning}}, Vol. \bibinfo{volume}{306},
  \bibinfo{pages}{19008--19031} (\bibinfo{publisher}{PMLR},
  \bibinfo{year}{2026}).
\newblock \urlprefix\url{https://proceedings.mlr.press/v306/cheng26l.html}.

\bibitem{OmniBench2024}
\bibinfo{author}{Li, Y.} \emph{et~al.} \emph{\bibinfo{title}{{OmniBench}:
  towards the future of universal omni-language models}}.
\newblock \emph{\bibinfo{booktitle}{Advances in Neural Information Processing
  Systems (NeurIPS)}}, Vol.~\bibinfo{volume}{38}, \bibinfo{pages}{34822--34851}
  (\bibinfo{publisher}{Curran Associates, Inc.}, \bibinfo{year}{2025}).
\newblock
  \urlprefix\url{https://proceedings.neurips.cc/paper_files/paper/2025/file/2c7c4a12a9dcf7dace9896d08154c705-Paper-Datasets_and_Benchmarks_Track.pdf}.

\bibitem{MMAU2024}
\bibinfo{author}{Sakshi, S.} \emph{et~al.} \emph{\bibinfo{title}{{MMAU}: a
  massive multi-task audio understanding and reasoning benchmark}}.
\newblock \emph{\bibinfo{booktitle}{International Conference on Learning
  Representations (ICLR)}}, \bibinfo{pages}{84929--84964}
  (\bibinfo{year}{2025}).
\newblock
  \urlprefix\url{https://proceedings.iclr.cc/paper_files/paper/2025/file/d36f208919582785db965fe648b9fe59-Paper-Conference.pdf}.

\bibitem{MMAUPro2025}
\bibinfo{author}{Kumar, S.} \emph{et~al.} \emph{\bibinfo{title}{{MMAU-Pro}: a
  challenging and comprehensive benchmark for holistic evaluation of audio
  general intelligence}}.
\newblock \emph{\bibinfo{booktitle}{Proceedings of the AAAI Conference on
  Artificial Intelligence}}, Vol.~\bibinfo{volume}{40},
  \bibinfo{pages}{22688--22697} (\bibinfo{publisher}{AAAI Press},
  \bibinfo{year}{2026}).

\bibitem{MuChoMusic2024}
\bibinfo{author}{Weck, B.} \emph{et~al.} \emph{\bibinfo{title}{{MuChoMusic}:
  evaluating music understanding in multimodal audio-language models}}.
\newblock \emph{\bibinfo{booktitle}{Proc. 25th International Society for Music
  Information Retrieval Conference (ISMIR)}}, \bibinfo{pages}{825--833}
  (\bibinfo{publisher}{ISMIR}, \bibinfo{year}{2024}).

\bibitem{MusicAVQA2022}
\bibinfo{author}{Li, G.} \emph{et~al.} \emph{\bibinfo{title}{Learning to answer
  questions in dynamic audio-visual scenarios}}.
\newblock \emph{\bibinfo{booktitle}{Proc. IEEE/CVF Conference on Computer
  Vision and Pattern Recognition (CVPR)}}, \bibinfo{pages}{19086--19096}
  (\bibinfo{publisher}{IEEE}, \bibinfo{year}{2022}).

\bibitem{MusicAVQA2_2024}
\bibinfo{author}{Liu, X.}, \bibinfo{author}{Dong, Z.} \&
  \bibinfo{author}{Zhang, P.} \emph{\bibinfo{title}{Tackling data bias in
  {MUSIC-AVQA}: crafting a balanced dataset for unbiased question-answering}}.
\newblock \emph{\bibinfo{booktitle}{Proc. IEEE/CVF Winter Conference on
  Applications of Computer Vision (WACV)}}, \bibinfo{pages}{4466--4475}
  (\bibinfo{publisher}{IEEE}, \bibinfo{year}{2024}).

\end{thebibliography}

\bmhead{Acknowledgements}
Vector icons in the schematics are from Servier Medical Art (CC BY 4.0), Icons8 Flat Color Icons (MIT) and Phosphor Icons (MIT). This study was funded by the National Natural Science Foundation of China (grant number 62471420) and the GuangDong Basic and Applied Basic Research Foundation (grant number 2025A1515012296), both awarded to L.L. The funders played no role in study design, data collection, analysis and interpretation of data, or the writing of this manuscript.

\bmhead{Author contributions}
P.Z. conceived the study with L.L., designed and implemented the encoding pipeline, Brain-Scope, Brain-MoE and Brain-AVQA, ran the experiments and analysed the data. B.T. and X.L. provided model infrastructure and contributed to the evaluation of the omni models. L.L. supervised the project. P.Z. and L.L. wrote the manuscript with input from all authors. All authors read and approved the final~manuscript.

\bmhead{Competing interests}
B.T. and X.L. are employees of Alibaba Group, which develops the Qwen model family evaluated in this work. P.Z. completed this work during an internship at Alibaba Token Hub, Alibaba Group. P.Z., B.T. and X.L. declare no other financial and no non-financial competing interests. L.L. declares no financial or non-financial competing interests.

\end{document}

\title[Supplementary Information]{Supplementary Information for: The Platonic brain bridge hypothesis: brain networks as architectural priors for multimodal large language models}
\author[1]{\fnm{Pengfei} \sur{Zhang}}
\author[2]{\fnm{Biao} \sur{Tian}}
\author[2]{\fnm{Xiangang} \sur{Li}}
\author*[1]{\fnm{Li} \sur{Liu}}\email{avrillliu@hkust-gz.edu.cn}

\affil[1]{\orgdiv{AI Thrust, Information Hub}, \orgname{The Hong Kong University of Science and Technology (Guangzhou)}, \orgaddress{\city{Guangzhou}, \country{China}}}
\affil[2]{\orgdiv{Alibaba Token Hub}, \orgname{Alibaba Group}, \orgaddress{\city{Beijing}, \country{China}}}

\maketitle

\section*{Supplementary Note 1: additional quantitative results}

\subsection*{Forward spectrum}

Modality structure. On the three brain-to-model bases (Ming-flash-omni-2.0, MiniCPM-o 4.5 and Qwen3-Omni, in that order), audio alone exceeded text alone by +0.081, +0.085 and +0.095, respectively. The increments were sub-additive: audio added between +0.042 and +0.062 to video but only +0.027 to +0.032 to video with subtitles, and subtitles added only +0.002, +0.009 and +0.008 (Supplementary Table 2).

Depth structure. Six of the seven models peaked at half depth and none at the last layer; the mean last-to-peak ratio was 0.947. The last-layer decline grew with a capability composite from three public benchmarks (Spearman $\rho = -0.786$, exhaustive permutation $P = 0.048$, $n = 7$), but after Holm correction over the family of six correlations no relation on this seven-model panel remained significant (smallest corrected $P = 0.29$). The best single layer reached 97.6\% to 99.3\% of the four-layer score.

Parcel-wise encoding maps of the seven models had pairwise Spearman $\rho = 0.987$ (range 0.964 to 0.9996; $n = 21$ pairs). In the 10 November 2026 snapshot of the post-challenge Algonauts 2025 leaderboard, the differences between submission 1014621 and the entry ranked second by mean score (1034419) were $-0.00233$, $+0.00149$, $-0.00032$ and $+0.00830$ for sub-01, sub-02, sub-03 and sub-05, respectively. The participant-mean difference was $+0.00178$, calculated before rounding. Peak relative depth was unrelated to model size (Spearman $\rho = -0.412$, exhaustive permutation $P = 0.571$, $n = 7$) and to the capability composite ($\rho = 0.000$, $P = 1.000$); the last-to-peak ratio ranged from 0.915 to 0.981. The capability composite correlated positively but not significantly with whole-cortex brain-likeness (partial Spearman +0.433 controlling for parameter count, exhaustive permutation $P = 0.38$, $n = 7$; mean parcel-wise partial correlation +0.129, 67.4\% of parcels positive, label permutation $P = 0.55$). An alternative statistical explanation of the last-layer decline, anisotropic last-layer states leaving less variance in a fixed number of components, predicts a decline at every depth; instead Nemotron-3-Nano-Omni peaks at quarter depth and the decline of Ming-flash-omni-2.0 is confined to the last quartile (0.3737 at 3/4 depth, 0.3750 at the peak, 0.3677 at the last layer).

Details summarized in the main text. Audio alone reached 87\%, 91\% and 94\% of the full-modality score on Ming-flash-omni-2.0, MiniCPM-o 4.5 and Qwen3-Omni, and every input subset scored below every subset containing it (paired parcel bootstrap, 10,000 resamples, $n = 1{,}000$ parcels). The brain-aligned sparse features used for cross-model matching numbered 185, 184 and 190 for Qwen3-Omni, Ming-flash-omni-2.0 and MiniCPM-o 4.5, 1.1 to 2.3 per cent of the autoencoder basis. The encoding maps of the seven models shared a component holding 97.87\% of total variance; after its removal, the first two principal components held 87.8\% and 9.2\% of the residual variance, respectively.

\subsection*{Brain-AVQA held-out and in-domain results}

Held-out gains on Brain-AVQA. On the 709 held-out questions, Brain-MoE on MiniCPM-o 4.5 exceeded the frozen base by +3.24 pp (paired exact McNemar test, two-sided $P = 0.0032$, $n = 709$); on consensus hard questions answered correctly by at most one frozen base the gain was +9.26 pp ($P = 2.7\times10^{-4}$, $n = 162$), and on questions both other bases missed it was +7.59 pp ($P = 0.070$, $n = 79$).

Same-question design and a sixth benchmark. Models whose experts and routing network were trained on all 2,362 questions, including these 709, gained $+2.68$, $+6.17$ and $+3.80$ percentage points more on all questions, on consensus hard questions and on leave-one-out hard questions respectively, and those three differences are the size of the training effect in each of the three strata. On a sixth public benchmark, AVQA\cite{Yang2022AVQA}, we evaluated MiniCPM-o 4.5 on 1,473 held-out questions from our eligible subset (Methods); accuracy rose from 0.8778 for the frozen base to 0.8968 for the brain arm (40 questions gained, 12 lost; $P = 1.28\times10^{-4}$).

In the in-domain fitting regime the base accuracies were 0.8213 (Qwen3-Omni), 0.7371 (Ming-flash-omni-2.0) and 0.6592 (MiniCPM-o 4.5), and consensus-hard gains were +12.69, +12.88 and +17.42 percentage points (pp), 2.3 to 3.5 times the full-set gains. In the same-question design on MiniCPM-o 4.5, models trained with the 709 questions gave +5.92/+15.43/+11.39 pp (all/consensus hard/leave-one-out hard) against +3.24/+9.26/+7.59 pp for models trained without them (leave-one-out hard: $P = 0.070$, $n = 79$); the leave-one-out hard comparison had 7/1 discordant pairs. On AVQA the gain was +1.90 pp ($P = 1.28\times10^{-4}$, $n = 1{,}473$). Held-out and training sets did not differ in base accuracy (Mann-Whitney $U$ tests, $P = 0.44$, 0.47 and 0.53) or consensus-difficulty composition ($\chi^2$ test, $P = 0.525$). The pooled real-minus-control routing hit-rate gap was +0.1513 in the half-episode split and +0.1518 in the whole-episode split ($P<2\times10^{-4}$), and the difference between splits for the real arm was +0.0062 ($P = 0.659$); routing hits were distributed across all seven networks.

\subsection*{Public-benchmark matrix and headroom dependence}
Random-arm gains over the base ranged from +0.20 to +8.76 pp across the 15 cells. Per-cell McNemar tests reached $P<0.05$ in nine cells, of which the three MMAU-Pro cells, two OmniBench cells and Qwen3-Omni $\times$ MMAU remained significant after Holm correction; the tied cell was MiniCPM-o 4.5 $\times$ MUSIC-AVQA v2.0 (discordant pairs 2/2). Base-level mean $\Delta_2$ were +0.0387 (Ming-flash-omni-2.0), +0.0337 (MiniCPM-o 4.5) and +0.0255 (Qwen3-Omni). Regression to the mean was bounded by simulating 5,000 draws with constant true gain and independent sampling of the two arms: the null distributions of $\rho$ had means $-0.098$ and $-0.003$ for the two levels, and the observed $-0.82$ and $-0.80$ fell outside both null distributions (simulation $P = 4\times10^{-4}$ and $2\times10^{-4}$). Shuffling benchmark labels only within each base (20,000 Monte Carlo draws) left the correlation significant ($P = 0.0026$). Leaving out any benchmark gave $\rho$ from $-0.687$ to $-0.939$, and any base from $-0.687$ to $-0.879$. Headroom and held-out size are collinear across benchmarks (Spearman $\rho = +0.90$, $n = 5$); in a joint standardized regression, effect size followed headroom (standardized coefficient $b = +0.555$), not held-out question count (+0.187), whereas evidence strength $-\log_{10}P$ followed held-out question count (+0.881), not headroom ($-0.060$). $\Delta_2$ correlated with log parameter count at $r = -0.021$. A quadratic fit in base accuracy (adjusted $R^2 = 0.723$) exceeded a linear fit (0.657), with positive curvature. Benchmark means over three bases: MMAU-Pro (headroom 0.431) and OmniBench (0.434) gave mean $\Delta_2$ of +0.0527 and +0.0450, MMAU (0.335) +0.0340, MuChoMusic (0.224) +0.0204 and MUSIC-AVQA v2.0 (0.158) +0.0112.

\subsection*{Repeated training seeds and random network-to-expert maps}
For MiniCPM-o 4.5 we added two further stage-two seeds (1 and 2) to the published seed-0 run on four of the five public benchmarks. $\Delta_2$ was positive in all 12 runs: OmniBench +6.83, +4.38 and +6.65 pp ($P = 3.8\times10^{-6}$, 0.0041 and $2.0\times10^{-5}$); MMAU +4.60, +2.80 and +4.80 pp ($P = 0.0011$, 0.059 and $9.0\times10^{-4}$); MMAU-Pro +5.31, +6.58 and +6.17 pp (all $P < 10^{-4}$); and MuChoMusic +1.65, +3.30 and +2.77 pp ($P = 0.25$, 0.016 and 0.073). Nine of the 12 differences reached $P < 0.05$. Each seed also redraws the media-key split, so runs are paired within a seed, and the MMAU runs used a 500-question split of the full 1,000-question test-mini set rather than the 490-question split of the main matrix, so their seed-0 value differs from the published one. Repeating stage one with a second seed gave $\Delta_2 = +6.13$ pp on OmniBench against +6.83 pp with the first seed; the brain arm itself moved by 0.70 pp, which is within noise ($P = 0.50$).

We also drew 20 of the 5,040 possible maps from brain networks to experts (seed 20260905, after removing the identity and a cyclic shift) and repeated the routing training under each of them on MiniCPM-o 4.5 in the in-domain fitting regime, with the experts held fixed and the routing network re-initialized and trained again. The real map was more accurate than all 20: 1.70 pp above their mean (s.d. 0.36 pp) and 0.93 pp above the best of them. The empirical $P$ is 0.048, the smallest value 20 draws can give. The routing hit rate of the real map was higher than that of every random map as well (0.3400 against at most 0.2202).

\subsection*{Pathways and feature ablation}
Content-pathway effects were +5.95 pp (MiniCPM-o 4.5 $\times$ MMAU-Pro, $P = 1.2\times10^{-12}$, $n = 2{,}203$), +7.31 pp (Ming-flash-omni-2.0 $\times$ MMAU-Pro, $P = 1.5\times10^{-16}$), +3.22 pp (Qwen3-Omni $\times$ MMAU-Pro, $P = 4.7\times10^{-8}$), +4.55 pp (MiniCPM-o 4.5 $\times$ OmniBench, $P = 0.0013$, $n = 571$) and +1.93 pp (Qwen3-Omni $\times$ OmniBench, $P = 0.052$); the two pathways sum exactly to the total effect. In the four-arm cells, content effects were +1.40 to +7.18 pp under the real prior and +1.93 to +5.95 pp under the shuffled prior; the smallest content effect (+1.40 pp) exceeded every prior effect except +2.28 pp in MiniCPM-o 4.5 $\times$ OmniBench. Under the reference criterion (all seven networks, $n = 1{,}000$ parcels) excess encoding damage was +0.00139, +0.00324 and +0.00891 ($P = 0.0022$, $8.4\times10^{-39}$ and $3.1\times10^{-73}$), and under the strict criterion for Qwen3-Omni ($n = 466$) +0.01054 ($P = 2.1\times10^{-40}$); the Holm-corrected $P$ for the Qwen3-Omni OmniBench downstream excess damage over five benchmarks was $3.7\times10^{-6}$, zeroing the matched control lowered accuracy only to 0.6077, and the pooled discordant pairs were 299/232.

\onecolumn
\section*{Supplementary Tables}
\begin{sitable}
\caption{Brain-likeness and depth structure of seven omni models}\label{tab:2}
\footnotesize
\begin{tabular*}{\textwidth}{@{\extracolsep{\fill}}lccccc@{}}
\toprule
Model & Parameters & $\bar r$ (four layers) & Best layer $\bar r$ & Peak depth & Last/peak \\
\midrule
Ming-flash-omni-2.0 & 100B & 0.3807 & 0.3750 & 0.50 & 0.981 \\
MiniCPM-o 4.5 & 9B & 0.3609 & 0.3582 & 0.50 & 0.915 \\
Nemotron-3-Nano-Omni & 31B & 0.3546 & 0.3463 & 0.25 & 0.949 \\
Qwen3-Omni Thinking & 30B & 0.3470 & 0.3441 & 0.50 & 0.938 \\
Qwen3-Omni Instruct$^{\ddagger}$ & 30B & 0.3470 & 0.3422 & 0.50 & 0.944 \\
Qwen2.5-Omni 7B & 7B & 0.3397 & 0.3338 & 0.50 & 0.947 \\
Qwen2.5-Omni 3B & 3B & 0.3367 & 0.3322 & 0.50 & 0.955 \\
\botrule
\end{tabular*}
{\par\vspace{3pt}\footnotesize\noindent $\bar r$ is the validation Pearson correlation with the participant-averaged response, averaged over 1,000 parcels (Schaefer-1000 parcellation, 512 principal components per layer concatenated over four layers, ridge regression, haemodynamic lag of two TRs). Best single-layer $\bar r$ is the maximum over the four relative depths fitted separately; last/peak ratio is the last-layer score divided by the peak score. The partial correlations with parameter count use total parameter counts of 100, 9, 31, 30.5, 30.5, 10.7 and 5.5 billion for the models in table order; the 2/4 and 3/4 scores of Qwen2.5-Omni 7B differ by 0.0002. $\ddagger$, the Instruct variant (Qwen3-Omni-30B-A3B-Instruct) is the Qwen3-Omni base used in all brain-to-model experiments (Brain-Scope and Brain-MoE).\par}
\end{sitable}

\begin{sitable}
\caption{Six input subsets of the modality ablation}\label{edt:1}
\footnotesize
\begin{tabular*}{\textwidth}{@{\extracolsep{\fill}}lrrrrrr@{}}
\toprule
Model & T & V & A & V+T & V+A & V+A+T \\
\midrule
Ming-flash-omni-2.0 & 0.250 & 0.317 & 0.331 & 0.348 & 0.379 & 0.381 \\
MiniCPM-o 4.5 & 0.244 & 0.311 & 0.329 & 0.330 & 0.352 & 0.361 \\
Qwen3-Omni (Instruct)$^{\S}$ & 0.246 & 0.299 & 0.341 & 0.338 & 0.357 & 0.365 \\
\botrule
\end{tabular*}
{\par\vspace{3pt}\footnotesize\noindent Whole-cortex brain-likeness $\bar r_{\mathcal{S}}$ (participant-averaged response) with only the listed input channels as input (T, subtitles; V, video; A, audio). Confidence intervals from a paired bootstrap over the 1,000 parcels (10,000 resamples) have half-widths of 0.0069 to 0.0095 on every cell. Every adjacent contrast excludes zero in all three models except A against V+T, for which the interval is $+0.0130$ to $+0.0215$ in Ming-flash-omni-2.0, $-0.0036$ to $+0.0055$ in MiniCPM-o 4.5 and $-0.0077$ to $+0.0002$ in Qwen3-Omni, so the two subsets are separated only in Ming-flash-omni-2.0. $\S$, separate float32 extraction at layers 8, 16, 24 and 48; see Methods. Increments and percentages quoted in Supplementary Note 1 are computed from unrounded values. Training rows vary by condition for Ming-flash-omni-2.0 (113,412 to 114,003) and are constant for the other two models; the validation split is identical in every cell ($n = 23{,}324$).\par}
\end{sitable}

\begin{sitable}
\caption{Cross-model matching of brain-aligned sparse features}\label{edt:3}
\footnotesize
\setlength{\tabcolsep}{3pt}
\begin{tabular*}{\textwidth}{@{\extracolsep{\fill}}lrrrrrr@{}}
\toprule
Model pair & Matched $|r|$ & Null maximum $|r|$ & Ratio & Random control $|r|$ & Ratio & Exact $P$ \\
\midrule
Qwen3-Omni, Ming-flash-omni-2.0 & 0.5257 & 0.0382 & 13.8 & 0.0407 & 12.9 & 1/12,045 \\
Qwen3-Omni, MiniCPM-o 4.5 & 0.5344 & 0.0392 & 13.6 & 0.0424 & 12.6 & 1/12,045 \\
Ming-flash-omni-2.0, MiniCPM-o 4.5 & 0.5068 & 0.0381 & 13.3 & 0.0372 & 13.6 & 1/12,045 \\
\botrule
\end{tabular*}
{\par\vspace{3pt}\footnotesize\noindent Each model contributes its brain-aligned features (185, 184 and 190 for Qwen3-Omni, Ming-flash-omni-2.0 and MiniCPM-o 4.5) on 13,383 shared stimulus grid points from 81 \textit{Friends} half-episode segments, matched one-to-one by the Hungarian algorithm. Shift null maximum is the largest mean matched $|r|$ over all 12,044 circular shifts of one series; random control is the mean over 30 size-matched random sets of non-brain-aligned features, none of which reached the observed value; $P$ is the exact circular-shift permutation over 12,045 shifts including the identity, Holm-corrected within the family to 3/12,045.\par}
\end{sitable}

\begin{sitable}
\caption{Robustness of the effective dimension of inter-model differences to four alternative explanations}\label{edt:4}
\footnotesize
\begin{tabular*}{\textwidth}{@{\extracolsep{\fill}}lp{6.5cm}@{}}
\toprule
Treatment & Participation ratio \\
\midrule
Raw residual matrix & 1.28 (PC1 87.8\%, PC2 9.2\%) \\
Per-model global gain regressed out & 1.21 \\
Per-model $z$-standardization & 1.18 \\
Aggregated to seven networks & 1.19 \\
Random 100/250/500 parcels & 1.29 $\pm$ 0.06 / 1.29 $\pm$ 0.03 / 1.28 $\pm$ 0.02 \\
Parcels split in half & 1.29 / 1.28 \\
Text-only baseline residualized & 1.17 \\
Text-only baseline residualized and $z$-standardized & 1.33 \\
Independent Gaussian noise (2,000 simulations) & 5.96 (95\% interval 5.93 to 5.98; none below the observed value) \\
\botrule
\end{tabular*}
\end{sitable}

\begin{sitable}
\caption{Three-arm decomposition into content and routing pathways}\label{edt:7}
\footnotesize
\setlength{\tabcolsep}{3pt}
\begin{tabular*}{\textwidth}{@{\extracolsep{\fill}}p{2.6cm}rp{2.9cm}p{2.9cm}p{2.7cm}@{}}
\toprule
Cell & $n$ & Total effect $\Delta_2$ & Content pathway & Routing pathway \\
\midrule
MiniCPM-o 4.5 $\times$ OmniBench & 571 & +0.0683 ($P = 3.8\times10^{-6}$, 55/16) & +0.0455 ($P = 0.0013$, 44/18) & +0.0228 ($P = 0.0072$, 17/4) \\
MiniCPM-o 4.5 $\times$ MMAU-Pro & 2,203 & +0.0531 ($P = 4.7\times10^{-10}$, 235/118) & +0.0595 ($P = 1.2\times10^{-12}$, 237/106) & $-0.0064$ ($P = 0.17$, 38/52) \\
Qwen3-Omni $\times$ OmniBench & 571 & +0.0140 ($P = 0.18$, 18/10) & +0.0193 ($P = 0.052$, 19/8) & $-0.0053$ ($P = 0.61$, 6/9) \\
Qwen3-Omni $\times$ MMAU-Pro & 2,203 & +0.0309 ($P = 1.9\times10^{-7}$, 119/51) & +0.0322 ($P = 4.7\times10^{-8}$, 120/49) & $-0.0014$ ($P = 0.78$, 25/28) \\
Ming-flash-omni-2.0 $\times$ MMAU-Pro & 2,203 & +0.0740 ($P = 2.5\times10^{-18}$, 260/97) & +0.0731 ($P = 1.5\times10^{-16}$, 274/113) & +0.0009 ($P = 0.91$, 38/36) \\
\botrule
\end{tabular*}
{\par\vspace{3pt}\footnotesize\noindent Total effect is BT $-$ RS, content pathway BS $-$ RS, routing pathway BT $-$ BS; parentheses give the paired exact McNemar $P$ and discordant pairs.\par}
\end{sitable}

\begin{sitable}
\caption{Three criteria of the matched ablation}\label{edt:8}
\footnotesize
\setlength{\tabcolsep}{3pt}
\begin{tabular*}{\textwidth}{@{\extracolsep{\fill}}lp{3.2cm}p{3.2cm}p{3.6cm}@{}}
\toprule
Base & Primary (visual and control excluded, $n = 709$) & Reference (all seven networks, $n = 1{,}000$) & Strict \\
\midrule
Ming-flash-omni-2.0 & +0.00128 ($P = 4.2\times10^{-5}$) & +0.00139 ($P = 0.0022$) & Not applicable \\
MiniCPM-o 4.5 & +0.00293 ($P = 1.2\times10^{-31}$) & +0.00324 ($P = 8.4\times10^{-39}$) & Not applicable \\
Qwen3-Omni & +0.01012 ($P = 4.1\times10^{-55}$) & +0.00891 ($P = 3.1\times10^{-73}$) & +0.01054 ($P = 2.1\times10^{-40}$, $n = 466$) \\
\botrule
\end{tabular*}
{\par\vspace{3pt}\footnotesize\noindent Excess damage is the mean parcel-wise encoding damage from zeroing brain-aligned features relative to the activation-matched control arm; $P$ from across-parcel Wilcoxon signed-rank tests.\par}
\end{sitable}

\begin{sitable}
\caption{Comparison of this work with the most related studies}\label{edt:9}
\footnotesize
\setlength{\tabcolsep}{3pt}
\begin{tabular*}{\textwidth}{@{\extracolsep{\fill}}>{\raggedright\arraybackslash}p{2.2cm}>{\raggedright\arraybackslash}p{2.0cm}>{\raggedright\arraybackslash}p{1.9cm}>{\raggedright\arraybackslash}p{2.3cm}>{\raggedright\arraybackslash}p{2.1cm}>{\raggedright\arraybackslash}p{2.1cm}>{\raggedright\arraybackslash}p{2.0cm}@{}}
\toprule
Work & Direction & Modality & Injected object & Capability validation & Attribution controls & Magnitude analysis \\
\midrule
TRIBE, TRIBE v2\cite{dAscoli2025,dAscoli2026} & Model to brain & Trimodal concatenation & None & None & Modality ablation & Encoding data-size scaling \\
MIRAGE\cite{Gokce2026} & Model to brain & Native omni & None & None & Native versus concatenated & None \\
Topoformer\cite{Binhuraib2025} & Brain principle to model & Text & Spatial querying and reweighting & GLUE, comparable to control BERT & Parameter- and training-matched non-topographic BERT & None \\
Topo-Omni\cite{AlKhamissi2026b} & Brain principle to model & Native omni & Topographic smoothness constraint & OmniBench, against a fine-tuned control & Cluster driving and suppression & None \\
Brain-guided language models\cite{Xiao2026} & Brain to model & Text & Task fMRI data & Reasoning benchmarks & Random signal, random direction & None \\
Brain-tuning\cite{Moussa2025,Policzer2025} & Brain to model & Speech, audiovisual & fMRI data & Speech probes; MUStARD and CMU-MOSEI & Stimulus- and LLM-tuned controls; objective/rank ablations (speech) & Data-size scaling (speech) \\
MiCRo\cite{AlKhamissi2026} & Brain principle to model & Text & Four cognitive domains & Multiple reasoning benchmarks & Same-architecture MoB; matched-data dense baseline; expert ablation & None \\
FPED, MoRE-Brain\cite{Ren2026,Wei2025} & Brain to image & Visual decoding & fMRI plus Yeo-7 / voxel-group experts & NSD reconstruction, not general omni capability & Fusion ablations (FPED); random-fMRI and bottleneck controls (MoRE) & None \\
Local intrinsic dimension, abstraction\cite{Yu2026,Cheng2026} & Encoding; brain tuning in Cheng et al. & Vision, language, speech & None (Yu); fMRI tuning (Cheng) & ImageNet associations; semantic/acoustic probes & Scale and architecture comparisons; random-feature controls & Dimensional geometry \\
This work & Bidirectional & Native omni & Yeo-7 partition and sparse-feature readout prior, frozen brain-learned subspace & Five public benchmarks and Brain-AVQA & 4-arm factorial; topology; joint permutation; matched ablation & Gain tracks headroom ($\rho = 0.80$) and low-dimensional inter-model differences (PR = 1.28) \\
\botrule
\end{tabular*}
{\par\vspace{3pt}\footnotesize\noindent Direction distinguishes encoding analyses from model interventions. Injected object: brain data means measured responses; organizational principle means a functional-network partition or topographic constraint. Validation lists the evaluated task, distinguishing public capability tests from brain reconstruction and observational associations. In grouped rows, semicolon-separated descriptions follow citation order unless explicitly labelled. Moussa et al. use stimulus-tuned controls for brain alignment and LLM-tuned controls for speech probes, including phoneme and phonetic-sentence-type classification; Policzer et al. report improvement on MUStARD but not on CMU-MOSEI. Yu et al. analyse ImageNet performance associations, whereas Cheng et al. also fine-tune WavLM on fMRI and test semantic/acoustic probes. Listed controls describe the published experiments and do not imply that they implement the exact factorial design used here.\par}
\end{sitable}

\begin{sitable}
\caption{Reliability of the Brain-AVQA dominant-network label}\label{edt:10}
\footnotesize
\begin{tabular*}{\textwidth}{@{\extracolsep{\fill}}lrr@{}}
\toprule
Quantity & Agreement & Cohen $\kappa$ \\
\midrule
Two disjoint participant pairs, split 1 & 0.3510 & 0.2397 \\
Two disjoint participant pairs, split 2 & 0.3476 & 0.2359 \\
Two disjoint participant pairs, split 3 & 0.3616 & 0.2521 \\
Mean over the three splits & 0.3534 & 0.2426 \\
Three participants against all four (leave-one-out) & 0.7593 & 0.7179 \\
Single participant pairs & 0.2853 & 0.1629 \\
All four participants identical & 0.0639 & \\
Chance level, seven networks & 0.1429 & \\
Permutation null, 1,000 relabellings & 0.1464 & \\
\botrule
\end{tabular*}
{\par\vspace{3pt}\footnotesize\noindent Labels were recomputed for each participant subset from the measured responses and compared window by window ($n = 2{,}362$). The bootstrap interval of the observed mean agreement (lower bound 0.328 to 0.343 over the three splits, 2,000 resamples) does not overlap the permutation null interval (upper bound 0.160 to 0.161), giving $z = 28.3$ to $29.4$. The released labels and all values above use a haemodynamic lag of three TRs (repetition times); mean split-half agreement is 0.3565, 0.3534 and 0.3458 at lags of two, three and four TRs. Reliability is therefore a group-level property of the four-participant mean and not a property of any single participant.\par}
\end{sitable}

\begin{sitable}
\caption{Dominant-network label distribution before and after network-level standardization}\label{edt:11}
\footnotesize
\begin{tabular*}{\textwidth}{@{\extracolsep{\fill}}lrrrrrrrr@{}}
\toprule
Stage & Vis & SomMot & DorsAttn & SalVentAttn & Limbic & Cont & Default & Ratio \\
\midrule
Before standardization & 595 & 423 & 258 & 212 & 59 & 313 & 502 & 10.1 \\
After standardization (released) & 374 & 369 & 288 & 249 & 323 & 322 & 437 & 1.76 \\
\botrule
\end{tabular*}
{\par\vspace{3pt}\footnotesize\noindent Counts over the 2,362 released windows. Standardization is a $z$-score of each network across the windows of the same episode half, so it removes amplitude differences between networks by construction. Ratio is the largest count divided by the smallest. The near-uniform released distribution is therefore an arithmetic consequence of that step and is not evidence about how strongly each network engages during passive viewing.\par}
\end{sitable}

\begin{sitable}
\caption{Six testable propositions of the Platonic brain bridge hypothesis}\label{sedtprop}
\footnotesize
\setlength{\tabcolsep}{3pt}
\begin{tabular*}{\textwidth}{@{\extracolsep{\fill}}p{1.5cm}p{4.5cm}p{3.2cm}p{3.1cm}@{}}
\toprule
Proposition & Formal statement & Test & Main result \\
\midrule
H1 Existence & $\bar r(M)>0$ for all seven models, with model ranking consistent across participants; a separate encoding system predicts responses to films without released response labels & Seven-model spectrum; rank correlation across four independently fitted participants; Algonauts 2025 out-of-distribution track, post-challenge leaderboard (Methods) & $\bar r$ = 0.337 to 0.381; Spearman $\rho = 0.958$; first by mean score among 27 post-challenge entries (10 November 2026) \\
H2 Structure & Modality: $\bar r_A>\bar r_V>\bar r_T$ and the increment of V+A over V is smaller than that of A over T; depth: peak layer $\neq$ last layer; carrier: cross-model matching of brain-aligned features exceeds controls & Modality ablation; 7 $\times$ 4 single-layer fits; Hungarian matching of SAE features with exact circular-shift permutation & Consistent in all three models; 0/7 peak at last layer; $|r|$ = 0.51 to 0.53, $P$ = 1/12,045 \\
H3 Injectability & A mapping from functional networks to expert modules and a brain prior incorporate network organization into forward computation and gating & Brain-MoE construction; two-stage training; SAE brain readout head & Injectable architecture (main-text Results) \\
H4 Effectiveness & $\Delta_1=\acc(\mathrm{MoE})-\acc(\mathrm{base})>0$ on held-out questions, generalizing across bases and benchmarks & Held-out, hard and out-of-distribution splits; 3 base $\times$ 5 benchmark matrix; per-question McNemar tests & +3.24 percentage points (Brain-AVQA held-out questions, MiniCPM-o 4.5); 15/15 positive \\
H5 Carrier & $\Delta_2=\acc(\mathrm{brain})-\acc(\mathrm{random})>0$; positive content main effect in four-arm design; real topology exceeds the rotated and uniform control topologies, which are indistinguishable & Question-level joint permutation; four-arm factorial; topology permutation on three bases and 20 random maps on MiniCPM-o 4.5; routing hit rate on held-out episodes & 14/15, $P = 3.2\times10^{-4}$; 14/14; 3/3 versus uniform and 2/3 versus rotated, and above all 20 random maps (in-domain); $\Delta_2>0$ in 12/12 seed runs \\
H6 Headroom & $\Delta_2$ correlates positively with base headroom $1-\acc(\mathrm{base})$, and inter-model differences are low-dimensional & Effective dimension of inter-model differences; rank correlation of cell-level effects with headroom & PR = 1.28; $\rho = 0.80$, $P = 3.0\times10^{-4}$ \\
\botrule
\end{tabular*}
{\par\vspace{3pt}\footnotesize\noindent $\bar r_{\mathcal{S}}$ is the brain-likeness with only modality subset $\mathcal{S}$ as input; $\Delta_1$ and $\Delta_2$ are the two effect sizes, measuring method effectiveness and the gain over capacity-matched random experts, respectively. SAE, sparse autoencoder; PR, participation ratio.}
\end{sitable}

\begin{sitable}
\caption{Single-layer brain-likeness of seven models at four relative depths}\label{sedtdepth}
\footnotesize
\begin{tabular*}{\textwidth}{@{\extracolsep{\fill}}lrrrrrr@{}}
\toprule
Model (layers) & 0.25 & 0.50 & 0.75 & 1.00 & Four layers & Best/four \\
\midrule
Ming-flash-omni-2.0 (32) & 0.3594 & \textbf{0.3750} & 0.3737 & 0.3677 & 0.3807 & 0.985 \\
MiniCPM-o 4.5 (36) & 0.3256 & \textbf{0.3582} & 0.3484 & 0.3276 & 0.3609 & 0.993 \\
Nemotron-3-Nano-Omni (52) & \textbf{0.3463} & 0.3409 & 0.3369 & 0.3288 & 0.3546 & 0.976 \\
Qwen2.5-Omni 3B (36) & 0.3164 & \textbf{0.3322} & 0.3286 & 0.3171 & 0.3367 & 0.987 \\
Qwen2.5-Omni 7B (28) & 0.3264 & \textbf{0.3338} & 0.3336 & 0.3162 & 0.3397 & 0.983 \\
Qwen3-Omni Instruct (48) & 0.3326 & \textbf{0.3422} & 0.3390 & 0.3231 & 0.3470 & 0.986 \\
Qwen3-Omni Thinking (48) & 0.3347 & \textbf{0.3441} & 0.3382 & 0.3227 & 0.3470 & 0.992 \\
\botrule
\end{tabular*}
{\par\vspace{3pt}\footnotesize\noindent Bold, peak; the last column is the ratio of the best single layer to the four-layer concatenation of the main protocol. For Qwen2.5-Omni 3B the 0.75 column is block 28 of 36 (relative depth 0.78).}
\end{sitable}

\begin{sitable}
\caption{Three-arm accuracies and paired comparisons of the in-domain topology control}\label{sedttopo}
\footnotesize
\setlength{\tabcolsep}{4pt}
\begin{tabular*}{\textwidth}{@{\extracolsep{\fill}}lrrr@{}}
\toprule
 & MiniCPM-o 4.5 & Qwen3-Omni & Ming-flash-omni-2.0 \\
\midrule
Real & 0.5318 & 0.7574 & 0.6359 \\
Rotated & 0.5169 & 0.7362 & 0.6287 \\
Uniform & 0.5127 & 0.7460 & 0.6232 \\
Real $-$ uniform (pp) & +1.91 & +1.14 & +1.27 \\
$P$; $n_{10}/n_{01}$ & $3.1\times10^{-4}$; 98/53 & 0.038; 92/65 & 0.0089; 77/47 \\
Real $-$ rotated (pp) & +1.48 & +2.12 & +0.72 \\
$P$; $n_{10}/n_{01}$ & 0.0055; 93/58 & $5.0\times10^{-4}$; 125/75 & 0.17; 78/61 \\
Rotated $-$ uniform $P$ & 0.50 & 0.079 & 0.30 \\
\botrule
\end{tabular*}
{\par\vspace{3pt}\footnotesize\noindent All three arms assemble the same stage-one brain-trained experts on all 2,362 questions and complete routing training under the real, rotated or uniform operator; $P$ is the two-sided per-question paired exact McNemar test and $n_{10}/n_{01}$ the discordant pairs (real correct and control wrong / real wrong and control correct).}
\end{sitable}

\begin{sitable}
\caption{Question counts, subset composition and held-out sizes of the five public benchmarks}\label{sedtbench}
\footnotesize
\setlength{\tabcolsep}{3pt}
\begin{tabular*}{\textwidth}{@{\extracolsep{\fill}}lp{2.3cm}p{1.6cm}rrrp{1.5cm}r@{}}
\toprule
Benchmark & Task & Modality & Source & Used & Coverage & Subset origin & Held-out $n$ \\
\midrule
OmniBench & General omni reasoning & Audio + image + text & 1,142 & 1,142 & 100.0\% & No filtering & 571 \\
MMAU test-mini & Audio understanding & Audio & 1,000 & 980 & 98.0\% & Question format & 490 \\
MMAU-Pro & Audio understanding (extended difficulty) & Audio & 5,305 & 4,405 & 83.0\% & MCQ eligibility & 2,203 \\
MuChoMusic & Music understanding & Audio & 1,187 & 851 & 71.7\% & Media availability & 425 \\
MUSIC-AVQA v2.0 & Joint audiovisual question answering & Audio + video & 10,819 & 350 & 3.2\% & Distractors and media & 179 \\
\botrule
\end{tabular*}
{\par\vspace{3pt}\footnotesize\noindent Sources: OmniBench\cite{OmniBench2024}, MMAU\cite{MMAU2024}, MMAU-Pro\cite{MMAUPro2025}, MuChoMusic\cite{MuChoMusic2024}, MUSIC-AVQA v2.0\cite{MusicAVQA2022,MusicAVQA2_2024}. Source denotes the referenced release or split: 1,000 questions for MMAU test-mini, all 5,305 mixed-format MMAU-Pro questions and the 10,819-question balanced MUSIC-AVQA v2.0 test split. Coverage is used divided by source questions. Question format denotes option/answer eligibility checks; MMAU-Pro additionally requires two to four options. MUSIC-AVQA items require three eligible distractors from the same question-type answer pool and available audiovisual media. These filters can change task composition; coverage does not denote content-independent sampling. All comparisons are paired on identical held-out questions.}
\end{sitable}

\begin{sitable}
\caption{Held-out results and two effect sizes on three bases $\times$ five public benchmarks}\label{sedtmatrix}
\footnotesize\setlength{\tabcolsep}{3pt}
\begin{tabular*}{\textwidth}{@{\extracolsep{\fill}}llrrrrrrr@{}}
\toprule
Base & Benchmark & $n$ & base & Brain $\Delta_1$ (pp) & Random $\Delta_1$ (pp) & $\Delta_2$ (pp) & $n_{10}/n_{01}$ & $P$ \\
\midrule
Ming & MMAU & 490 & 0.6551 & +2.86 & +0.20 & +2.65 & 20/7 & 0.019 \\
Ming & MMAU-Pro & 2,203 & 0.5679 & +9.08 & +1.68 & +7.40 & 260/97 & $2.5\times10^{-18}$$^{\dagger}$ \\
Ming & MuChoMusic & 425 & 0.7529 & +2.59 & +0.24 & +2.35 & 21/11 & 0.11 \\
Ming & MUSIC-AVQA v2.0 & 179 & 0.7542 & +2.79 & +1.12 & +1.68 & 3/0 & 0.25 \\
Ming & OmniBench & 571 & 0.5377 & +10.51 & +5.25 & +5.25 & 67/37 & 0.0042$^{\dagger}$ \\
MiniCPM & MMAU & 490 & 0.6918 & +5.31 & +2.24 & +3.06 & 28/13 & 0.028 \\
MiniCPM & MMAU-Pro & 2,203 & 0.5565 & +11.98 & +6.67 & +5.31 & 235/118 & $4.7\times10^{-10}$$^{\dagger}$ \\
MiniCPM & MuChoMusic & 425 & 0.7906 & +2.59 & +0.94 & +1.65 & 17/10 & 0.25 \\
MiniCPM & MUSIC-AVQA v2.0 & 179 & 0.8771 & +0.56 & +0.56 & 0.00 & 2/2 & 1.00 \\
MiniCPM & OmniBench & 571 & 0.5482 & +11.73 & +4.90 & +6.83 & 55/16 & $3.8\times10^{-6}$$^{\dagger}$ \\
Qwen3 & MMAU & 490 & 0.6490 & +8.37 & +3.88 & +4.49 & 28/6 & $2.0\times10^{-4}$$^{\dagger}$ \\
Qwen3 & MMAU-Pro & 2,203 & 0.5824 & +11.85 & +8.76 & +3.09 & 119/51 & $1.9\times10^{-7}$$^{\dagger}$ \\
Qwen3 & MuChoMusic & 425 & 0.7835 & +6.82 & +4.71 & +2.12 & 13/4 & 0.049 \\
Qwen3 & MUSIC-AVQA v2.0 & 179 & 0.8939 & +3.35 & +1.68 & +1.68 & 3/0 & 0.25 \\
Qwen3 & OmniBench & 571 & 0.6130 & +5.95 & +4.55 & +1.40 & 18/10 & 0.18 \\
\botrule
\end{tabular*}
{\par\vspace{3pt}\footnotesize\noindent One cell per row; $n$, held-out questions; base, frozen-base accuracy (headroom is $1-$ base); brain and random columns give the gain of each arm $\Delta_1$ over the base in percentage points (pp); $\Delta_2$, brain minus random (pp); $n_{10}/n_{01}$, questions answered correctly only by the brain arm and only by the random arm; $P$, two-sided paired exact McNemar test of brain versus random; $\dagger$, Holm-corrected $P<0.05$ within the family of 15 cells.}
\end{sitable}

\begin{sitable}
\caption{Four-arm held-out accuracies and factorial effects of seven complete $2\times2$ cells and one cell run with two splits}\label{sedtfourarm}
\footnotesize\setlength{\tabcolsep}{2pt}
\begin{tabular*}{\textwidth}{@{\extracolsep{\fill}}lrrrrrrrrrr@{}}
\toprule
Cell & $n$ & BT & BS & RT & RS & Content & Content & Prior & Prior & $\gamma$ \\
 & & & & & & $\mid$real & $\mid$shuffled & $\mid$brain & $\mid$random & \\
\midrule
MiniCPM $\times$ OmniBench & 571 & 0.6655 & 0.6427 & 0.5937 & 0.5972 & +0.0718 & +0.0455 & +0.0228 & $-0.0035$ & +0.0263 \\
MiniCPM $\times$ MMAU & 490 & 0.7449 & 0.7347 & 0.7082 & 0.7143 & +0.0367 & +0.0204 & +0.0102 & $-0.0061$ & +0.0163 \\
Ming $\times$ MMAU & 490 & 0.6837 & 0.6776 & 0.6592 & 0.6571 & +0.0245 & +0.0204 & +0.0061 & +0.0020 & +0.0041 \\
Qwen3 $\times$ MMAU & 490 & 0.7327 & 0.7286 & 0.6980 & 0.6878 & +0.0347 & +0.0408 & +0.0041 & +0.0102 & $-0.0061$ \\
Qwen3 $\times$ OmniBench & 571 & 0.6725 & 0.6778 & 0.6585 & 0.6585 & +0.0140 & +0.0193 & $-0.0053$ & 0.0000 & $-0.0053$ \\
MiniCPM $\times$ MMAU-Pro & 2,203 & 0.6764 & 0.6827 & 0.6296 & 0.6232 & +0.0468 & +0.0595 & $-0.0064$ & +0.0064 & $-0.0127$ \\
Qwen3 $\times$ MMAU-Pro & 2,203 & 0.7009 & 0.7022 & 0.6732 & 0.6700 & +0.0277 & +0.0322 & $-0.0014$ & +0.0032 & $-0.0045$ \\
\midrule
Ming $\times$ OmniBench$^{\dagger}$ & 474 & 0.6793 & 0.6646 & 0.6139 & 0.6181 & +0.0654 & +0.0464 & +0.0148 & $-0.0042$ & +0.0190 \\
\botrule
\end{tabular*}
{\par\vspace{3pt}\footnotesize\noindent BT, BS, RT and RS denote (brain experts, real prior), (brain experts, shuffled prior), (random experts, real prior) and (random experts, shuffled prior); Content$\mid$real = BT $-$ RT, Content$\mid$shuffled = BS $-$ RS, Prior$\mid$brain = BT $-$ BS, Prior$\mid$random = RT $-$ RS, and $\gamma$ is the interaction of the saturated two-factor model; the four arms of a cell share identical held-out questions and base predictions. $\dagger$, in this cell the BT and RS runs and the BS and RT runs used two different media-key splits (571 held-out questions each); values are computed on the 474 questions held out in both splits, on which base predictions are identical, so the arms were not trained on identical training sets and the cell is not counted among the seven complete cells.}
\end{sitable}

\clearpage
\begin{sitable}
\caption{Leave-one-film-out validation of the network-organized expert pool on Movie10}\label{sedtpool}
\footnotesize
\begin{tabular*}{\textwidth}{@{\extracolsep{\fill}}llrrrr@{}}
\toprule
Participant & Held-out film & Routed & Equal-weight & Primary & Routed $-$ equal \\
\midrule
sub-01 & Bourne & 0.1929 & 0.1882 & 0.1668 & +0.0048 \\
sub-01 & Figures & 0.2619 & 0.2566 & 0.2354 & +0.0053 \\
sub-01 & Life & 0.2184 & 0.2088 & 0.1794 & +0.0096 \\
sub-01 & Wolf & 0.2146 & 0.2088 & 0.1962 & +0.0057 \\
sub-01 & mean & 0.2219 & 0.2156 & 0.1945 & +0.0064 \\
\midrule
sub-02 & Bourne & 0.1530 & 0.1473 & 0.1332 & +0.0058 \\
sub-02 & Figures & 0.2462 & 0.2394 & 0.2174 & +0.0068 \\
sub-02 & Life & 0.2026 & 0.1935 & 0.1666 & +0.0092 \\
sub-02 & Wolf & 0.1963 & 0.1904 & 0.1775 & +0.0059 \\
sub-02 & mean & 0.1995 & 0.1926 & 0.1737 & +0.0069 \\
\midrule
sub-03 & Bourne & 0.1936 & 0.1830 & 0.1705 & +0.0106 \\
sub-03 & Figures & 0.2516 & 0.2401 & 0.2260 & +0.0116 \\
sub-03 & Life & 0.2259 & 0.2132 & 0.1841 & +0.0127 \\
sub-03 & Wolf & 0.2092 & 0.1992 & 0.1868 & +0.0100 \\
sub-03 & mean & 0.2201 & 0.2089 & 0.1918 & +0.0112 \\
\midrule
sub-05 & Bourne & 0.1475 & 0.1417 & 0.1309 & +0.0058 \\
sub-05 & Figures & 0.2049 & 0.1996 & 0.1903 & +0.0054 \\
sub-05 & Life & 0.1635 & 0.1554 & 0.1354 & +0.0081 \\
sub-05 & Wolf & 0.1771 & 0.1721 & 0.1656 & +0.0050 \\
sub-05 & mean & 0.1732 & 0.1672 & 0.1555 & +0.0061 \\
\botrule
\end{tabular*}
{\par\vspace{3pt}\footnotesize\noindent Parcel-mean Pearson correlation on the held-out film (1,000 parcels). Routed, experts mixed with weights organized by hemisphere and Yeo-7 network; equal-weight, mean over model families; primary, a single expert. All settings were chosen on the other three films, and no out-of-distribution response was used (Methods). Differences were computed before rounding. Film names are the Movie10 film keys.\par}
\end{sitable}

\section*{Supplementary Data}

\noindent\textbf{Supplementary Data 1 | Official Algonauts 2025 out-of-distribution leaderboard snapshot, source data and figure-update package.}
Provided as a separate archive (\texttt{Supplementary\_Data\_1\_Leaderboard\_20261110.zip}). The archive contains the raw Codabench API response (\texttt{codabench\_leaderboard\_raw.json}), retrieval and audit metadata (\texttt{implementation\_codabench\_metadata.json} and \texttt{leaderboard\_figure\_numeric\_qa.json}), normalized 27-submission leaderboard tables (\texttt{leaderboard\_normalized.csv} and \texttt{leaderboard\_normalized.json}), replacement vector panels (\texttt{updated\_panels/}), immutable pre-update baseline figures (\texttt{baselines/}), the standalone figure-regeneration and numerical-QA script (\texttt{update\_leaderboard\_figures\_20261110.py}), and documented final-stage replay code and verification records (\texttt{leaderboard\_reproducibility/}).
\vspace{6pt}

\clearpage
\begin{multicols}{2}
\bibliography{refs}
\end{multicols}